\documentclass{article}
\usepackage{amssymb,epsf,rotate}

\newcommand{\be}{\begin{equation}}
\newcommand{\ee}{\end{equation}}
\newcommand{\bea}{\begin{eqnarray}}
\newcommand{\eea}{\end{eqnarray}}
\newcommand{\dekdp}{KD$_2$PO$_4$}
\newcommand{\kdp}{KH$_2$PO$_4$}

\newcommand{\e}{\varepsilon_6}
\newcommand{\eps}{\varepsilon}
\newcommand{\s}{\sigma_6}

\newcommand{\et}{\eta^{(1)}}
\newcommand{\dd}{\delta}
\newcommand{\ds}{\displaystyle}

\newcommand{\btu}{\blacktriangle}
\newcommand{\eee}{\epsfxsize 2cm\epsffile}

\begin{document}
\begin{titlepage}

\title{
Influence of shear stress $\sigma_6$
on the phase transition and physical properties of KD$_2$PO$_4$
type ferroelectrics
}

\author{
I.V.Stasyuk, R.R.Levitskii, I.R.Zachek, A.P.Moina, A.S.Duda\\
\it Institute for Condensed Matter Physics, \\
\it 790011, 1 Svientsitskii St., Lviv, Ukraine\\
{\tt alla@icmp.lviv.ia}
}

\maketitle


\begin{abstract}
{\small
Within the four-particle cluster approximation for the proton ordering
model we study the influence of shear stress $\sigma_6$
on the phase transition, static dielectric, elastic and thermal properties
of deuterated KD$_2$PO$_4$ type ferroelectrics. Thermodynamic
potentials and physical characteristics of the crystals in the
presence of stress $\sigma_6$ are calculated. Numerical analysis of the
obtained results is performed, and the set of the theory parameters
providing the best fit to the available experimental data is found.
The $T_{\rm C}-\sigma_6$ phase diagram is constructed; temperature
dependences of the calculated quantities are studied, possible stress
dependences are indicated. Influence of $\sigma_6$ stress is analogous to
that of electric field conjugate to the order parameter.}

\end{abstract}
\end{titlepage}

\renewcommand{\theequation}{\arabic{section}.\arabic{equation}}

\sloppy
\section{Introduction}

More that sixty years have passed since the discovery of the
ferroelectricity in the  KH$_2$PO$_4$ crystals. During those
years, a huge number of papers has been devoted to the studies of
the phase transitions in the crystals of the KH$_2$PO$_4$ family and of
their physical properties. The most prominent peculiarity of these studies
is a strong connection between theory and experiment, which is thought to
be an important source of the obtained progress in microscopic
understanding of their properties. In this aspect, the high pressure
studies of these crystals come extremely important. External
pressure changes the internal structure of the system, altering thereby
the molecular potentials in the crystal and its dielectric and thermal
properties. And shear stresses give rise to shear piezoelectric lattice
strains induced also by external electric fields via the
piezoelectric effect; that allows one to consistently explore the
electric, electromechanic and thermal characteristics of the crystals.

A  microscopic model of strained KH$_2$PO$_4$ type crystals has been
proposed in \cite{a1,a2,a3}. According to this model, external  mechanical
stresses give rise to additional internal fields, linear in strains and
(in the case of diagonal components of stress tensor) mean values of
quasispins.  Influence of stresses of different symmetries on energies of
deuteron configurations was studied.

In \cite{a1,a2,a3} there were also explored the effects
of $\sigma_{12} = \sigma_{xx} - \sigma_{yy}$ stress on the transition
temperature to the ferroelectric phase and the relations between the
applied stress $\sigma_{xx} - \sigma_{yy}$ and the induced strain
$\varepsilon_{12} = \varepsilon_{xx} - \varepsilon_{yy}$ taking into
account the deuteron rearrangement  on hydrogen bonds. In \cite{a4,a5}
using the model proposed in \cite{a1,a2,a3}, within the cluster
approximation taking into account the short-range  and long-range
interactions and stress $\sigma_{12}$, the dielectric, piezoelectric, and
thermal characteristics of KD$_2$PO$_4$ were calculated and studied.
It has been shown that the stress $\sigma_{12}$ can lead to a phase
transition to the monoclinic phase.

Influence of hydrostatic pressure and uniaxial pressure
$-p = \sigma_3$ on the physical properties of the KD$_2$PO$_4$ type
crystals was investigated in \cite{a6,a7,a8}.  We showed that under the
proper choice of the theory parameters, a satisfactory description of
the experimental data for the pressure dependences of spontaneous
polarization, longitudinal static dielectric permittivity and transition
temperature was obtained. Using the Glauber model \cite{a9} the
expressions for the real and imaginary parts of the longitudinal dynamic
dielectric permittivity of KD$_2$PO$_4$ were found, and their temperature
and frequency dependences at different values of hydrostatic pressure were
calculated.

It is also interesting to study the influence of the shear stress $\s$
giving rise to a shear strain $\varepsilon_6 = \varepsilon_{xy}$,
which transforms after the irreducible representation $B_2$. After this
representation, polarization $P_3$ transforms too; hence,
the influence of external stress $\sigma_6 = \sigma_{xy}$ is similar to
that of electric field $E_3$.  Importance of this study results also from
the fact that spontaneous polarization $P_3$ in KD$_2$PO$_4$  is
accompanied by spontaneous shear strain $\varepsilon_6$.

In \cite{a10} the dielectric, piezoelectric, and elastic
properties of  KH$_2$PO$_4$ with taking into account the
strain $\varepsilon_6$ induced by the piezoelectric effect are studied
within the modified Slater model.

Experimental measurements of dielectric, electromechanic, and elastic
characteristics of K(H$_{1-x}$D$_x)_2$PO$_4$ were performed in
several works. Thus, in \cite{a11} the temperature dependences of the
dielectric permittivities of free and clamped KH$_2$PO$_4$ crystal,
piezoelectric constants $d_{36}$, $e_{36}$,
$h_{36}$ and elastic constants $c^P_{66}$, $c^E_{66}$ and
$s^P_{66}$, $s^E_{66}$ were  reported.  Temperature dependences of the
piezoelectric module $d_{36}$ of the KH$_2$PO$_4$ crystal at direct and
inverse  piezoelectric effect are given in \cite{a12} and \cite{a13},
respectively.  Temperature dependence of elastic constant $c^E_{66}$ for
KH$_2$PO$_4$ is presented in \cite{a14}. Experimental data for
the temperature dependence of   $d_{36}$, $\varepsilon_{33}$ and
$s^E_{66}$ for a partially deuterated crystal with $x = 0.89$ are
given in \cite{a15,a16}.

In this paper we study the influence of the shear stress $\sigma_6$
on the phase transition and  physical properties of deuterated
ferroelectric crystals of the KH$_2$PO$_4$ family.  We shall also
compare the obtained results for thermodynamic, dielectric, elastic, and
piezoelectric characteristics of the crystals with the
corresponding experimental data.

\section{The crystal Hamiltonian}
\renewcommand{\theequation}{\arabic{section}.\arabic{equation}}
\setcounter{equation}{0}

We consider a system of deuterons moving on  Ž--D-$\dots$-Ž bonds
in deuterated crystals of the \dekdp\ type. The
primitive cell of  such a crystal is composed of
two neighbouring PO$_4$ tetrahedra together with four hydrogen bonds
attached to one of them ($"A"$ type tetrahedra). Hydrogen bonds going to
another ($"B"$ type) tetrahedron belong to four  nearest structural
elements surrounding it (see the figure below).

\begin{figure}[hbt]
\begin{center}
\leavevmode
\epsfxsize=6.5cm
\epsffile{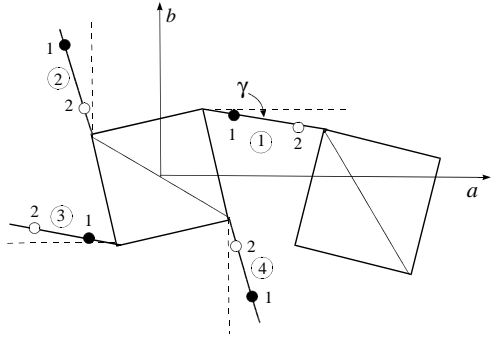}
\end{center}
\caption{\small
\small Primitive cell of the \dekdp\ crystal.
Numbers in circles correspond to hydrogen bonds; 1,2 are the deuteron
equilibrium sites. By dashed lines we show the directions of hydrogen
bonds in an unstrained crystal in the paraelectric phase. $a$,
$b$ are the axes of an undeformed $I\bar42d$ cell;
$\gamma$ is an angle between the directions of hydrogen bonds in strained
and unstrained crystals.}
\end{figure}

Hamiltonian of a deuteron subsystem of the
\dekdp\ type crystals taking into account the short-range and long-range
interactions in the presence of the shear stress
$\s=\sigma_{xy}$ giving rise to the strain $\e=\varepsilon_{xy}$ and of
the external electric fields
$E_i(i=1,2,3)$ applied along the crystallographic axes
$(a,b,c)$ reads
\bea
&&
\hspace{-5ex}
\hat H_i=\frac{\bar vN}{2}c_{66}^{E0}\e^2-\bar vNe_{36}^{0}E_{3}\e-
\frac{\bar vN}{2}\chi_{ii}^0E_i^2+{}\label{2.1}\\
&&
\hspace{-5ex}
\quad {} + \sum_{q'f'qf}J_{ff'}(qq')
\frac{\langle\sigma_{qf}\rangle}{2}\frac{\langle\sigma_{q'f'}\rangle}{2}-
\sum_{qf}[2\mu F_f^z(6)+\mu_{fi}E_i]\frac{\sigma_{qf}}{2}+
\hat H_{\rm short}. \nonumber
\eea
The first term in the right hand side of (\ref{2.1})  corresponds
to that part of the  elastic energy which does not depend on hydrogen
arrangement ($c_{66}^{E0}$ is the so called ``seed'' elastic constant); the
second term in (\ref{2.1}) is the energy of interaction between polarization
induced by  the piezoelectric effect  due to the strain
$\e$ (not taking into account the hydrogen subsystem contribution) and field
$E_3$ ($e_{36}^{0}$ is the seed coefficient of the piezoelectric stress); the
third term is the energy due to polarization induced by external field
independently of the hydrogen configuration ($\chi_{33}^0$ is the seed
dielectric susceptibility).  The last term in
(\ref{2.1}) describes the short range configurational interactions
between deuterons around the "A" and "B" type tetrahedra. Two eigenvalues
of Ising spin $\sigma_{qf}=\pm 1$ are assigned to two equilibrium
positions of a deuteron on the $f$-th bond in the $q$-th unit cell
($\sigma_{qf}=\pm 1$).
$\bar v=v/{k_{\rm B}}$, $v$ is the primitive cell volume; ${k_{\rm B}}$
is the Boltzmann constant.  $F_f^i$ are internal fields, arising  in
strained crystals, which include the effective long-range interaction
between deuterons (taken into account in the mean field approximation) and
 the additional internal fields related to the strain $\e$ \cite{a3}:
\bea
&& 2\mu F^z(6)=2\mu F_{1}^z(6)=-2\mu F_{2}^{z}(6)=-2\mu F_{3}^{z}(6)= 2\mu
F_{4}^{z}(6)=\nonumber\\ &&\quad
{}=2\nu_c\eta^{(1)z}(6)-2\psi_6\varepsilon_6,\nonumber\\ && 2\mu
F_{1}^{xy}(6)=2\nu_1\eta_1^{(1)x,y}(6)+2\nu_3\eta_3^{(1)x,y}(6) +
\nonumber\\
&&\quad {}
+2\nu_2[\eta_2^{(1)x,y}(6)+\eta_4^{(1)x,y}(6)]-2\psi_6\e,\nonumber\\
&& 2\mu
F_{2}^{xy}(6)=2\nu_2[\eta_1^{(1)x,y}(6)+\eta_3^{(1)x,y}(6)]+ \nonumber\\
&&
\label{2.2}
\quad {}+
2\nu_1\eta_2^{(1)x,y}(6)+2\nu_3\eta_4^{(1)x,y}(6)-2\psi_6\e,\\
&& 2\mu
F_{1}^{xy}(6)=2\nu_3\eta_1^{(1)x,y}(6)+2\nu_1\eta_3^{(1)x,y}(6)
+\nonumber\\
&&\quad {}
+2\nu_2[\eta_2^{(1)x,y}(6)+\eta_4^{(1)x,y}(6)]-2\psi_6\e,\nonumber\\
&& 2\mu
F_{4}^{xy}(6)=2\nu_2[\eta_1^{(1)x,y}(6)+\eta_3^{(1)x,y}(6)]+ \nonumber\\
&&\quad {}+
2\nu_3\eta_2^{(1)x,y}(6)+2\nu_1\eta_4^{(1)x,y}(6)-2\psi_6\e\nonumber,
\eea
where
\bea
&&
\!\!
\eta^{(1)z}(6)=\langle\sigma_{q1}\rangle^z=\langle\sigma_{q2}\rangle^z=
\langle\sigma_{q3}\rangle^z=\langle\sigma_{q4}\rangle^z, \quad\!\!
\eta_f^{(1)x,y}(6)=\langle\sigma_{qf}\rangle^{x,y};
\nonumber\\
&&\!\!
\nu_c=\nu_1+2\nu_2+\nu_3;
\nu_1=\frac{J_{11}}4, \nu_2=\frac{J_{12}}4, \nu_3=\frac{J_{13}}4;
\nonumber
\eea
$J_{ff'}=\sum_{{\bf R}_q-{\bf R}_{q'}}J_{ff'}(qq')$ is the Fourier transform
of the long-range deuteron-deuteron interaction, $\psi_6$ is the so called
deformation potential; $\mu=e\dd$ is the dipole moment of
a hydrogen bond; $\dd$ is the D-site distance.

In (\ref{2.1}) and (\ref{2.2}) we do not take into account piezoelectric
shear strains $\eps_4$ or $\eps_5$ induced by electric fields $E_1$ and
$E_2$, respectively.

The Hamiltonian is usually  chosen such
that to reproduce the energy levels of the Slater-type model for KDP (see,
for instance \cite{Blinc_Zeks}) -- the Slater energies $\eps$,
$w$, and  $w_1$ ($\eps\ll w\ll w_1$), determined by  the energies of
up-down $\eps_s$, lateral $\eps_a$, single-ionized $\eps_1$, and
 double-ionized $\eps_0$ deuteron configurations.
 \[
\varepsilon=\varepsilon_a-\varepsilon_s,\quad
w=\varepsilon_1-\varepsilon_s,\ \ w_1=\varepsilon_0-\varepsilon_s.
\]
  If $\e=0$,
configurations with two hydrogens in potential wells being close to upper
and lower oxygens of a given PO$_4$ group  (each PO$_4$ tetrahedron
is oriented such that two of its edges are parallel to the $ab$ plane) and
with the hydrogens on the two other bonds being close to the  neighboring
tetrahedra,  have the same energy $\eps_s$, assumed to be the lowest.
Correspondingly, lateral configurations with two hydrogens close to an
upper and a lower oxygens are four-fold degenerated; single-ionized with
only one (or three) hydrogens close to a given group are eight-fold
degenerated, and double-ionized with four hydrogens (or without any) are
twice degenerated.

However, the strain $\s$ splits certain deuteron configurations
of the conventional Slater-Takagi model. In Table below  we
present all possible deuteron configurations and their energies.
Since the system is not any longer symmetric with respect to a reflection
in the $ab$ plane $\sigma_h$ or to the reflection with $\pi/4$ rotation
around the $c$-axis $S_4$ (both operations change the sign of polarization
and strain $\e$), the up-down configurations ($i=1,2$) split to two
different levels, and lateral configurations split to two groups of twice
degenerated levels each ($i=5,6$) and ($i=7,8$). Configurations within each
group are symmetric with respect to $\pi/2$  rotation $C_2$. Similarly,
single-ionized configuration are divided into two groups, within each
the $c$-component of dipole moments assigned to a configuration are
directed up ($i=9,10,11,12$) or down ($i=13,14,15,16$).

Here we assume that the strain $\e$ alters the  energies of deuteron
configurations only by splitting the degenerated levels due to lowering of
the system symmetry. From geometric point of view, this happens mostly
because the angle between perpendicular in the paraelectric phase hydrogen
bonds is changed, and the PO$_4$ groups are distorted, whereas in the
hydrostatic pressure case the changes are usually attributed to
pressure-induced changes in the D-site distance $\delta$
\cite{a6,Blinc-pressure}.

\renewcommand{\arraystretch}{0.7}
\renewcommand{\tabcolsep}{0.2pt}
\begin{table}[p]
\begin{tabular}{|c|c|c|c|c|c|c|c|c|c|}
\hline
$i$ & & & $E_{i6}$ &$i$ & & & $E_{i6}$ \\
\hline
1& \eee{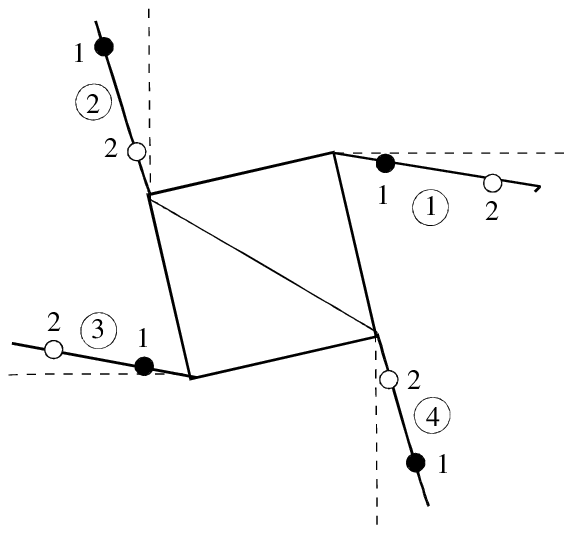} & $ ++++{}$ & $\eps_s-\delta_{s6}\eps_6$ &
9 & \eee{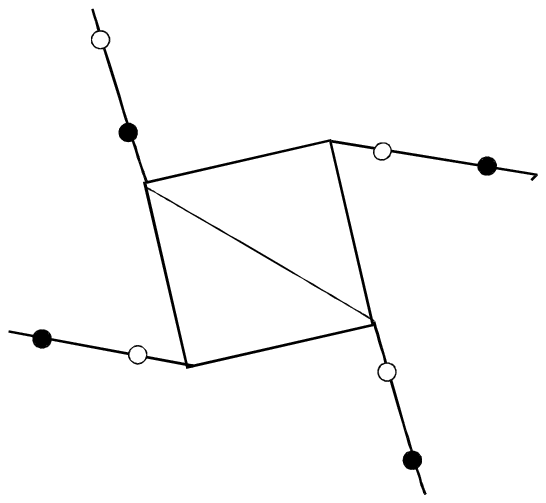}& $ ---+{}$ & $\eps_1-\delta_{16}\eps_6$ \\
2& \eee{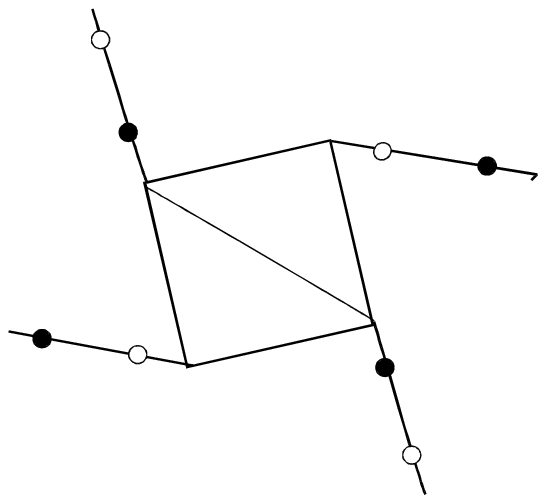} & $ ----{}$ & $\eps_s+\delta_{s6}\eps_6$ &
10& \eee{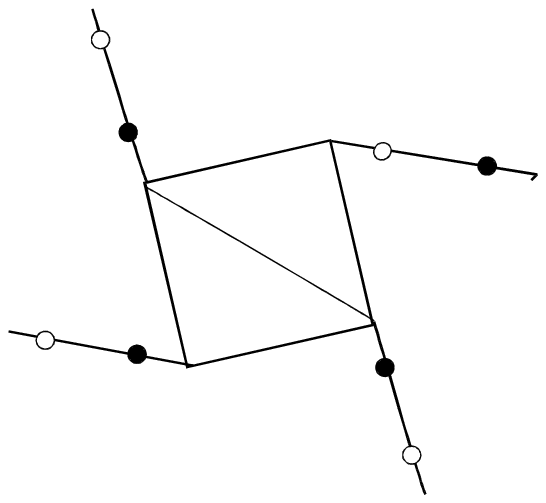}&$ --+-{}$ &
\\
\cline{1-4}
3& \eee{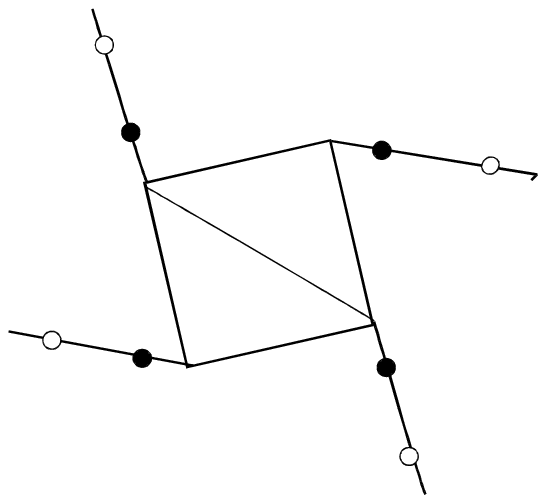} &$ +-+-{}$ & $\eps_0$ &
11& \eee{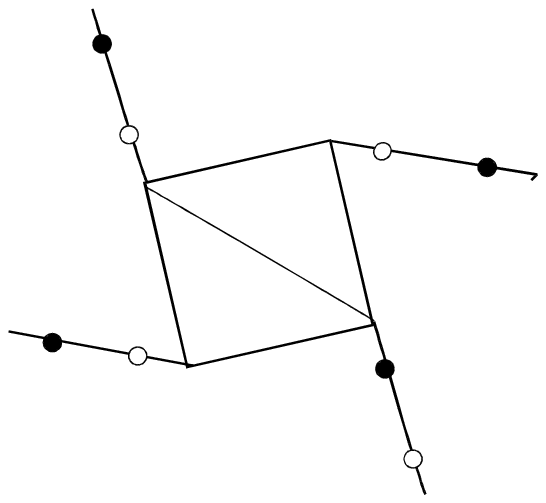}& $ -+--{}$ &\\
4& \eee{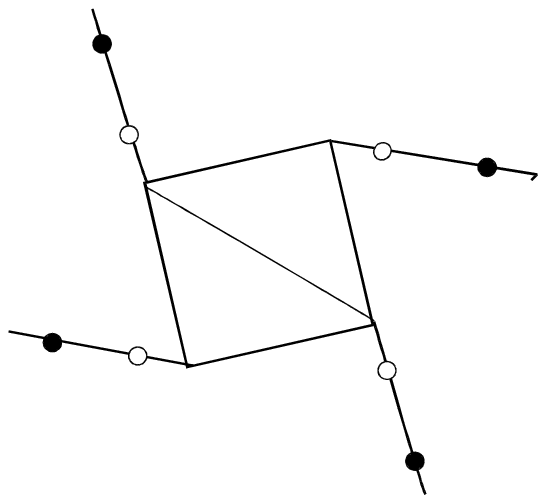} &$ -+-+{}$ & $\eps_0$ &
12& \eee{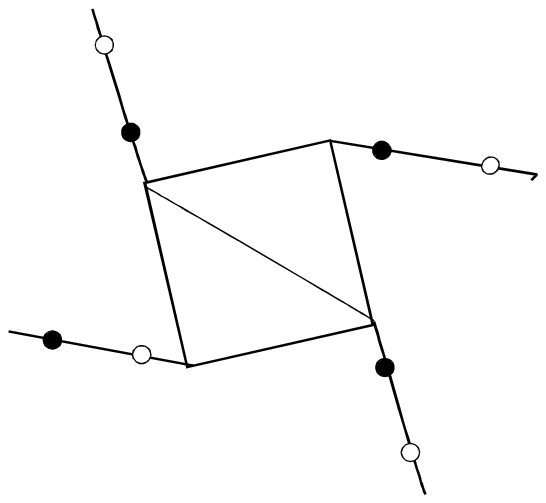} &$ +---{}$ & \\
\hline
5& \eee{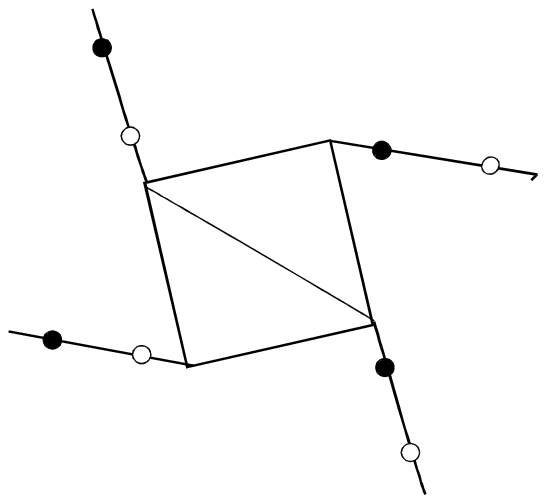} &$ ++--{}$ & $\eps_a+\delta_{a6}\eps_6$ &
13& \eee{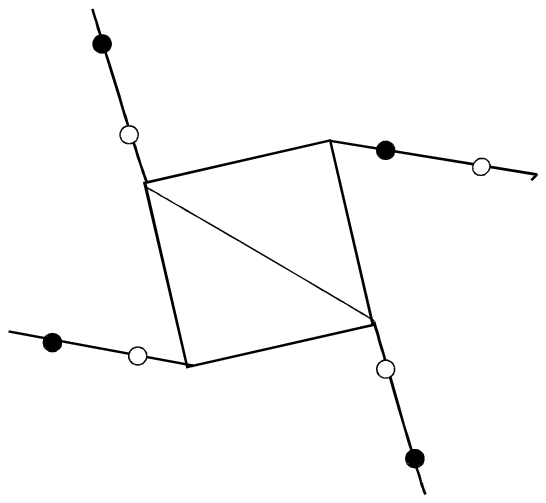}&$ ++-+{}$ & $\eps_1+\delta_{16}\eps_6$ \\
6& \eee{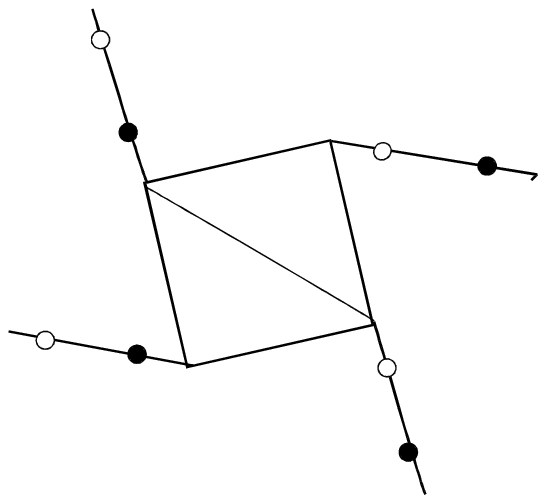} &$ --++{}$ & $\eps_a+\delta_{a6}\eps_6$ &
14& \eee{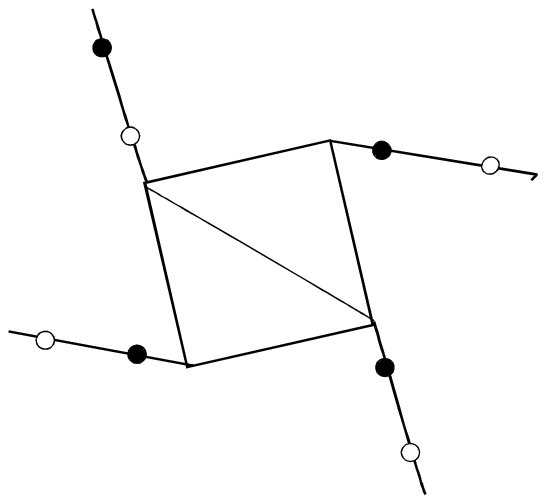} &$ +++-{}$ &\\
\cline{1-4}
7& \eee{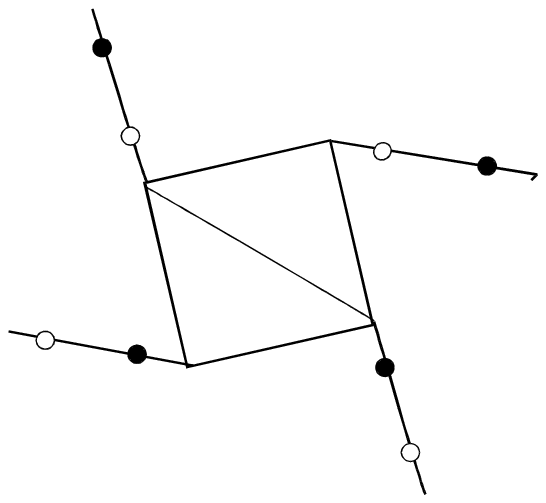} &$ -++-{}$ & $\eps_a-\delta_{a6}\eps_6$ &
15& \eee{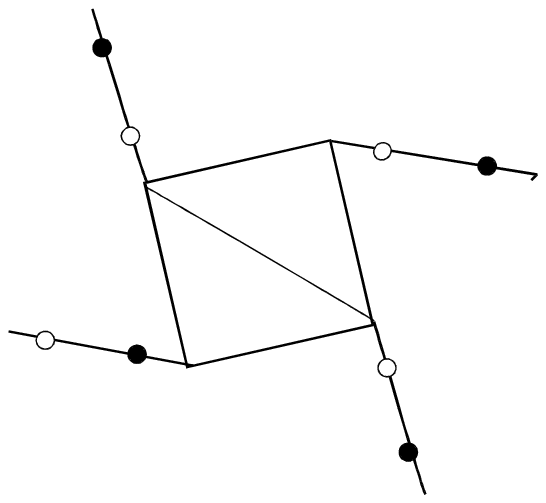}&$ -+++{}$ & \\
8& \eee{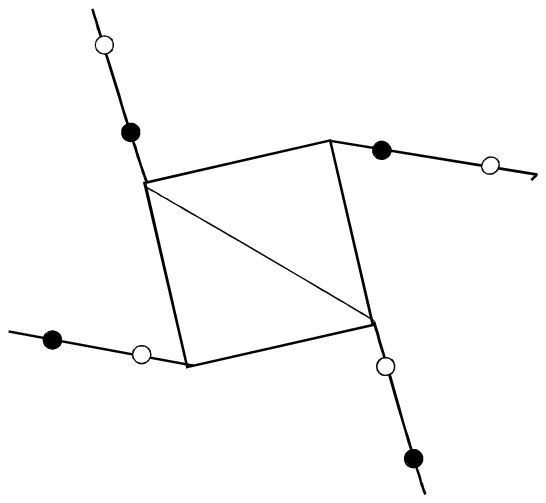} &$ +--+{}$ & $\eps_a-\delta_{a6}\eps_6$ &
16& \eee{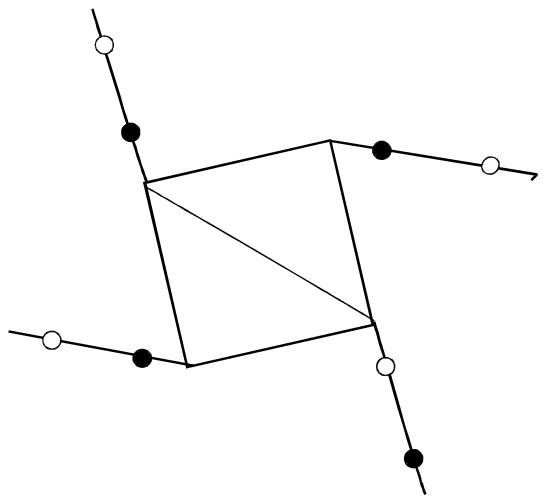} &$ +-++{}$ &\\
\hline
\end{tabular}
\label{splitting}
\end{table}
\renewcommand{\arraystretch}{1}
\renewcommand{\tabcolsep}{1pt}

To rewrite the energies of deuteron configurations $E_{i6}$ in
terms of pseudospins, we associate the  configuration operator $\hat N_i$
with the configuration $i$ according to the following rule: each operator
is a product of four factors, one per each hydrogen bond, each factor
being equal to $\frac{1}{2}(1+\sigma_{qf})$ if deuteron is in the first
minimum at the $f$-th bond and $\frac{1}{2}(1-\sigma_{qf})$ otherwise.
Then the Hamiltonian of the short-range  interactions takes the form
\bea
&&
 \hspace{-5ex}
 \lefteqn{\hat H_{\rm short}(6)=
\sum_{q=1}^{N}\sum_{i=1}^{16}[\hat N_i^A(q)+\hat N_i^B(q)]=}\\
&&
\hspace{-5ex}
{}= \sum_{q=1}^{N}\left\{
\frac \e4\left(-\delta_{s6}+2\delta_{16}\right)
\sum_{f=1}^4\frac{\sigma_{qf}}{2}+\nonumber\right.\\
&&
\hspace{-5ex}
-
\e (\delta_{s6}{+}2\delta_{16})\!\!
\left[\frac {\sigma_{q1}}2\frac {\sigma_{q2}}2\frac {\sigma_{q3}}2+
      \frac {\sigma_{q1}}2\frac {\sigma_{q2}}2\frac {\sigma_{q4}}2+
      \frac {\sigma_{q1}}2\frac {\sigma_{q3}}2\frac {\sigma_{q4}}2+
      \frac {\sigma_{q2}}2\frac {\sigma_{q3}}2\frac {\sigma_{q4}}2
\right]{+}\!\nonumber\\
&&
\hspace{-5ex}
{}+(V+\delta_6\varepsilon_6)\left [
\frac {\sigma_{q1}}2\frac {\sigma _{q2}}2+
\frac {\sigma_{q3}}2\frac {\sigma _{q4}}2\right]+
(V-\delta_6\varepsilon_6)\left[\frac {\sigma _{q2}}2\frac {\sigma
_{q3}}2+\frac {\sigma _{q4}}2\frac {\sigma _{q1}}2\right]+
\nonumber
\\
&&
\hspace{-5ex}\left.
{}+U\left [\frac {\sigma _{q1}}2\frac {\sigma _{q3}}2+
\frac {\sigma _{q2}}2\frac {\sigma _{q4}}2\right ]+
\label{2.4p}
 \Phi\frac {\sigma _{q1}}2\frac {\sigma _{q2}}2
\frac {\sigma _{q3}}2\frac {\sigma_{q4}}2\right\}+
\sum_{q=1}^{N}\sum_{i=1}^{16}\hat N_i^B(q)
.\nonumber
\eea
Here the following notations are used
\bea
&&V=V_{12}=V_{23}=V_{34}=V_{41}=-\frac12w_1,\nonumber\\
&&U=V_{13}=V_{24}=\frac12w_1-\varepsilon,\quad
\Phi=4\varepsilon+2w_1-8w,\nonumber\\
&&\varepsilon=\varepsilon_a-\varepsilon_s,\ \
w=\varepsilon_1-\varepsilon_s,\ \ w_1=\varepsilon_0-\varepsilon_s,\nonumber
\eea
where $\varepsilon_s$, $\varepsilon_a$, $\varepsilon_1$, $\varepsilon_0$
are energies of up-down, lateral, single-ionized and double-ionized
deuteron
configurations in an unstrained crystal, respectively.

The further calculations
  will be performedd in the four-particle cluster approximation.
The cluster Hamiltonian in the presence of the stress $\s$ and
electric field $E_i$ reads
  \bea
&&
\label{2.4}
\hspace{-5ex}\hat H^{iA}_{q4}(6)=
\frac e4\left(-\delta_{s6}+2\delta_{16}\right)
\sum_{f=1}^4\frac{\sigma_{qf}}{2}+\\
&&
\hspace{-5ex}
-
\e (\delta_{s6}{+}2\delta_{16})\!\!
\left[\frac {\sigma_{q1}}2\frac {\sigma_{q2}}2\frac {\sigma_{q3}}2+
      \frac {\sigma_{q1}}2\frac {\sigma_{q2}}2\frac {\sigma_{q4}}2+
      \frac {\sigma_{q1}}2\frac {\sigma_{q3}}2\frac {\sigma_{q4}}2+
      \frac {\sigma_{q2}}2\frac {\sigma_{q3}}2\frac {\sigma_{q4}}2
\right]{+}\!\nonumber\\
&&
\hspace{-5ex}
{}+(V+\delta_6\varepsilon_6)\left [
\frac {\sigma_{q1}}2\frac {\sigma _{q2}}2+
\frac {\sigma_{q3}}2\frac {\sigma _{q4}}2\right]+
(V-\delta_6\varepsilon_6)\left[\frac {\sigma _{q2}}2\frac {\sigma
_{q3}}2+\frac {\sigma _{q4}}2\frac {\sigma _{q1}}2\right]+
\nonumber
\\
&&
\hspace{-5ex}
{}+U\left [\frac {\sigma _{q1}}2\frac {\sigma _{q3}}2+
\frac {\sigma _{q2}}2\frac {\sigma _{q4}}2\right ]+
 \Phi\frac {\sigma _{q1}}2\frac {\sigma _{q2}}2
\frac {\sigma _{q3}}2\frac {\sigma_{q4}}2
-
\sum_f\frac{z_{f6}^i}\beta\frac{\sigma _{qf}}2.\nonumber
\eea
In
(\ref{2.4})
\begin{eqnarray}
&&z^z_6=z^z_{16}=z^z_{26}=z^z_{36}=z^z_{46}=\beta[-\Delta^z_6+2\mu
F^z(6)+\mu_3E_3];\nonumber\\
&&z^x_{16}=\beta [-\Delta^x_6+2\mu F_1^x(6)+\mu_\perp\cos\gamma E_1],
\nonumber\\
&&
z^x_{26}=\beta [-\Delta^x_6+2\mu F_2^x(6)-\mu_\perp\sin\gamma E_1],
\nonumber\\
&&z^x_{36}=\beta [-\Delta^x_6+2\mu F_3^x(6)-\mu_\perp\cos\gamma E_1],
\nonumber\\
&&
z^x_{46}=\beta [-\Delta^x_6+2\mu F_2^x(6)+\mu_\perp\sin\gamma E_1];\\
&&z^y_{16}=\beta [-\Delta^y_6+2\mu F_1^y(6)+\mu_\perp\sin\gamma E_2],
\nonumber\\
&& z^y_{26}=\beta [-\Delta^y_6+2\mu F_2^y(6)-\mu_\perp\cos\gamma E_2],
\nonumber\\
&&z^y_{36}=\beta [-\Delta^y_6+2\mu F_3^y(6)-\mu_\perp\sin\gamma E_2],
\nonumber\\
&& z^y_{46}=\beta [-\Delta^y_6+2\mu F_4^y(6)+\mu_\perp\cos\gamma E_2];
\nonumber
\end{eqnarray}
$\mu_3$ and $\mu_\perp$ are the longitudinal and transverse effective
dipole moments of primitive cells, created by displacements and
polarization of heavy ions which are triggered by deuteron ordering.

We took into account the following relations which are obeyed in the
presence of stress $\sigma_6$ and fields $E_i$
\bea
&&\hspace {-3.8ex}
\mu_{13} = \mu_{23} = \mu_{33} = \mu_{43} =
\mu_3,\\
&& \hspace {-3.8ex}\mu_{11} =  \mu_1\cos\gamma, \quad \mu_{21} = -
\mu_{\perp}\sin\gamma, \quad 
\mu_{12} = \mu_{\perp}\sin\gamma, \;\;
- \mu_{22} =  \mu_\perp\cos\gamma;\nonumber\\
&&\hspace {-3.8ex} \mu_{31} = - \mu_{\perp}\cos\gamma, \quad \mu_{41} =
\mu_{\perp}\sin\gamma, \;\;
\mu_{32} = - \mu_{\perp}\sin\gamma, \quad
\mu_{42} = \mu_\perp\cos\gamma. \nonumber
\eea

 Using the
condition
\be \langle\sigma_{qf}\rangle^i =
\frac{{\rm Sp}\{\sigma_{qf} {\rm e}^{-\beta\hat H_{q4}^{iA}(6)}\}}{{\rm
Sp}\{{\rm e}^{-\beta\hat H_{q4}^{iA}(6)}\}} = \frac{{\rm Sp}\{\sigma_{qf}
{\rm e}^{-\beta\hat H_{qf}^{i}(6)}\}}{{\rm Sp}\{{\rm e}^{-\beta\hat
H_{qf}^{i}(6)}\}},
\ee
where the single-particle deuteron Hamiltonians are
\be
\hat H^i_{qf}(6) = - \frac{{\bar z}^i_6}{\beta}\frac{\sigma_{qf}}{2},
\quad
{\bar z}^i_{f6} = - \beta\Delta^i_6 + z^i_{f6},
\ee
we calculate the single-particle distribution functions of
 deuterons and exclude the parameters $\Delta^i_6$
\be
\label{2.10}
\eta^{(1)z}(6) = \frac{m^z(6)}{D^z_6},
\ee
where
\bea
&&m^z(6) = \sinh (2z^z_6 +\beta\delta_{s6}\e)+
         2b \sinh (z^z_6  -\beta\delta_{16}\e),\quad\nonumber\\
&&
D^z_6 = \cosh (2z^z_6 +\beta\delta_{s6}\e) +
     4b \cosh (z^z_6 -\beta\delta_{16}\e)+ d + aa_6 +
a/a_6,\nonumber\\
&&z^z_6 = \frac12\ln\frac{1+\eta^{(1)z}(6)}{1-\eta^{(1)z}(6)} +
\beta\nu_c(6)
\eta^{(1)z}(6) - \beta\psi_6\varepsilon_6 + \frac{\beta\mu_3E_3}{2},
\\
&& a = \exp(-\beta\varepsilon), \; b = \exp(- \beta w), \;
d = \exp(- \beta w_1), \; a_6 = \exp(- \beta\delta_6\varepsilon_6),
\nonumber\eea
also
\be
\label{2.13}
\eta^{(1)\alpha}_f(6) = \frac{m^{\alpha}_f(6)}{D^{\alpha}(6)}, \qquad
(\alpha = x, y),
\ee
where
\begin{eqnarray}
&&
\hspace{-5ex}
m_{{}^1_3}^\alpha(6){=}\sinh(\frac{A_1^\alpha(6)}{2}+\beta\delta_{s6}\e)
{+}d\sinh\frac{A_2^\alpha(6)}{2}{\pm}
 \nonumber\\&&\quad\pm
 aa_6\sinh\frac{A_3^\alpha(6)}{2}{\pm}
\frac a{a_6}\sinh\frac{A_4^\alpha(6)}{2}+\nonumber\\
&&
\hspace{-5ex}
\quad{}
+b\left[\pm\sinh(\frac{A_5^\alpha(6)}{2}-\beta\delta_{16}\e)\mp
\sinh(\frac{A_6^\alpha(6)}{2}-\beta\delta_{16}\e)+
\right.\nonumber\\&&\quad \left.
+\sinh(\frac{A_7^\alpha(6)}{2}-\beta\delta_{16}\e)+
\sinh(\frac{A_8^\alpha(6)}{2}-\beta\delta_{16}\e)\right],\nonumber\\
&&
\hspace{-5ex}
  m_{{}^2_4}^\alpha(6){=}\sinh(\frac{A_1^\alpha(6)}{2}+\beta\delta_{s6}\e)
{-}d\sinh\frac{A_2^\alpha(6)}{2}\pm
\nonumber\\&&\quad\pm
aa_6
\sinh\frac{A_3^\alpha(6)}{2}{\mp}
\frac a{a_6}\sinh\frac{A_4^\alpha(6)}{2}+\nonumber\\
&&
\hspace{-5ex}
\quad{}
+b\left[\sinh(\frac{ A_5^\alpha(6)}{2}-\beta\delta_{16}\e)+
\sinh(\frac{ A_6^\alpha(6)}{2}-\beta\delta_{16}\e)\pm
\right.\nonumber\\&&\quad \left.\pm
\sinh(\frac{A_7^\alpha(6)}{2}-\beta\delta_{16}\e)\mp
\sinh(\frac{A_8^\alpha(6)}{2}-\beta\delta_{16}\e)\right]\\
&&
\hspace{-5ex}
D^\alpha(6){=}\cosh(\frac{A_1^\alpha(6)}{2}+\beta\delta_{s6}\e)
{+}d\sinh\frac{A_2^\alpha(6)}{2}{+}
\nonumber\\&&\quad+
aa_6\cosh\frac{A_3^\alpha(6)}{2}{+}
\frac a{a_6}\cosh\frac{A_4^\alpha(6)}{2}+\nonumber\\
&&\quad{}+b\left[\cosh(\frac{A_5^\alpha(6)}{2}-\beta\delta_{16}\e)+
\cosh(\frac{A_6^\alpha(6)}{2}-\beta\delta_{16}\e)+
\right.\nonumber\\&&\quad \left.
+\cosh(\frac{A_7^\alpha(6)}{2}-\beta\delta_{16}\e)+
\cosh(\frac{A_8^\alpha(6)}{2}-\beta\delta_{16}\e)\right] \nonumber
\end{eqnarray}
Here the following notations are used
\begin{eqnarray*}
&&A_{{}^1_2}^\alpha(6)=z_1^\alpha(6){\pm}z_2^\alpha(6)+
z_3^\alpha(6)\pm z_4^\alpha(6);\\
&& A_{{}^3_4}^\alpha(6)=z_1^\alpha(6)\pm z_2^\alpha(6)-
z_3^\alpha(6)\mp z_4^\alpha(6);\\
&&A_{{}^5_6}^\alpha(6)=\pm z_1^\alpha(6)+z_2^\alpha(6)\mp
z_3^\alpha(6)+z_4^\alpha(6);\\
&& A_{{}^7_8}^\alpha(6)=z_1^\alpha(6)\pm z_2^\alpha(6)+
z_3^\alpha(6)\mp z_4^\alpha(6);\nonumber
\end{eqnarray*}
and
\begin{eqnarray}
z^x_{16}&=&\frac12\ln{\frac{1+\eta_{1}^{(1)x}(6)}{1-\eta_{1}^{(1)x}(6)}}+
\beta \nu_1\eta_{1}^{(1)x}(6)+\beta \nu_3\eta_{3}^{(1)x}(6)+\nonumber\\
&+&\beta \nu_2\eta_{2}^{(1)x}(6)+\beta \nu_2\eta_{4}^{(1)x}(6)-
\beta\psi_6\varepsilon_6+\frac{\beta\mu_\perp\cos\gamma E_1}2,\nonumber\\
z^x_{36}&=&\frac12\ln{\frac{1+\eta_{3}^{(1)x}(6)}{1-\eta_{3}^{(1)x}(6)}}+
\beta \nu_3\eta_{1}^{(1)x}(6)+\beta \nu_1\eta_{3}^{(1)x}(6)+\nonumber\\
&+&\beta \nu_2\eta_{2}^{(1)x}(6)+\beta \nu_2\eta_{4}^{(1)x}(6)-
\beta\psi_6\varepsilon_6-\frac{\beta\mu_\perp\cos\gamma E_1}2,\\
z^x_{26}&=&\frac12\ln{\frac{1+\eta_{2}^{(1)x}(6)}{1-\eta_{2}^{(1)x}(6)}}+
\beta \nu_2\eta_{1}^{(1)x}(6)+\beta
\nu_2\eta_{3}^{(1)x}(6)+\nonumber\\
&+&\beta \nu_1\eta_{2}^{(1)x}(6)+\beta \nu_3\eta_{4}^{(1)x}(6)-
\beta\psi_6\varepsilon_6-\frac{\beta\mu_\perp\sin\gamma E_1}2\nonumber,\\
z^x_{46}&=&\frac12\ln{\frac{1+\eta_{4}^{(1)x}(6)}{1-\eta_{4}^{(1)x}(6)}}+
\beta \nu_2\eta_{1}^{(1)x}(6)+\beta \nu_2\eta_{3}^{(1)x}(6)+\nonumber\\
&+&\beta \nu_3\eta_{2}^{(1)x}(6)+\beta \nu_1\eta_{4}^{(1)x}(6)-
\beta\psi_6\varepsilon_6+\frac{\beta\mu_\perp\sin\gamma E_1}2\nonumber; \\
z^y_{16}&=&\frac12\ln{\frac{1+\eta_{1}^{(1)y}(6)}{1-\eta_{1}^{(1)y}(6)}}+
\beta \nu_1\eta_{1}^{(1)y}(6)+\beta \nu_3\eta_{3}^{(1)y}(6)+\nonumber\\
&+&\beta \nu_2\eta_{2}^{(1)y}(6)+\beta \nu_2\eta_{4}^{(1)y}(6)-
\beta\psi_6\varepsilon_6-\frac{\beta\mu_\perp\sin\gamma E_2}2,\nonumber\\\
\label{2.16}
z^y_{36}&=&\frac12\ln{\frac{1+\eta_{3}^{(1)y}(6)}{1-\eta_{3}^{(1)y}(6)}}+
\beta \nu_3\eta_{1}^{(1)y}(6)+\beta \nu_1\eta_{3}^{(1)y}(6)+\nonumber\\
&+&\beta \nu_2\eta_{2}^{(1)y}(6)+\beta\nu_2\eta_{4}^{(1)y}(6)-
\beta\psi_6\varepsilon_6+\frac{\beta\mu_\perp\sin\gamma E_2}2,\\
z^y_{26}&=&\frac12\ln{\frac{1+\eta_{2}^{(1)y}(6)}{1-\eta_{2}^{(1)y}(6)}}+
\beta \nu_2\eta_{1}^{(1)y}(6)+\beta
\nu_2\eta_{3}^{(1)y}(6)+\nonumber\\
&+&\beta \nu_1\eta_{2}^{(1)y}(6)+\beta  \nu_3\eta_{4}^{(1)y}(6)-
\beta\psi_6\varepsilon_6-\frac{\beta\mu_\perp\cos\gamma E_2}2,\nonumber\\
z^y_{46}&=&\frac12\ln{\frac{1+\eta_{4}^{(1)y}(6)}{1-\eta_{4}^{(1)y}(6)}}+
\beta \nu_2\eta_{1}^{(1)y}(6)+\beta
\nu_2\eta_{3}^{(1)y}(6)+\nonumber\\
&+&\beta \nu_3\eta_{2}^{(1)y}(6)+\beta
\nu_1\eta_{4}^{(1)y}(6)-
\beta\psi_6\varepsilon_6+\frac{\beta\mu_\perp\cos\gamma E_2}2.\nonumber
\end{eqnarray}

\section{
Elastic,  piezoelectric, and dielectric properties of the \dekdp\ type
crystals  under the mechanic stress $\sigma_6$}

Influence of the stress $\sigma_6$ on the characteristics
 of the \dekdp\ type crystals will be considered using the
thermodynamic potential per primitive cell, which
within the four-particle cluster approximation reads
\setcounter{equation}{0}
\bea
\label{3.1}
&& g_{1E}(6) =
\frac{\bar v}{2}c^{E0}_{66}\varepsilon^2_6 - {\bar v}e^0_{36}\varepsilon_6E_3
- \frac{\bar v}{2}\chi^{\varepsilon 0}_{33}E^2_3 + 2T\ln 2 + \\
&&\quad{}+ 2{\nu}_c[\eta^{(1)z}(6)]^2 -
2T\ln[1- (\eta^{(1)z}(6))^2] - 2T\ln D^z_6- {\bar v}\sigma_6\varepsilon_6.
\nonumber
\eea
Let us mention the equation for the extremum of the thermodynamic
potential $g_{1E}$ with respect to $\et$
 coincides with Equation
(\ref{2.10}).

From the thermodynamic equilibrium conditions
\[
\frac{1}{\bar v}\left( \frac{\partial g_{1E}(6)}{\partial\varepsilon_6}
\right)_{E_{3},\sigma_6} = 0, \qquad
\frac{1}{\bar v}\left( \frac{\partial g_{1E}(6)}{\partial E_3}
\right)_{\sigma_6} = - P_3
\]
we obtain
\bea
&&
\hspace{-4ex}
\sigma_6 = c^{E0}_{66}\varepsilon_6 -
e^0_{36}E_3 + \frac{4\psi_6}{\bar v}
\frac{m^z(6)}{D^z_6}
+\frac{2\delta_{a6}}{\bar v D_6^z}M_{a6}-
 \frac{2\delta_{s6}}{\bar v D_6^z}M_{s6}+
 \frac{2\delta_{16}}{\bar v D_6^z}M_{16},
\nonumber\\
&&
\label{3.2}
 \hspace{-4ex}
 P_3 = e^0_{36}\varepsilon_6 + \chi^{\varepsilon 0}_{33}E_3 +
2\frac{\mu}{v}\frac{m^z(6)}{D^z_6},
\eea
where
\[
M_{a6}=aa_6-\frac{a}{a_6},\;
M_{s6}=\sinh (2z^z_6 +\beta\delta_{s6}\e),\;
M_{16}=4b \sinh (z^z_6 -\beta\delta_{16}\e).
\]
 From (\ref{3.2}) we find the electric field
\be
\label{3.4}
E_3 = -h^0_{36}\varepsilon_6 + k^{\varepsilon 0}_{33}\left( P_3 -
2\frac{\mu}{v}\frac{m^z(6)}{D^z_6}\right),
\ee
where $h^0_{36} = {e^0_{36}}/{\chi^{\varepsilon 0}_{33}}, \quad
k^{\varepsilon 0}_{33} = {1}/{\chi^{\varepsilon 0}_{33}}$.
Substituting the expression (\ref{3.4}) into (\ref{3.2}), we get
\bea
&& \label{3.5}
\sigma_6 = c^{P0}_{66}\varepsilon_6 -
h^0_{36}\left(P_3 -2\frac{\mu}{v}
\frac{m^z(6)}{D^z_6}\right)
+ \frac{4\psi_6}{\bar v}\frac{m^z(6)}{D^z_6} +\nonumber\\
&&
\quad+\frac{2\delta_{a6}}{\bar v D_6^z}M_{a6}-
 \frac{2\delta_{s6}}{\bar v D_6^z}M_{s6}+
 \frac{2\delta_{16}}{\bar v D_6^z}M_{16},
\eea
where $c^{P0}_{66} = c^{E0}_{66} + e^0_{36}h^0_{36}$.

Expressions (\ref{3.4}) and (\ref{3.5}) can also be obtained from the
conditions
\[
\frac{1}{\bar v}\left( \frac{\partial f}{\partial\varepsilon_6}
\right)_{P_3} = \sigma_6, \qquad
\frac{1}{\bar v}\left( \frac{\partial f}{\partial P_3}
\right)_{\varepsilon_6} = E_3,
\]
whereas the free energy $f$ is
\be
\label{3.7}
f(6) = g_{2E}(6) - {\bar v}P_3E_3.
\ee
Substituting  the expressions (\ref{3.1})
and (\ref{3.4}) into (\ref{3.7}), we obtain
\bea
&&
\hspace{-4ex}
f(6) = \frac{\bar v}{2}c^{P0}_{66}\varepsilon^2_6 -
{\bar v}h^0_{36}\varepsilon_6P_3 + \frac{\bar v}{2}k^0_{33}P^2_3 -
\frac{\bar v}{2}k^0_{33}\left( 2\frac{\mu}{\bar v}\eta^{(1)z}(6)\right)^2 +
\\
&&\quad{}+ 2T\ln 2 + 2{\nu}_c[\eta^{(1)z}(6)]^2 - 2T\ln[1-
(\eta^{(1)z}(6))^2]
- 2T\ln D^z_6 \nonumber.
\eea

Let us study now the influence of the stress $\s$ on  piezoelectric,
dielectric and elastic properties of \dekdp.

From the expression for the mean value of quasispin
$\langle\sigma_{qf}\rangle$ (\ref{2.10}) it follows that
\[
\left(\frac{\partial \et(6)}{\partial\e}\right)_{E_3}=
\frac{\beta\theta_6}{D_6-2\varphi_6^\eta\varkappa_6},
\]
where
\begin{eqnarray*}
&&
\varkappa_6 = \cosh (2z_6 +\beta\delta_{s6}\e)
 + b\cosh (z_6 -\beta\delta_{16}\e) - \eta^{(1)}(6)m(6), \\
&&
r_6=\delta_{s6} \cosh (2z_6 +\beta\delta_{s6}\e)
-2b\delta_{16}\cosh(z_6 -\beta\delta_{16}\e)+\\
&&\quad
+\eta^{(1)z}(6)
\left[ -\delta_{s6}M_{s6}+\delta_{a6}M_{a6}+\delta_{16}M_{16}\right],\\
&&
\varphi^\eta_6 = \frac{1}{1- (\eta^{(1)}(6))^2} + \beta\nu_c,
\theta_6=-2\varkappa_6\psi_6+r_6.
\end{eqnarray*}
  Hence, the coefficient of piezoelectric stress $e_{36}$ is
\bea
&& e_{36} = - \left( \frac{\partial\sigma_6}{\partial E_3}
\right)_{\varepsilon_6} =  \left(
\frac{\partial P_3}{\partial\varepsilon_6}\right)_{E_3}
= e^0_{36} + \frac{2\mu_3}{v}
\frac{\beta\theta_6}{D_6-2\varphi_6^\eta\varkappa_6}.
\eea
In the paraelectric phase at $\s=0$, the coefficient of piezoelectric
stress $e_{36}$ equals
\be
 e_{36}^+ =
 e^0_{36} +\frac{\mu_3}{v}
\frac{2\beta[-2(1+b)\psi_6+\delta_{s6}-2b\delta_{16}]}
 {-1+2b+2a+d-2\beta\nu_c(1+b)}.
\ee

Differentiating the expression for polarization (\ref{3.2}) with  respect
to the field $E_3$ at constant strain $\e$ we obtain dielectric
susceptibility of a clamped crystal
\be
\chi^{\varepsilon}_{33} = \left(\frac{\partial P_3}{\partial
E_3}\right)_{\varepsilon_6} = \chi^0_{33} + {\bar v}\frac{\mu^2}{v^2}\frac1T
\frac{2\varkappa_6}{D_6- 2\varkappa^a_6\varphi^eta_6}.
\ee

Using the relations
\begin{eqnarray*}
&& (D_6 - 2\varkappa_6\varphi^\eta_6)
\left(\frac{\partial \et(6)}{\partial\e}\right)_{P_3}=
\beta\theta_6+{\beta\mu_3}\varkappa_6
\left(\frac{\partial E_3}{\partial\e}\right)_{P_3},\\
&&  (D_6 - 2\varkappa_6\varphi^\eta_6)
\left(\frac{\partial \et(6)}{\partial\s}\right)_{E_3}=
\beta\theta_6
\left(\frac{\partial \e}{\partial\s}\right)_{E_3},
\end{eqnarray*}
where $\ds s_{66}^E=\left(\frac{\partial P_3}{\partial \sigma_6}\right)$
 is the crystal compliance at constant electric field,
  we obtain   the constant of piezoelectric stress $h_{36}$:
   \be
   h_{36} = - \left( \frac{\partial
E_3}{\partial\varepsilon_6}\right)_{P_3} = - \left(
\frac{\partial\sigma_6}{\partial P_3}\right)_{\varepsilon_6} =
\frac{e_{36}}{\chi^{\varepsilon}_{33}}
\ee
and the coefficient of the piezoelectric strain
\be
d_{36}=e_{36}s_{66}^E.
\ee

 From (\ref{3.2}) and (\ref{3.5}) we get the expressions for the
elastic constant $c^E_{66}$ at constant field
\bea
&&\hspace{-5ex}c^E_{66}=\left(
\frac{\partial\sigma_6}{\partial\varepsilon_6}\right)_{E_3} = c^{E0}_{66}
+ \frac{8\beta\psi_6}{{\bar v}}
  \frac{-\varkappa_6\psi_6+r_6}{D_6 -
2\varkappa_6\varphi^\eta_6} -
\frac{4\varphi^\eta_6r_6^2}{{\bar v}TD_6(D_6 -
2\varkappa_6\varphi^\eta_6)}- \nonumber\\
 &&\hspace{-5ex} -
\frac{2\beta}{{\bar v}D_6} \!\left[
\delta_{s6}^2\cosh (2z_6 {+}\beta\delta_{s6}\e)+
\delta_{a6}^2(ac_6{+}\frac{a}{c_6})+
4b \delta_{16}^2\cosh (z_6 {-}\beta\delta_{16}\e)\right]+\!\nonumber\\
&&
\label{3.9}
\hspace{-5ex}+
\frac{2\beta}{{\bar v}D_6}
\left[ -\delta_{s6}M_{s6}+\delta_{a6}M_{a6}+\delta_{16}M_{16}\right]^2;
\eea
and for the
elastic constant $c^P_{66}$ at  constant polarization
\be
c^P_{66} = c^E_{66} + e_{36}h_{36}.
\ee
In the paraelectric phase at $\s=0$
when $\eta^{(1)}(6)=0$, the renormalized elastic constant for
an unstrained crystal  is
\begin{eqnarray}
\label{3.11}&&
c^{E+}_{66} = c^{E0}_{66} -
\frac{4\psi_6}{\bar v T}\frac{-2(1+b)\psi_6+\delta_{s6}}
{\bar v T[-1 + 2b + 2a - 2\beta\bar\nu_c(1+b)]} +\\
&&\nonumber\quad{}+
\frac{4\psi_6\delta_{s6}}{\bar v D_6^+T}-
 \frac{2}{\bar v T(1+4b+2a)}(\delta^2_{s6}+2a\delta_{a6}^2+\delta_{16}).
\end{eqnarray}

Hence, the relations (\ref{3.2})--(\ref{3.5}) take the form
\bea
&&\sigma_6 = c^E_{66}\varepsilon_6 - e_{36}E_3, \qquad
\sigma_6 = c^P_{66}\varepsilon_6 - h_{36}P_3, \nonumber\\
&&
\label{3.15}
  P_3 = e_{36}\varepsilon_6 + \chi^{\varepsilon}_{33}E_3, \qquad
E_3 = - h_{36}\varepsilon_6 + k^{\varepsilon}_{33}P_3,
\eea
where $k^{\varepsilon}_{33} = 1/\chi^{\varepsilon}_{33}$ is
the inverse dielectric susceptibility. These relations can be also
obtained within the phenomenological approach, but we preferred to
have the microscopic expressions for
$c^E_{66}$, $e_{36}$, $\chi^{\varepsilon}_{33}$, $h_{36}$.

From the system of equations (\ref{3.15}) at $E_3 =
{}$const 
\bea
&& c^P_{66}\left(\frac{\partial\varepsilon_6}{\partial\sigma_6}\right)_{E_3}
-h_{36}\left(\frac{\partial P_3}{\partial\sigma_6}\right)_{E_3} = 1,
\nonumber\\
&& -h_{36}\left(\frac{\partial\varepsilon_6}{\partial\sigma_6}\right)_{E_3} +
k^{\varepsilon}_{33}\left(\frac{\partial P_3}{\partial\sigma_6}\right)_{E_3}
=0
\eea
we find the expressions for the coefficient of piezoelectric strain
\be
d_{36} = \left(\frac{\partial P_3}{\partial\sigma_6}\right)_{E_3} =
\frac{h_{36}}{c^P_{66}k^{\varepsilon}_{33} - h^2_{36}} =
\frac{e_{36}}{c^P_{66}- e_{36}h_{36}}
\ee
and compliances at constant field
\be
s^E_{66} = \left(\frac{\partial\varepsilon_6}{\partial\sigma_6}\right)_{E_3}
= \frac{k^{\varepsilon}_{33}}{c^P_{66}k^{\varepsilon}_{33} - h^2_{36}} =
\frac{1}{c^P_{66}- e_{36}h_{36}} = \frac{1}{c^E_{66}}.
\ee

From  (\ref{3.15})  at $P_3={}$const we get
\[
c^E_{66}\left(\frac{\partial\varepsilon_6}{\partial\sigma_6}\right)_{P_3} -
e_{36}\left(\frac{\partial E_3}{\partial\sigma_6}\right)_{P_3} = 1,
\]
\[
e_{36}\left(\frac{\partial\varepsilon_6}{\partial\sigma_6}\right)_{P_3} +
\chi^{\varepsilon}_{33}\left(\frac{\partial E_3}
{\partial\sigma_6}\right)_{P_3} = 0;
\]
hence the constant of piezoelectric strain is
\be
g_{36} = - \left(\frac{\partial E_3}{\partial\sigma_6}\right)_{P_3} =
\frac{e_{36}}{\left(c^E_{66} + \frac{e^2_{36}}{\chi^{\varepsilon}_{33}}
\right)\chi^{\varepsilon}_{33}} = \frac{h_{36}}{c^E_{66} + e_{36}h_{36}} =
\frac{h_{36}}{c^P_{66}},
\ee
and compliance at constant polarization is
\be
s^P_{66} = \left(\frac{\partial\varepsilon_6}{\partial\sigma_6}\right)_{P_3}
= \frac{1}{c^E_{66} + \frac{e^2_{36}}{\chi^{\varepsilon}_{33}}} =
\frac{1}{c^P_{66}}.
\ee

Differentiating the system (\ref{3.15}) with respect to the field $E_3$
at constant stress, we get
\be
c^E_{66}\left(\frac{\partial\varepsilon_6}{\partial E_3}\right)_{\sigma_6}
-e_{36} = 0, \quad
- e_{36}\left(\frac{\partial\varepsilon_6}{\partial E_3}\right)_{\sigma_6}
+\left(\frac{\partial P_3}{\partial E_3}\right)_{\sigma_6} =
\chi^{\varepsilon}_{33}.
\ee
Since
\[
d_{36} = \left(\frac{\partial\varepsilon_6}{\partial E_3}\right)_{\sigma_6},
\]
then the dielectric susceptibility at $\sigma_6 ={}$const is
\be
\chi^{\sigma}_{33} = \left(\frac{\partial P_3}{\partial E_3}
\right)_{\sigma_6} = \chi^{\varepsilon}_{33} + e_{36}d_{36}.
\ee
Hence, we obtained the microscopic expressions for $c^E_{66}$, $e_{36}$,
and $\chi^{\varepsilon}_{33}$; the other characteristics are
 expressed via them.

Let us calculate now the transverse
dielectric susceptibility of the \dekdp\ type  crystals in the presence of
 stress $\sigma_6$ defined as
\bea
\chi_{11}^{\varepsilon} &=& \frac{\mu_\perp\cos\gamma}{v}\left(
\frac{\partial\eta_1^{(1)x}(6)}{\partial E_1} -
\frac{\partial\eta_3^{(1)x}(6)}{\partial E_1}\right) +
\nonumber\\
&+& \frac{\mu_\perp\sin\gamma}{v}\left(-\frac{\partial\eta_2^{(1)x}(6)}
{\partial E_1} + \frac{\partial\eta_4^{(1)x}(6)}{\partial E_1}\right).
\eea
Using the relations (\ref{2.13})--(\ref{2.16}) we obtain the following
system of equations
\bea
&&
\hspace{-5ex}
\beta\mu_\perp(\cos\gamma\varkappa_6^a -M_{a6}\sin\gamma)=
\nonumber\\
&&
\hspace{-5ex}
=
 (D_6 - \varkappa_6^a\varphi_6^a)
 \frac{\partial\left(\eta_1^{(1)x}(6)-\eta_3^{(1)x}(6)\right)}
 {\partial E_1}
 -M_{a6}\varphi_6^a
 \frac{\partial\left(\eta_4^{(1)x}(6)-\eta_2^{(1)x}(6)\right)}
{\partial E_1}
\nonumber\\
&&
\hspace{-5ex}
\beta\mu_\perp(M_{a6}\cos\gamma -\sin\gamma\varkappa_6^a) =
\nonumber\\
&&\hspace{-5ex} =
 (D_6 - \varkappa_6^a\varphi_6^a)\!
 \frac{\partial\left(\eta_4^{(1)x}(6)-\eta_2^{(1)x}(6)\right)}
{\partial E_1}-
  M_{a6}\varphi_6^a
   \frac{\partial\left(\eta_1^{(1)x}(6)-\eta_3^{(1)x}(6)\right)}
 {\partial E_1}
,
\nonumber
\eea
where the  following notations are used
\[
\varkappa_6^a = aa_6 + \frac{a}{a_6} + 2b\cosh (z_6-\beta\delta_{16}\e),
\quad 
\varphi_6^a = \frac{1}{1 - (\eta^{(1)}(6))^2} + \beta(\nu_1 - \nu_3).
\]
In the  result
\bea
\chi_{11}^{\varepsilon} &=& {\bar v}\frac{\mu_\perp^2}{v^2}
\frac1T
\frac{(D_6 - \varphi_6^a\varkappa_6^a)\varkappa_6^a - \varphi_6^aM_{a_6}^2}
{(D_6 - \varphi_6^a\varkappa_6^a)^2 - (\varphi_6^aM_{a6})^2} + \nonumber\\
&-& {\bar v}\frac{\mu_\perp^2}{v^2}\frac1T \frac{D\varkappa_6}
{(D_6 - \varphi_6^a\varkappa_6^a)^2 - (\varphi_6^aM_{a6})^2}\sin 2\gamma.
\eea

\section{Influence of stress $\s$
on thermal properties of the \dekdp\
type  crystals}

Molar entropy  of the deuteron subsystem of
\dekdp\ type crystals under stress $\sigma_6$ reads
\setcounter{equation}{0}
\bea
S_6 = - R\left( \frac{\partial f_6}{\partial T}\right)_{P_3,\varepsilon_6}&=&
R \left\{ 2\ln2 + 2\ln [1 - (\eta^{(1)}(6))^2] + 2\ln D_6 + \right.\nonumber\\
&+&
\label{4.1}
\left. 4T\varphi_6^T\eta^{(1)}(6) + \frac{2M_6}{D_6} \right\},
\eea
where $R$ is the gas constant, and
\bea
&&\varphi_6^T = - \frac{1}{T^2}({\bar \nu}_c \eta^{(1)}(6) -
\psi_6\varepsilon_6), \\
&&
 M_6 = 4b\frac{w}{T}\cosh z_6 + d\frac{w_1}{T} + \left(ac_6 + \frac{a}{c_6}
\right)\frac{\varepsilon}{T} + \left(ac_6 - \frac{a}{c_6}\right)
\frac{\delta_6\varepsilon_6}{T}. \nonumber
\eea

The molar specific heat of the deuteron subsystem of
\dekdp\ type crystals can be calculated by differentiating the
entropy
 \be \label{4.3} \Delta
C_6^\sigma = T\left( \frac{\partial S}{\partial T}\right)_{\sigma} =
\Delta C_6^{\varepsilon} + q_6^P\alpha_6,
\ee
where $\Delta C_6^{\varepsilon}$ is the molar specific heat  at
constant strain
\be
\label{4.4}
\Delta C_6^{\varepsilon} = q_6^{P,\varepsilon} + q_6^{\varepsilon}
p_6^{\sigma}.
\ee
Using the relations (\ref{4.1}) we obtain
\bea
&&
q_6^{P,\varepsilon} {=} \left(\frac{\partial S_6}{\partial T}
\right)_{P_3,\varepsilon_6} = {}\\
&& {} =\frac{2R}{D_6}\left\{ 2T\varphi_6^T
[2\varkappa_6T\varphi_6^T + 2(q_6 - \eta^{(1)}(6)M_6)] +
N_6 - \frac{M_6^2}{D_6} \right\}, \nonumber\\
&&
 q_6^{\varepsilon} = \left(\frac{\partial S_6}{\partial P_3}
\right)_{\varepsilon_6,T} = \frac{v}{\mu_3}\frac{2RT}{D_6}\varphi_6^\eta
\{ 2\varkappa_6T\varphi_6^T + [q_6 - \eta^{(1)}(6)M_6]\} \nonumber
\eea
is the polarization heat at given $\varepsilon_6$,
\begin{eqnarray*}
&&
\hspace{-4ex}
  q_6^P = \left(\frac{\partial S_6}{\partial\varepsilon_6}\right)_{P_3,T} =
\frac{2R}{D_6}\left\{ 2T\varphi_6^T
(-2\varkappa_6\psi_6 + r_6) -
2[q_6 - \eta^{(1)}(6)M_6]
\psi_6 \right. \nonumber\\
&&
\hspace{-3ex}
\left. \qquad-\lambda_6+\frac{M_6}{D_6}
(-\delta_{s6}M_{s6}+\delta_{a6}M_{a6}+\delta_{16}M_{16})
\right\},
\end{eqnarray*}
is the deformation heat at given $P_3$, where
\bea
&&
\hspace{-5ex}
N_6 =\frac{1}{T^2}\!\left[\!
({\varepsilon+\delta_{a6}\e})^2aa_6+
({\varepsilon-\delta_{a6}\e})^2\frac{a}{a_6}+
4 b{w}^2\cosh(z^z_6 -\beta\delta_{16}\e)
+\right.\nonumber\\
&&
 \hspace{-5ex}
{} + {w_1}^2 d-
\left({\delta_{s6}\e}\right)^2 \cosh(2z_6 +\beta\delta_{as}\e)+
\nonumber\\
&&
\hspace{-5ex}
\left. {}+
 \left({\delta_{16}\e}\right)^2 4 b\cosh(z_6 -\beta\delta_{16}\e)
 +
\left({\delta_{16}\e w}\right)8 b\cosh(z_6 -\beta\delta_{16}\e)\right],
\nonumber\\
&&
\hspace{-5ex}
q_6 = \frac{1}{T}\!\left[
-\delta_{s6}\e\cosh(2z_6 +\beta\delta_{as}\e)+
 \delta_{16}\e 2 b\cosh(z_6 -\beta\delta_{16}\e)+\right.
 \nonumber\\
&&   \qquad
 \left.{}+  2bw \cosh(z_6 -\beta\delta_{16}\e) \right],
\nonumber \\
&&
\hspace{-5ex}
\lambda_6=\frac{1}{T}\!\left[\!
-\delta_{s6}^2\e\sinh(2z_6 +\beta\delta_{as}\e)
+
 \delta_{16}^2\e 4 b\cosh(z_6 -\beta\delta_{16}\e)
\!+\right.\nonumber\\
&&
\hspace{-5ex}
\left. +
\delta_{a6}\! \left(\!aa_6(\varepsilon + \delta_{a6}\varepsilon_6) -
\frac{a}{a_6}
(\varepsilon - \delta_{a6}\varepsilon_6)\!\right)+
  \delta_{16}w\e 4 b\sinh(z_6 -\beta\delta_{16}\e)\right].
  \nonumber
\eea
In (\ref{4.3}) and (\ref{4.4}) $p_6^\sigma = \left({\partial
P_3}/{\partial T}
\right)_{\sigma,E_3}$ is the pyroelectric coefficient, and $\alpha_6 =
\left({\partial \varepsilon_6}/{\partial T}\right)_{\sigma}$ is the
thermal expansion coefficient.

From relations (\ref{3.15})  we get
\be
p_6^\sigma = p_6^\varepsilon + e_{36}\alpha_6,
\ee
where
\be
p_6^{\varepsilon} = \frac{\mu_3}{v}\frac2T\frac{2\varkappa_6T\varphi_6^T +
[q_6 - \eta^{(1)}(6)M_6]}{D_6-2\varkappa_6\varphi_6^\eta},
\ee
and the thermal expansion coefficient reads
\be
\alpha_6 = \frac{-p_6 + h_{36}p_6^{\varepsilon}}{c_{66}^E},
\ee
where $p_6 = \left( \frac{\partial \sigma_6}{\partial
T}\right)_{p_3,\varepsilon_6} = q_6^p$ is the thermal pressure.

\section{Discussion}

Before going into the discussion of the proposed in previous sections
theory, let us note that, strictly speaking, this theory can be valid for
a completely deuterated KD$_2$PO$_4$ crystal only, whereas the
vast majority of experimental data concerns the crystals,
deuterated partially (see \cite{a11,a12,a13,a14,a15,a16}).
Nevertheless, as the relaxational character of dielectric dispersion in
 K(H$_{1-x}$D$_x)_2$PO$_4$ crystals implies, tunneling effects in them are
 most likely suppressed by the short-range correlations between hydrogens
\cite{a17,a18,a19a}. So at least in the case of high
deuteration, we are allowed to neglect tunneling and apply the theory to
partially deuterated crystals as well. Hereafter we shall consider the
crystal with nominal deuteration $x=0.89$ with the transition temperature
$T_{C0}=210.7$~K, for which experimental data for the
 relevant elastic, piezoelectric, and dielectric characteristics are
 available.

 As the values of the theory parameters $\eps$ and $w$
 we choose those found in \cite{a19,a20}, which
at $\sigma_6=0$ provide a satisfactory agreement of the calculated
curves with the experimental data for spontaneous polarization,
specific heat, static and dynamic dielectric permittivities of the
crystal.

To determine the deformation
potentials  and ``seed'' quantities, we use the
experimental data of \cite{a15,a16} for the temperature
dependences of the coefficient of piezoelectric strain
$d_{36}$, dielectric susceptibility $\chi_{33}$, and
compliance $s_{66}^E$ at $\sigma_6 = 0$.  Using the known
relations between  dielectric, elastic and
piezoelectric characteristics and having the values of
$d_{36}$, $\chi_{33}^\sigma$, and $s_{66}^E$ we can calculate
the piezoelectric constants $e_{36}$, $h_{36}$, $g_{36}$, elastic
characteristics $c_{66}^P$, $c_{66}^E$, $s_{66}^P$, and dielectric
susceptibility $\chi_{33}^\varepsilon$.

Analogous calculations were carried out also for $x=0$ using the data of
\cite{a12,a13,a14}. Let us mentions that the experimental
values of $d_{36}$ or $c_{66}^E$ do not agree with
the results presented in
\cite{a11}.

The deformation potentials $\psi_6$ and $\delta_{s6}$, and the
parameter of the long-range interaction $\nu$  were found from the
condition that the transition temperature at $\s=0$ was
$T_{\rm C0}=210.7$~K, and that the best description of the $e_{36}$,
$s_{66}^E$, and  $\eps_{33}^\sigma$ temperature curves
was obtained. Let us note that we get the best fit if the
following constraint
\be
\label{aa}
\delta_{s6}=325+2\psi_6\exp(-\frac{w}{T_{C0}})
\ee
is held. By changing $\psi_6$ (or $\delta_{s6}$) we can slightly alter the
theoretical slopes $\partial h_{36}/\partial T$ and
$\partial  g_{36}/\partial T$ at
$T>T_{C0}$. So, fitting the theoretical temperature dependences of
the piezomodules $h_{36}$ and $g_{36}$ to experimental points, we choose
the values of $\psi_6$ and $\delta_{s6}$. It should, however, be noted that
we can vary $\delta_{s6}$  rather strongly (from 0 up to the adopted in this
paper value), provided Eq.~\ref{aa} is obeyed, and still obtain a fair
description of  $h_{36}$ and $g_{36}$.

The adopted value of the long-range interaction parameter is much lower than
the one used in the theories where spontaneous strain is not taken into
account \cite{a19,a20,Vaks1}. Low $\nu$ allows us to describe the
paraelectric Curie constant and ferroelectric spontaneous polarization
with the same value of the effective dipole moment $\mu_3$ 
describes response of heavy ions to deuteron ordering), strictly
determined by the experimental saturation polarization.

The values of $\delta_{a6}$ and $c_{66}^{E0}$ were chosen by fitting to
experimental data the calculated $s_{66}^E(T)$ and $c_{66}^P(T)$
dependences. Moderate values of $\delta_{16}$ (splitting of the
single-ionized deuteron configurations) do not perceptibly affect the
calculated curves, and its larger values only make the fitting worse.
Hence, for the sake of simplicity we neglect this splitting, taking
$\delta_{16}=0$.

The ``seed'' $e_{36}^0$ and $\chi_{33}^{\varepsilon 0}$ are merely the
high temperature limits of experimental temperature dependences $e_{36}$ and
$\chi_{33}^{\varepsilon}$.

The adopted values of the theory parameters are presented in
Table~\ref{parameters}.

\begin{table}[hbt]
\caption{The theory parameters.}
\small
\begin{center}
\begin{tabular}{|ccc|cccc|cccc|}
\hline
  $ \eps$
& $ w$
& $ \nu_c$
& $\psi_6$
& $\delta_{s6}$
& $\delta_{a6}$
& $\delta_{16}$
&  $c_{66}^{E0}\cdot 10^{-10}$
& $\mu_3/v$
& $e_{36}^{0}$
& $\chi_{33}^{(0)}$\\
\hline
\multicolumn{3}{|c|}{(K)}  &
\multicolumn{4}{c|}{(K)}  &
 (dyn/cm$^2)$
& ($\mu$C/cm$^2$) &
(esu/cm$^2$) & \\
\hline
88.3 & 778 & 35.976
& $-250$ & 1750 & $-187.5$ & 0
& 7.0 &  6.21 & $0.42\cdot 10^4$ & 0.4\\
\hline
\end{tabular}
\end{center}
\label{parameters}
\end{table}

To find the order parameter $\eta^{(1)}$ and strain
$\varepsilon_6$ we minimize the thermodynamic potential
$g_{1E}(6)$ with respect to $\eta^{(1)}$  and
determine $\varepsilon_6$ from Eq.~(\ref{3.2}).

Let us demonstrate how the presented theory desribes the physical
characteristics of the crystal related to the strain $\e$ and polarization
$P_3$ at $\s=0$. As one can see, the theoretical results are in a good
quantitative agreement with the experimental data in the paraelectric
phase.

\begin{figure}[hbt]
\begin{center}
\leavevmode
\epsfysize=5cm
\rotate[r]{\epsffile{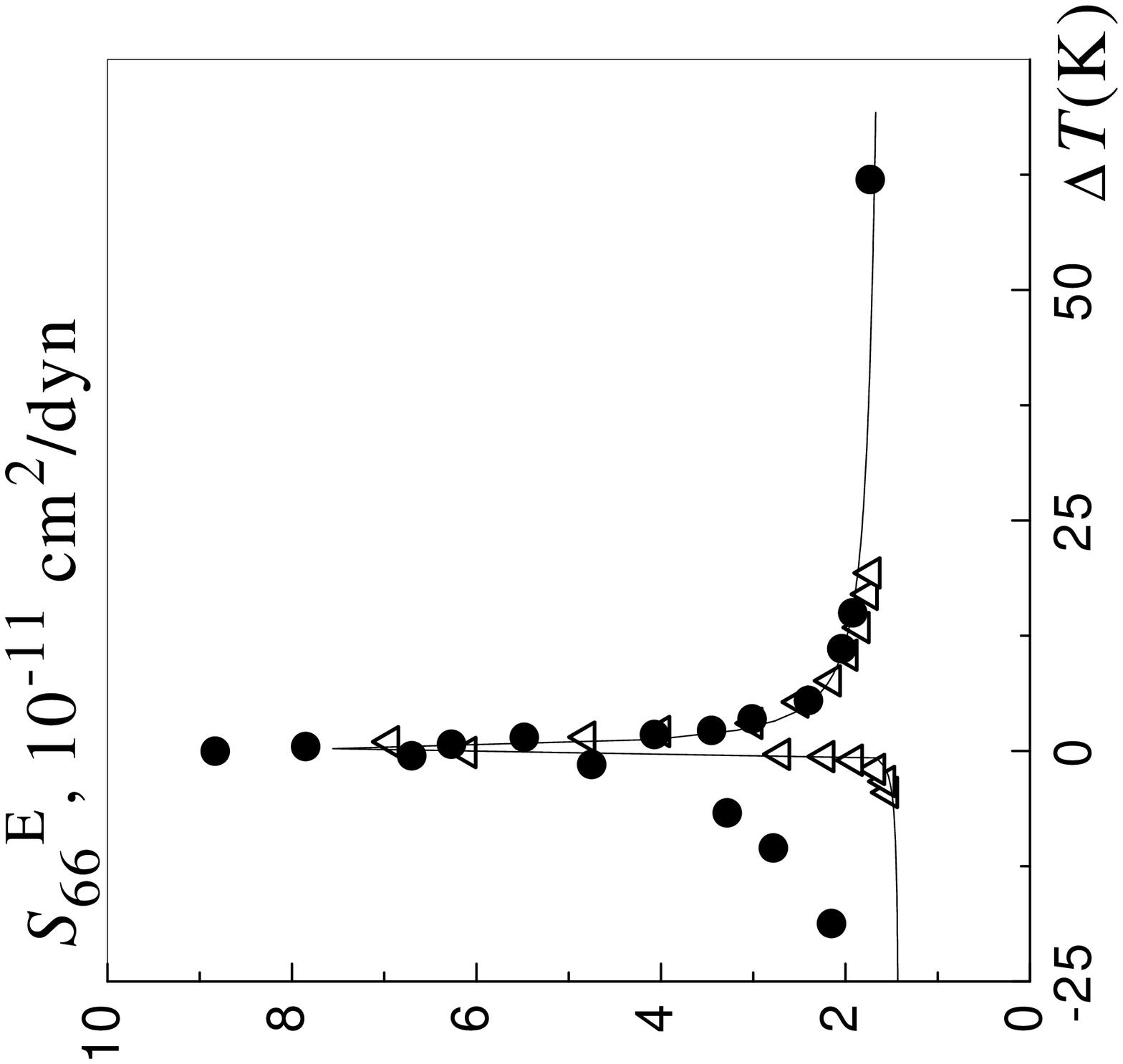}}
\hspace{0.5cm}
\epsfysize=5.5cm
{\epsffile{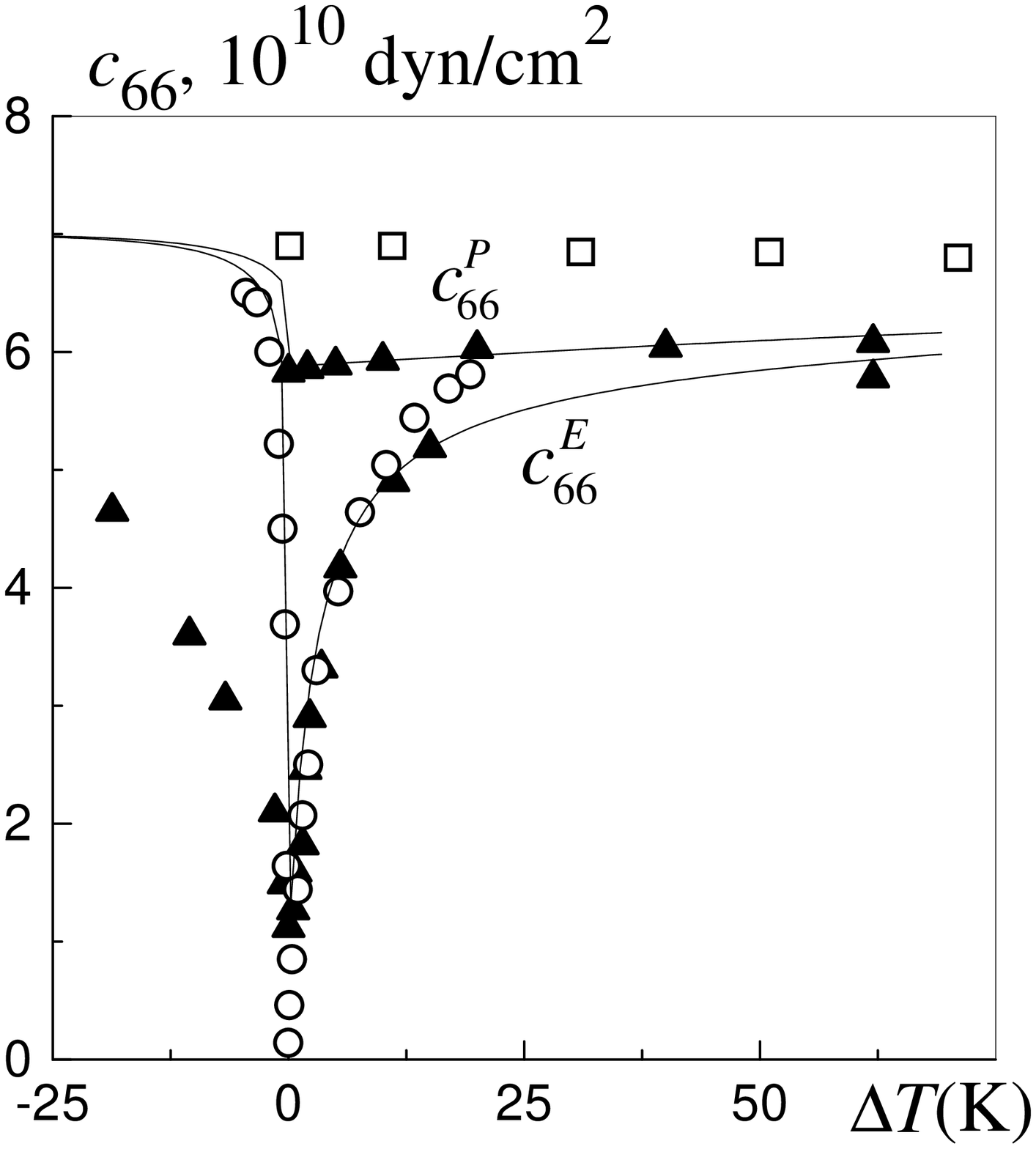}}
\end{center}
\caption{\small
 Temperature dependence of compliance $s_{66}^E$: $\bullet$ --
\protect\cite{a16},
$\triangle$ -- $s_{66}^E = 1/c_{66}^E$ \protect\cite{a14} and elastic
constants $c_{66}^P$ and  $c_{66}^E$: $\btu$ -- $c_{66}^E = 1/s_{66}^E$
\protect\cite{a16}, $c_{66}^P =
\frac{1}{S_{66}^E} +
{d_{36}^2}/{S_{66}^{E2}\chi_{33}^\varepsilon}$ \protect\cite{a15,a16};
$\circ$ -- \protect\cite{a14}; $\square$ -- \protect\cite{a11}.}
 \label{se_ex}
\end{figure}

In fig.~\ref{se_ex} we plot the dependence of the compliance
$s_{66}^E$ and elastic constants $c_{66}^E$ and $c_{66}^P$
of the K(H$_{0.11}$D$_{0.89}$)$_2$PO$_4$ crystal on $\Delta T =
T-T_{\rm C}$ at $\sigma_6=0$.
 Here we also presented
the experimental data for $x=0.89$ \cite{a16} and $x=0.00$
\cite{a11,a14}. As one can see, the theoretical results are in a good
quantitative agreement with the experimental data
of \cite{a16} in the paraelectric phase. The compliance $s_{66}^E$
at $T \to T_{\rm C}$ has an anomalous increase, whereas the elastic
constant $c_{66}^E$ vanishes at the  Curie point. The elastic constant
$c_{66}^P$ at $T < T_{\rm C}$ is nearly constant with temperature, only
has an about  6\% decrease at $T = T_{\rm C}$, and slightly increases in
the paraelectric phase.  That accords with the conclusions of  \cite{a23},
where it was shown that the ferroelectric phase transition in
KH$_2$PO$_4$ did not affect the magnitude of $s_{66}^P$, and hence of
$c_{66}^P$.  According to the data of \cite{a11}, at $x=0.0$ the elastic
constant $c_{66}^P$ slightly decreases with temperature in the
paraelectric phase.

\begin{figure}[hbt]
\begin{center}
\leavevmode
\epsfysize=5cm
\rotate[r]{\epsffile{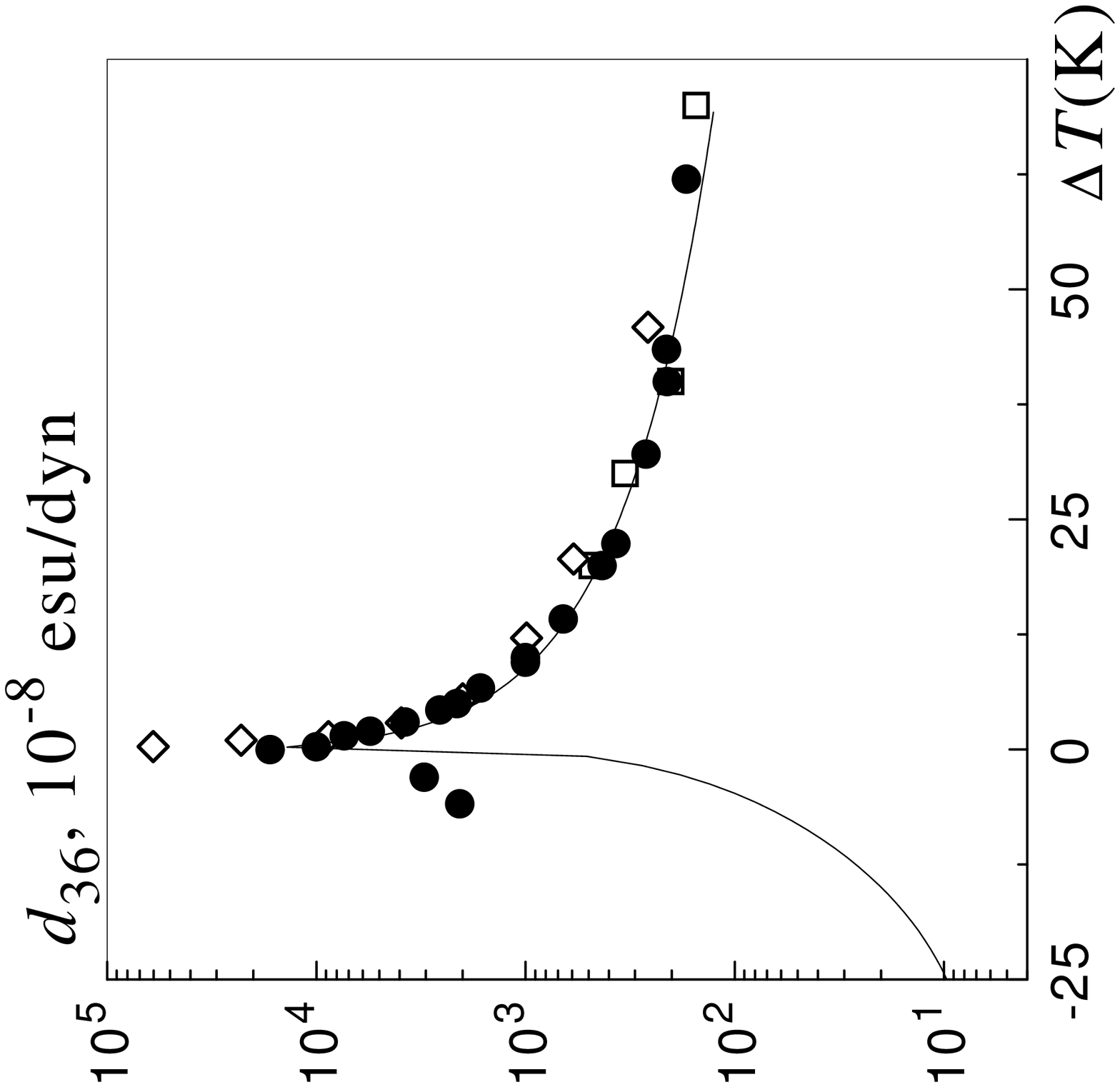}}
\hspace{0.5cm}
\epsfysize=5cm
\rotate[r]{\epsffile{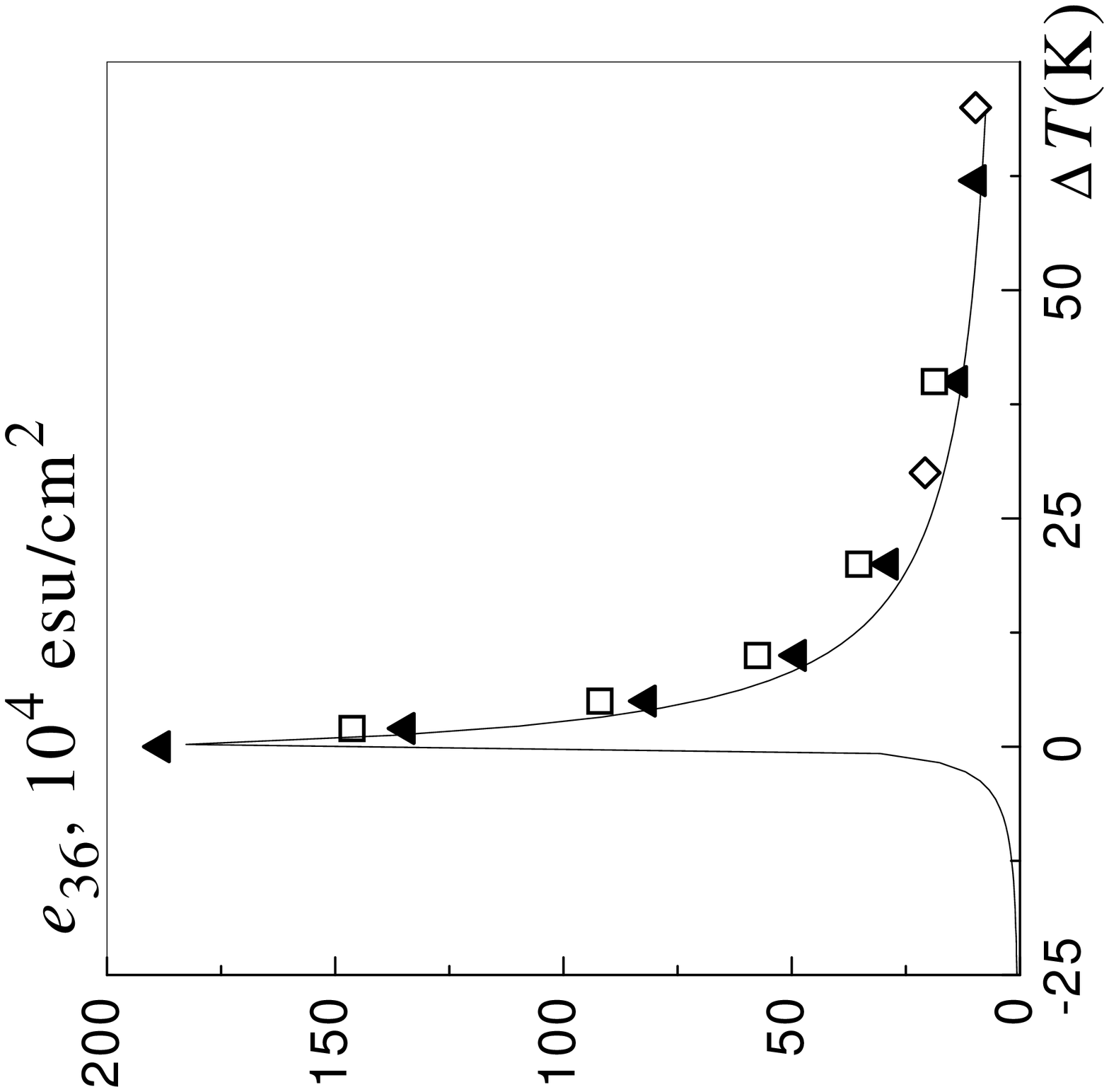}}
\end{center}
\caption{\small
Temperature dependence of coefficient of piezoelectric strain
$d_{36}$ (left, experimental points taken from
\protect\cite{a15} - $\bullet$ ($x=0.89$), \protect\cite{a12,a13} --
$\diamond$ ($x = 0.00$), \protect\cite{a11} -- $\square$ ($x = 0.00$))
and coefficient of piezoelectric stress $e_{36}$ (right,   $\btu$ --
$e_{36} = \frac{d_{36}\protect\cite{a15}}{S_{66}^E \protect\cite{a16}}$;
$\diamond$ -- \protect\cite{a12,a13}, $\square$ -- \protect\cite{a11}).}
 \label{d_e_ex}
\end{figure}

Temperature dependence of the coefficient of piezoelectric strain $d_{36}$
and coefficient of piezoelectric stress $e_{36}$ of the
K(H$_{0.11}$D$_{0.89}$)$_2$PO$_4$ crystal at $\sigma_6=0$ is shown in
fig.~\ref{d_e_ex}. A satisfactory agreement of the
presented theory with the experimental data of \cite{a15} for $d_{36}$ and
with the data obtained from the formula $e_{36} = d_{36}
\cite{a15}/s_{66}^E \cite{a16}$ is observed. At $T \to T_{\rm C}$
coefficients $d_{36}$ and $e_{36}$ sharply increase. In
the ferroelectric phase, the calculated coefficients $d_{36}$ and $e_{36}$
sharply decrease with a rate much higher than in the paraelectric phase.

\begin{figure}[hbt]
\begin{center}
\leavevmode
\epsfysize=5cm
\rotate[r]{\epsffile{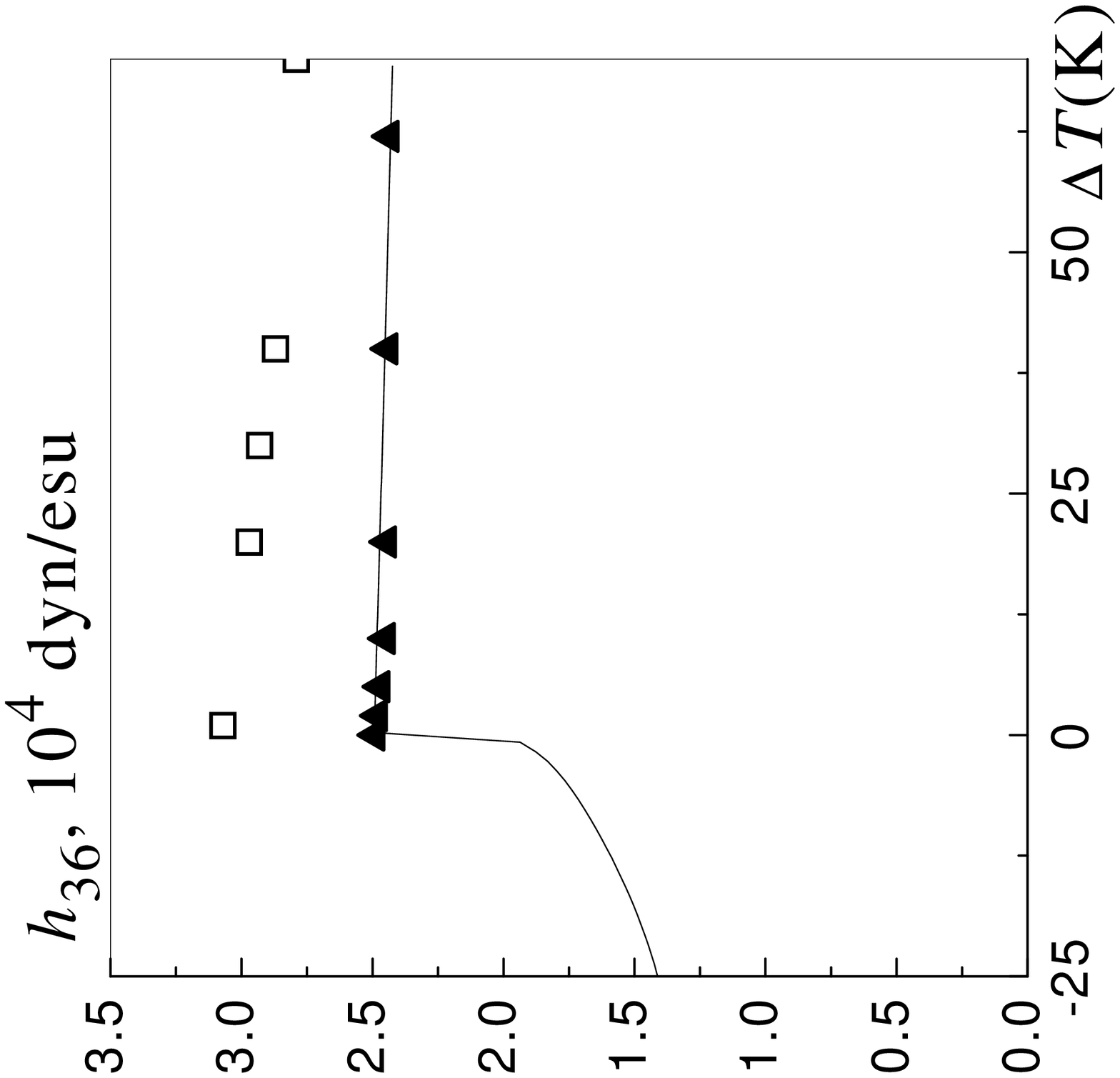}}
\hspace{0.5cm}
\epsfysize=5cm
\rotate[r]{\epsffile{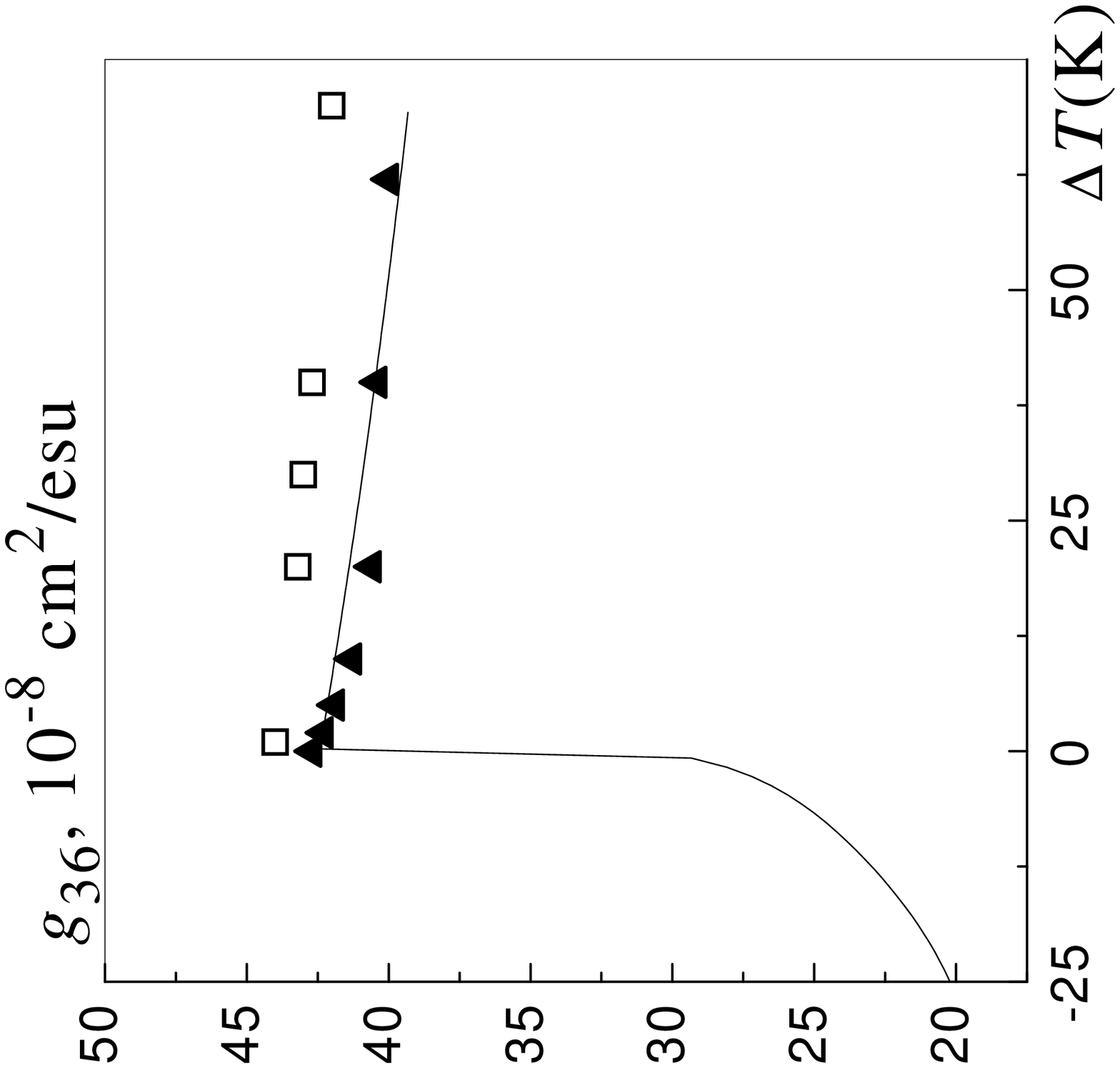}}
\end{center}
\caption{\small
Temperature dependence of constant of piezoelectric stress $h_{36}$
(left, $\btu$ -- $h_{36} =
{d_{36}\protect\cite{a15}}/{S_{66}^E\chi_{33}^\varepsilon
\protect\cite{a16}}$, $\square$ -- \protect\cite{a11}) and constant of
piezoelectric strain $g_{36}$ (right, $\btu$ -- $g_{36} =
{d_{36}}/{\chi_{33}^\varepsilon}$
\protect\cite{a5}, $\square$ -- \protect\cite{a11}).}
 \label{g_h_ex}
\end{figure}

Temperature dependence of the calculated constant of
piezoelectric stress $h_{36}$ and constant of  piezoelectric
strain $g_{36}$ at $\sigma_6=0$ is shown in fig.~\ref{g_h_ex} along with
the obtained using the experimental data of \cite{a14,a15} values of
$h_{36}$ and $g_{36}$ for $x=0.89$ and the data of \cite{a11} for
$x=0.00$.  The piezoelectric constants $h_{36}$ or $g_{36}$ do not have
singularities at the ferroelectric transition; therefore they are called
``true'' piezoelectric constants of the crystals.

\begin{figure}[hbt]
\begin{center}
\leavevmode
\epsfysize=5cm
\rotate[r]{\epsffile{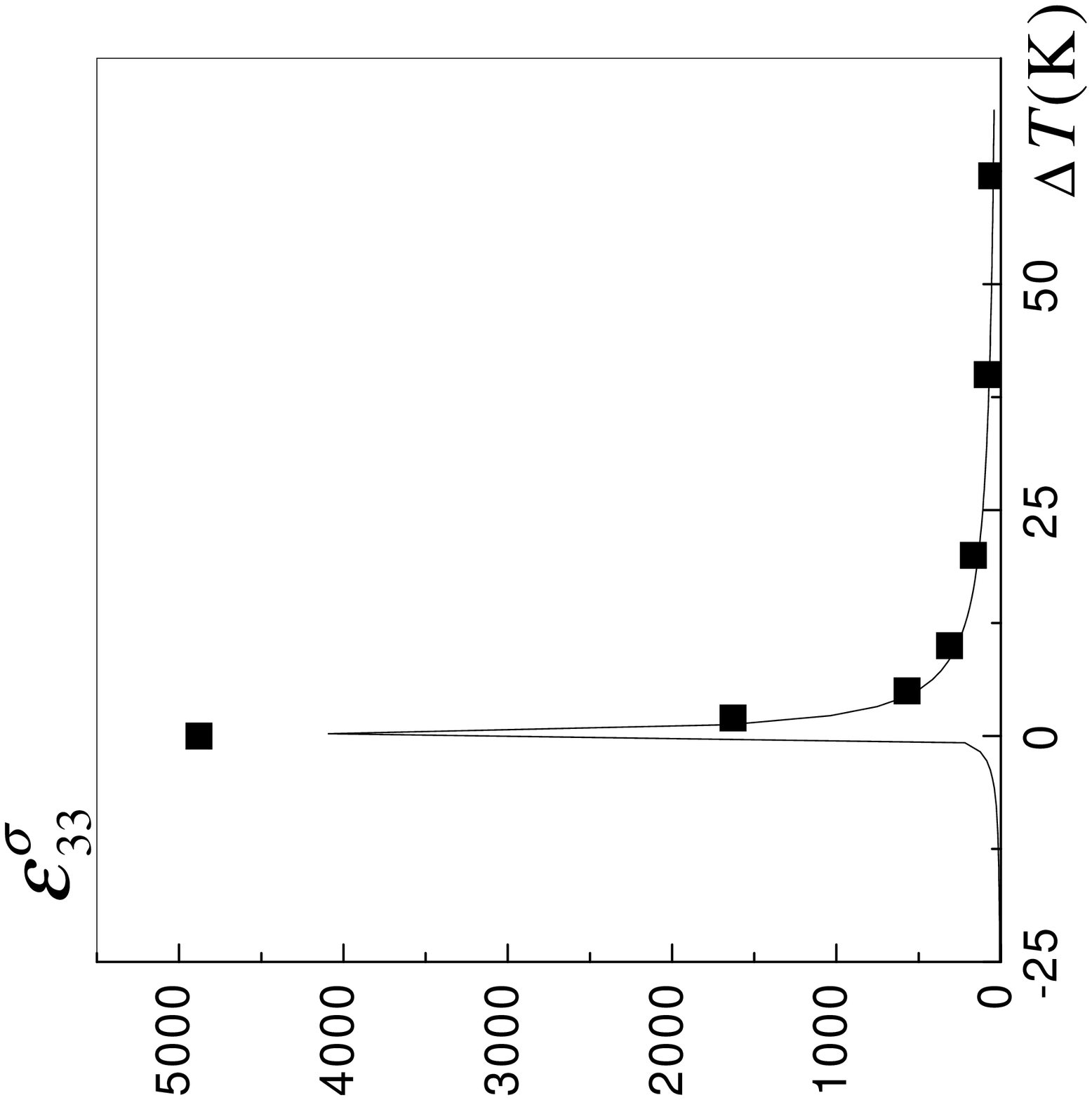}}
\hspace{0.5cm}
\epsfysize=5cm
\rotate[r]{\epsffile{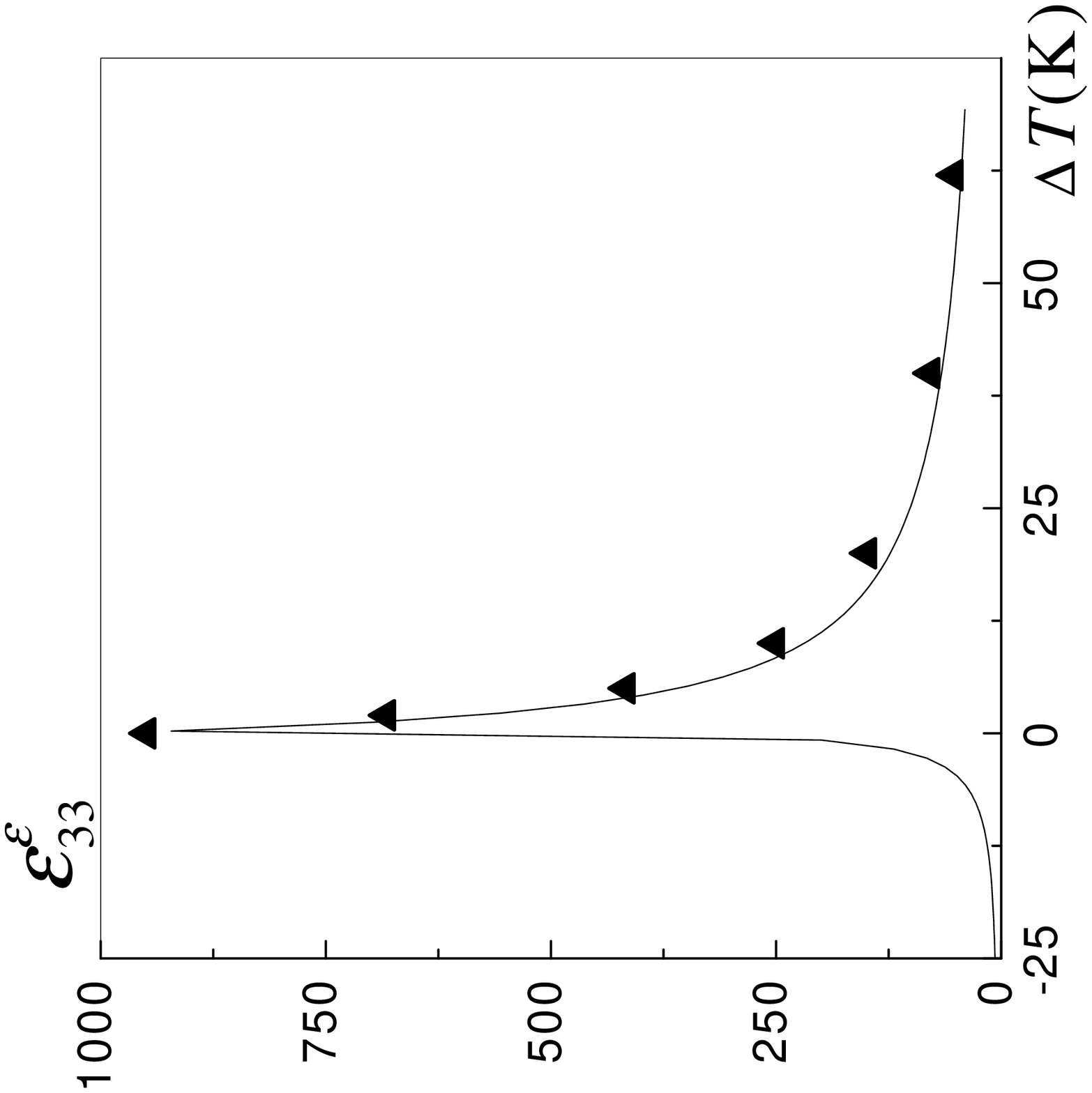}}
\end{center}
\caption{\small
Temperature dependence of dielectric permittivity of a free crystal
$\varepsilon_{33}^\sigma$ (left, $\blacksquare$ -- \protect\cite{a15}) and
dielectric permittivity of a clamped crystal
$\varepsilon_{33}^\varepsilon$ (right, $\btu$ --
$\varepsilon_{33}^\varepsilon =
\varepsilon_{33}^\sigma - \frac{d_{36}^2}{S_{66}^E}$
\protect\cite{a15,a16}).}
 \label{epsl}
\end{figure}

In fig.~\ref{epsl} we plot the temperature dependences of the dielectric
permittivities of a free ($\varepsilon_{33}^\sigma$)
 and clamped ($\varepsilon_{33}^\varepsilon$) crystals
 at $\sigma_6 =0$ along with the
experimental data of \cite{a15} for $\varepsilon_{33}^\sigma$.
The ``experimental'' data for the dielectric permittivity
$\varepsilon_{33}^\varepsilon$ were calculated using the
relation
$\varepsilon_{33}^\varepsilon = \varepsilon_{33}^\sigma - 4\pi
e_{36}d_{36}$.

\clearpage

The presented graphs indicate that one can describe the temperature
peculiarities of these dielectric, piezoelectric, and elastic properties of
\dekdp\ attributing those pecularities to deuteron subsystem only, with the
heavy ions lattice counterpart considered as background and temperature
independent one. However, since we cannot unambigously set the
values of $\psi_6$ and $\delta_{s6}$, we are not in position to determine
what are the relative weights in these peculiarities of piezoelectric
coupling, described by the parameter $\psi_6$, and of the short-range
up-down deuteron configurations splitting, induced by strain $\s$ and
described by the parameter $\delta_{s6}$.

\begin{figure}[hbt]
\begin{center}
\leavevmode
\epsfxsize=\textwidth
{\epsffile{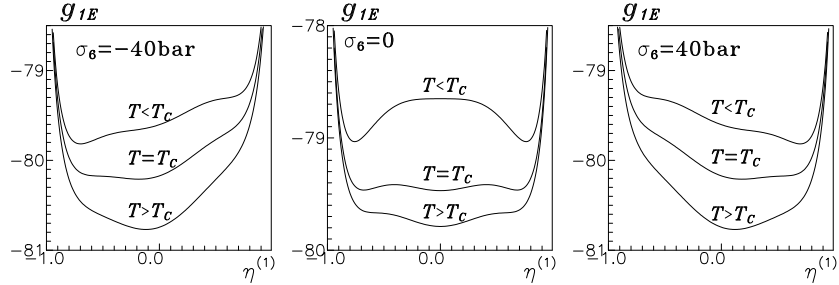}}
\end{center}
\caption{\small
Dependences of the thermodynamic potential  $g_{1E}$ on the order
parameter $\eta^{(1)}$ at different values of stress $\sigma_6$. }
 \label{xx} \end{figure}

In fig.~\ref{xx} the  dependences of the thermodynamic potential
$g_{1E}$ on the order parameter $\eta^{(1)}$ at different values of
temperature and stress $\sigma_6$ are presented. The behavior of
$g_{1E}$ at zero stress $\sigma_6$ is usual for the first order
phase transition: a little below the  transition point $g_{1E}$ has three
minina -- one at $\eta^{(1)}=0$ and two symmetric ones
$\pm\eta^{(1)}\neq 0$.  The last two are of the same depth and lower
than the one at $\eta^{(1)}=0$.  At $T=T_{\rm C}$ all minima are of the
same depth (criterion of the phase transition), and at $T>T_{\rm C}$ the
central minimum becomes the deepest. At higher temperatures the minima of
$g_{1E}$ at $\eta^{(1)}\neq 0$ disappear.

Under stress $\sigma_6$, the dependence of $g_{1E}(\eta^{(1)})$ becomes
asymmetric; the lower values of the  thermodynamic
potential are at those values  of the order parameter,
the sign of which coincide with the sign of the stress.

\begin{figure}[p]
\begin{center}
\leavevmode
\epsfysize=0.8\textheight
\epsffile{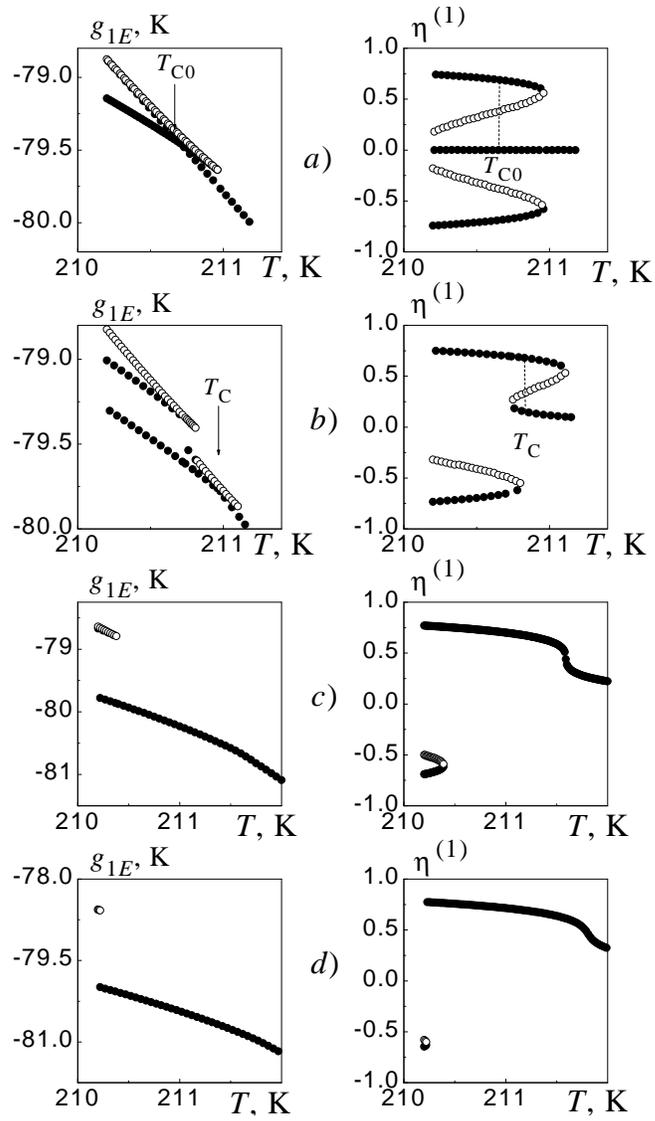}
\end{center}
\caption{\small
Temperature dependences of the thermodynamic potential and solutions of
equation for the order parameter at different values of stress
$\sigma_6$:  a) 0~bar, b) 40~bar, c) 77~bar, d) 100~bar. $\circ$ and
$\bullet$ correspond to maxima  and minima of the thermodynamic potential,
respectively.  } \label{st} \end{figure}

In fig.~\ref{st} we plot the dependences of the solutions of the equation for
the thermodynamic potential $g_{1E}$  extremum and the corresponding
values of  $g_{1E}$ on temperature at different
values of stress $\s$ in the vicinity of the transition point.

Equation (\ref{2.10}) may have up to
five different  solutions,  three of which correspond to
minimuma, and two correspond to maxima  of the  thermodynamic
potential. One of the minimuma -- the central one -- is at small values of
the order parameter, and the sign of $\et$ coincides with the sign
of the stress $\s$. The two other minima, if they exist,
are nearly symmetric (absolutely symmetric at
$\s=0$), and, as we have already mentioned,  that
minimum, the sign of which coincides with the sign of
the stress is deeper.

Let $\et_1<0$, $\et_2\approx 0$, and $\et_3>0$ be the possible
solutions of Equation (\ref{2.10}), corresponding to the  minimuma
of $g_{1E}$. As one can see (fig.~\ref{st}), at $\s>0$
\begin{eqnarray*} &&
g_{1E}(\et_3)<g_{1E}(\et_2)~{\rm at}~T<T_{\rm C},\\ &&
g_{1E}(\et_3)=g_{1E}(\et_2)~{\rm at}~T=T_{\rm C},
({\rm the~phase~transition~criterion}) \\
&& g_{1E}(\et_3)>g_{1E}(\et_2)~{\rm at}~T>T_{\rm C}.
\end{eqnarray*}

The temperature of the first order phase transition at which
the branches of the thermodynamic potential intersect, increases with
stress $\s$, and the values of $\et_2$ at the transition point
increase, whereas $\et_3$ decrease. A decrease in the magnitude of jump in
the order parameter $\delta \et=\et_3-\et_2$ means that the order of
the phase transition is moving towards to the second order.
At certain stress $\s^*$ $\delta \et$ turns to zero -- there is a critical
point where the second order phase transition takes place (at $T^*$).
Further increase in stress smears off the phase transition and results in
continuous and smooth temperature dependence of the order parameter.  Such
behavior is typical for the first order phase transitions in
ferroelectrics to which the electric field conjugate to spontaneous
polarization is applied \cite{a22}.

\begin{figure}[h]
\begin{center}
\leavevmode
\epsfysize=0.7\textwidth
\rotate[r]{\epsffile{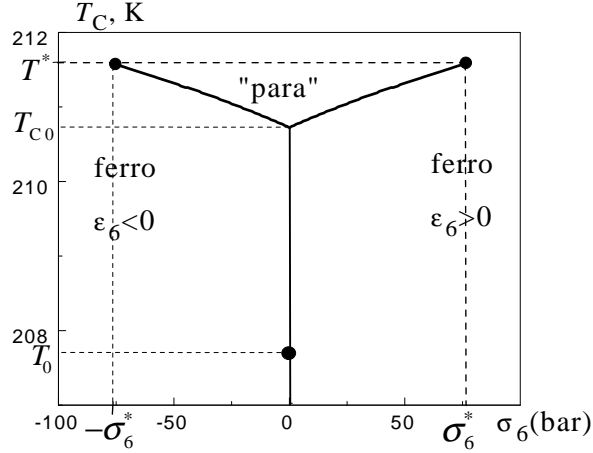}}
\end{center}
\caption{\small
$T_{\rm C}-\sigma_6$ phase diagrams of a
KD$_2$PO$_4$ crystal.} \label{pd} \end{figure}

Corresponding $T_{\rm C}-\s$ phase diagram is depicted in
fig.\ref{pd}. Only the stable phases
corresponding to absolute minima of thermodynamic potential are showed. The
following  temperatures are indicated:  $T_{\rm C0}=210.7$~K is the
temperature of the first order phase transition at
$\s=0$; $T^*=211.6$~K is the temperature of the second order phase
transition at the critical point at stress $\s^*=75$~bar; $T_{0}=210.8$~K
is the Curie temperature of a free crystal $\s=0$, at which the compliance
$s_{66}^E$ and the longitudinal static dielectric permittivity
$\varepsilon_{33}^\sigma$ diverge.  As one can see, the  diagram is
symmetric with respect to a change $\s\to-\s$.  An increase in the
transition temperature with stress $\s$ is practically linear with the slope
$\partial T_{\rm C}/\partial |\s|=21.6$~K/kbar, which is much faster with
than with hydrostatic or uniaxial $p=-\sigma_3$ pressure
\cite{Romanyuk-big}.

The observed ``Y''-shaped form of the phase diagram is not unique but
typical for the systems in fields conjugate to the order parameter.  The
$T_{\rm C}-E_3$ phase diagram of KH$_2$PO$_4$ of the same topology was
obtained within the phenomenological approach without invoking any
microscopic model by Schmidt \cite{Schmidt}; an increase in the transition
temperature with $E_3$ and cease of the transition at certain critical
$E_3^*$ were observed experimentally in \dekdp\ by
Sidnenko and Gladkii \cite{Sidnenko}.

Let us note that in order to observe the dependence of temperature phase
transition on stress $\s$ experimentally, one should apply an external
stress to the {\it paraelectric} crystal and cool it below the transition
point.  In absence of external fields or stresses, crystal at the
transition can with 50:50 probability have positive of negative
spontaneous strain.  However, since in the presence of the stress $\s$ the
minimum of the thermodynamic potential  at $\et$ of the same sign as that
of $\s$ is deeper than the opposite minimum, the sign of the strain below
the transition point will coincide with the sign of stress (similarly, as
the direction of polarization in a crystal coincides with the direction of
external field).  The same picture will take place if the stress is
applied to {\it spontaneously polarized and strained} crystal, provided
that this stress induces the strain of the same sign as that of the
spontaneous strain. However, if stress is applied to {\it spontaneously
polarized and strained} crystal such that the signs of the stress induced
and spontaneous strains are opposite, the system appears in a local minimum
of the potential (for instance, at $\s>0$
$g_{1E}(\et_1)>g_{1E}(\et_3)$), that is, in a metastable state.
To predict theoretically  when the transition to a stable state
(which requires switching of polarization and strain) is impossible.
Temperature of such a transition will essentially depends on the
experimental conditions. Nevertheless, we can maintain that this
temperature is not higher than the  temperature of absolute instability
(the point where the local minimum disappears), and, as one can
see at Figure~\ref{st},  the  temperature of absolute instability
decreases with stress $\s$.

\begin{figure}[p]
\begin{center}
\leavevmode
\epsfysize=0.8\textheight
\epsffile{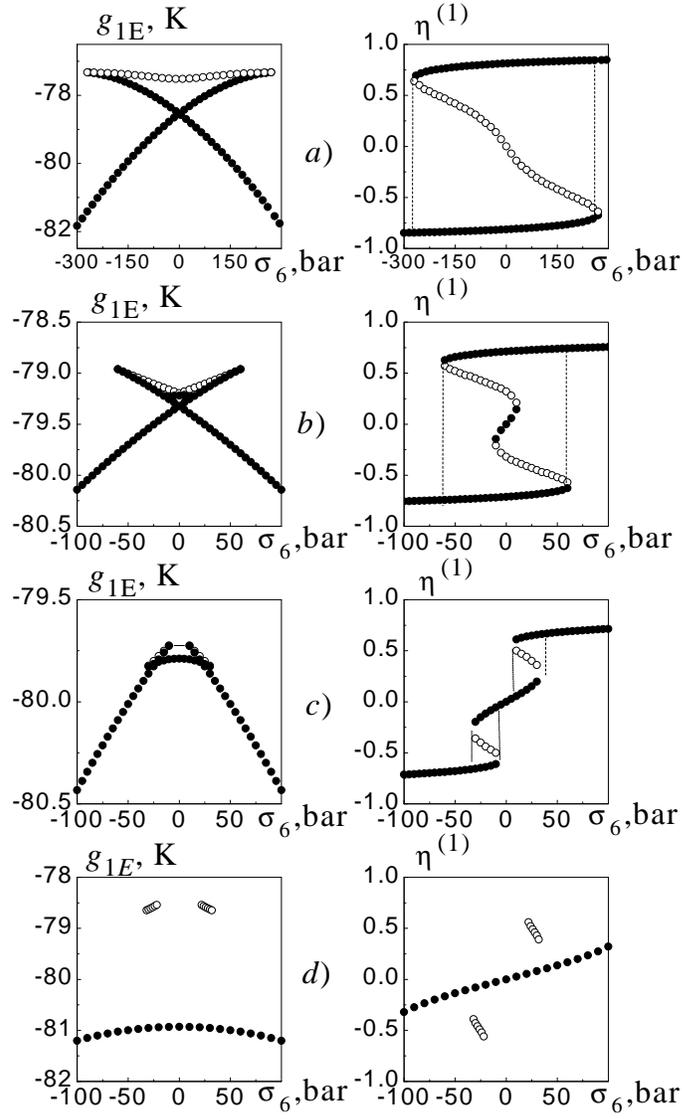}
\end{center}
\caption{\small
Stress dependences of the thermodynamic potential and solutions of
equation for the order parameter at different values of
temperature: a) 209~K, b) 210.5~K, c) 211~K, d) 212~K. $T_{\rm
C0}=210.7$~K. $\circ$ and $\bullet$ correspond to maxima  and minima of
the thermodynamic potential, respectively.} \label{sp} \end{figure}

Let us consider now the dependences of the values of the order parameter
corresponding to minima of the thermodynamic potential on stress $\s$ at
different values of temperature in the vicinity of the transition point,
depicted in fig.~\ref{sp}. At $T<T_{\rm C0}$, the branches of the
thermodynamic potential $g_{1E}(\et_1<0)$ and $g_{1E}(\et_3>0)$ intersect,
and the solutions of Equation (\ref{2.10}) $\et_1$ and $\et_3$ can exist
simultaneously. Experimentally, on changing stress $\s$ a regular
hysteresis  loop $P_3-\s$ should be observed.

At $T_{\rm C0}<T<T^*$, only one of the non-zero minima and the central
minimum of the thermodynamic potential ($\et_1$ and $\et_2$ or
$\et_3$ and $\et_2$) can coexist. Experiment should reveal a double
hysteresis loop. At temperatures higher than the critical
$T^*$, the dependences $g_{1E}(\s)$ and $\et(\s)$ are smooth,
and no jump in the order parameter is observed.  Let us note
that such a sequence of the hysteresis loops -- a single loop  at
$T<T_{\rm C0}$, a double loop at $T_{\rm C0}<T<T^*$, and a gradual change
 at $T>T^*$ -- was obtained yet by Merz for BaTiO$_3$ in electric field
  \cite{Merz} and by Sidnenko and Gladkii \cite{Sidnenko}  for \dekdp\
  in the field $E_3$, and is, apparently, typical for the phase order
  phase transitions in ferroelectrics under fields conjugate to the order
  parameter.

\begin{figure}[hbt]
\begin{center}
\leavevmode
\epsfysize=5cm
\rotate[r]{\epsffile{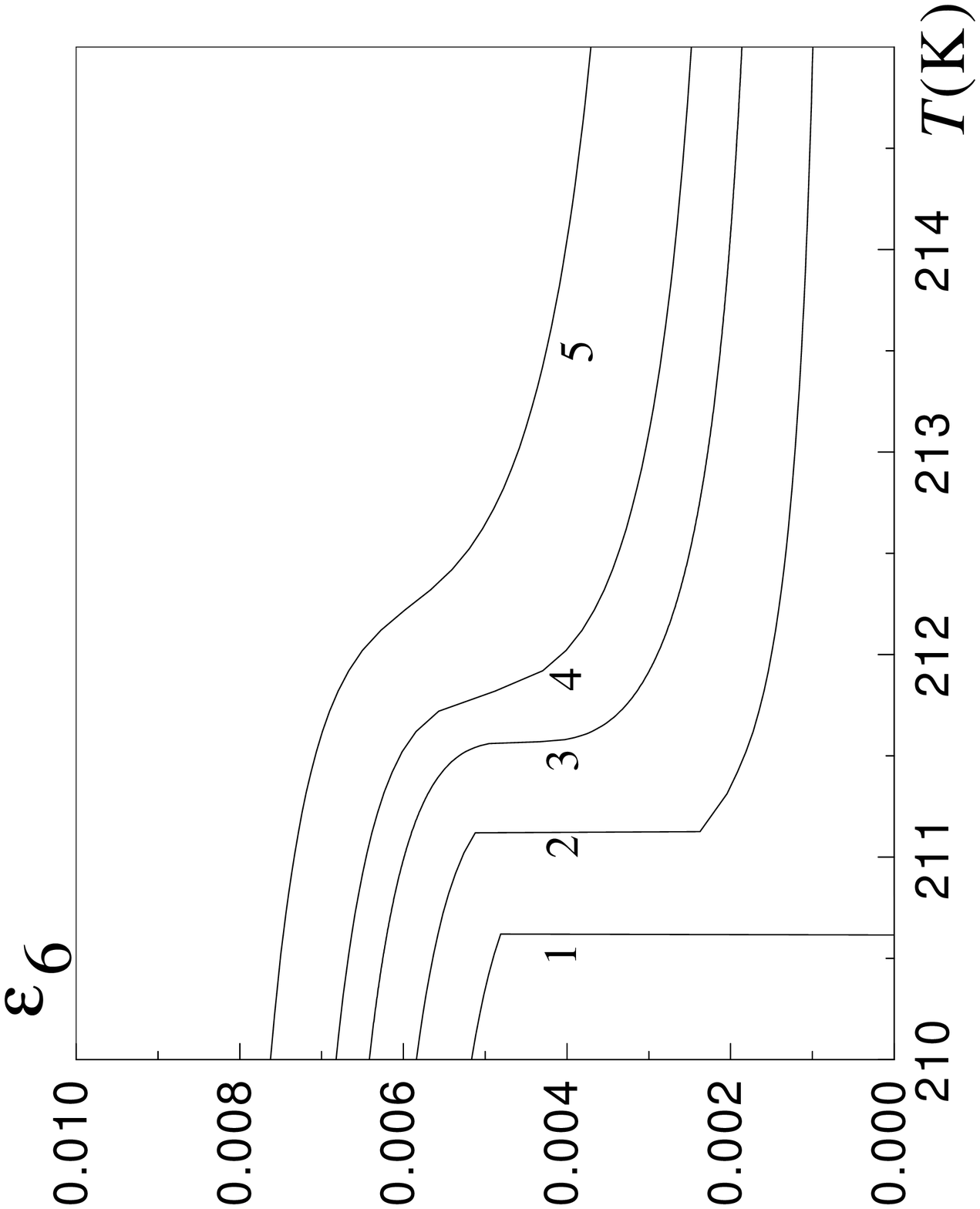}}
\hspace{0.5cm}
\epsfysize=5cm
\rotate[r]{\epsffile{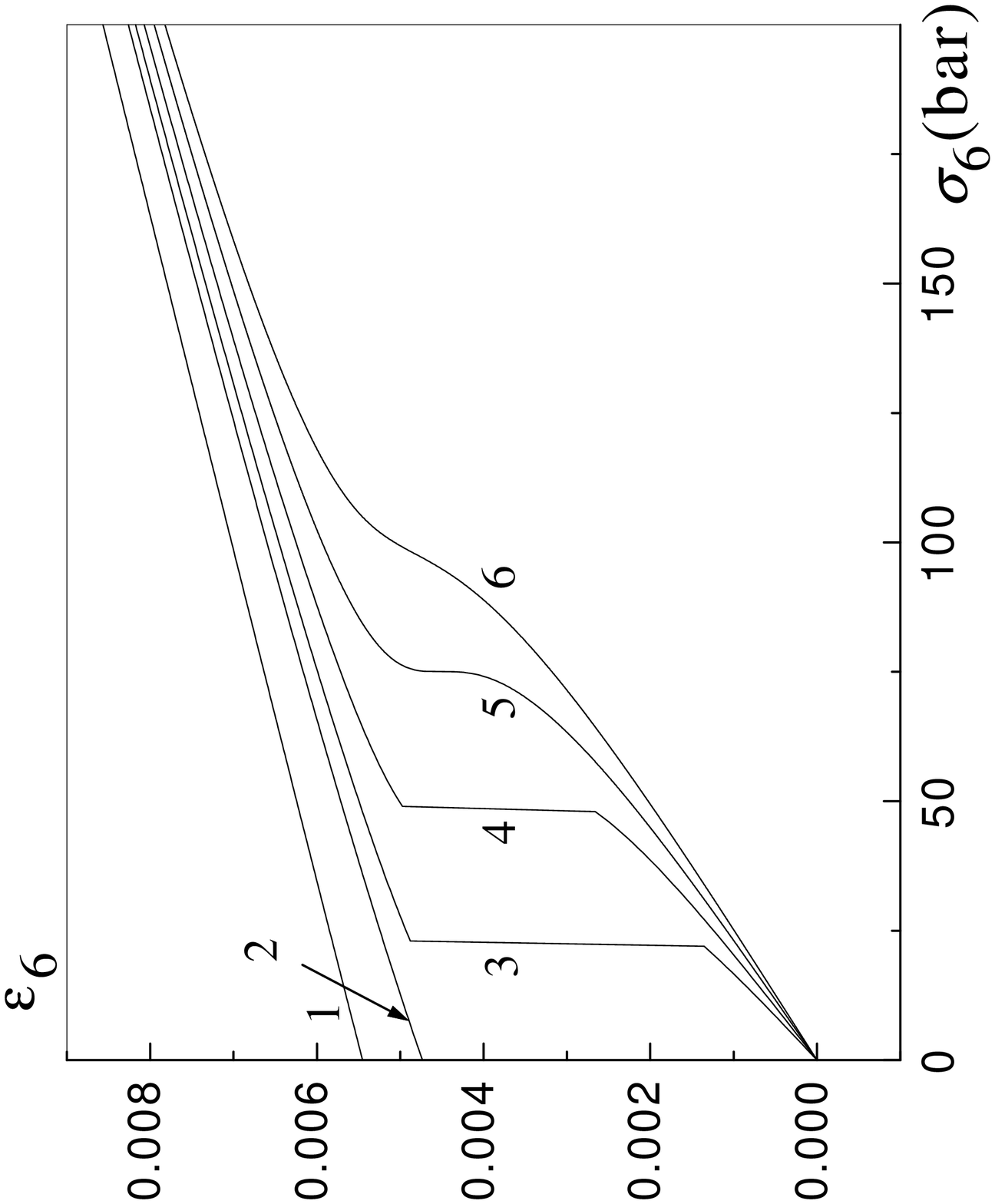}}
\end{center}
\caption{\small
Dependences of strain  $\varepsilon_6$ on temperature at
different values of stress $\sigma_6$ (bar) (left): 1 -- 0; 2 -- 40; 3 --
40; 4 -- 100;   5 -- 150 and on stress $\sigma_6$ at different
values of temperature $T$(K) (right):  1 -- 209.0; 2 -- $T_{\rm
C0}=210.7$; 3 -- 211; 4 -- 211.3; 5 -- $T^*=211.57$; 6 -- 211.8.}
\label{eps} \end{figure}
\begin{figure}[hbt]
\begin{center}
\leavevmode
\epsfysize=5cm
\rotate[r]{\epsffile{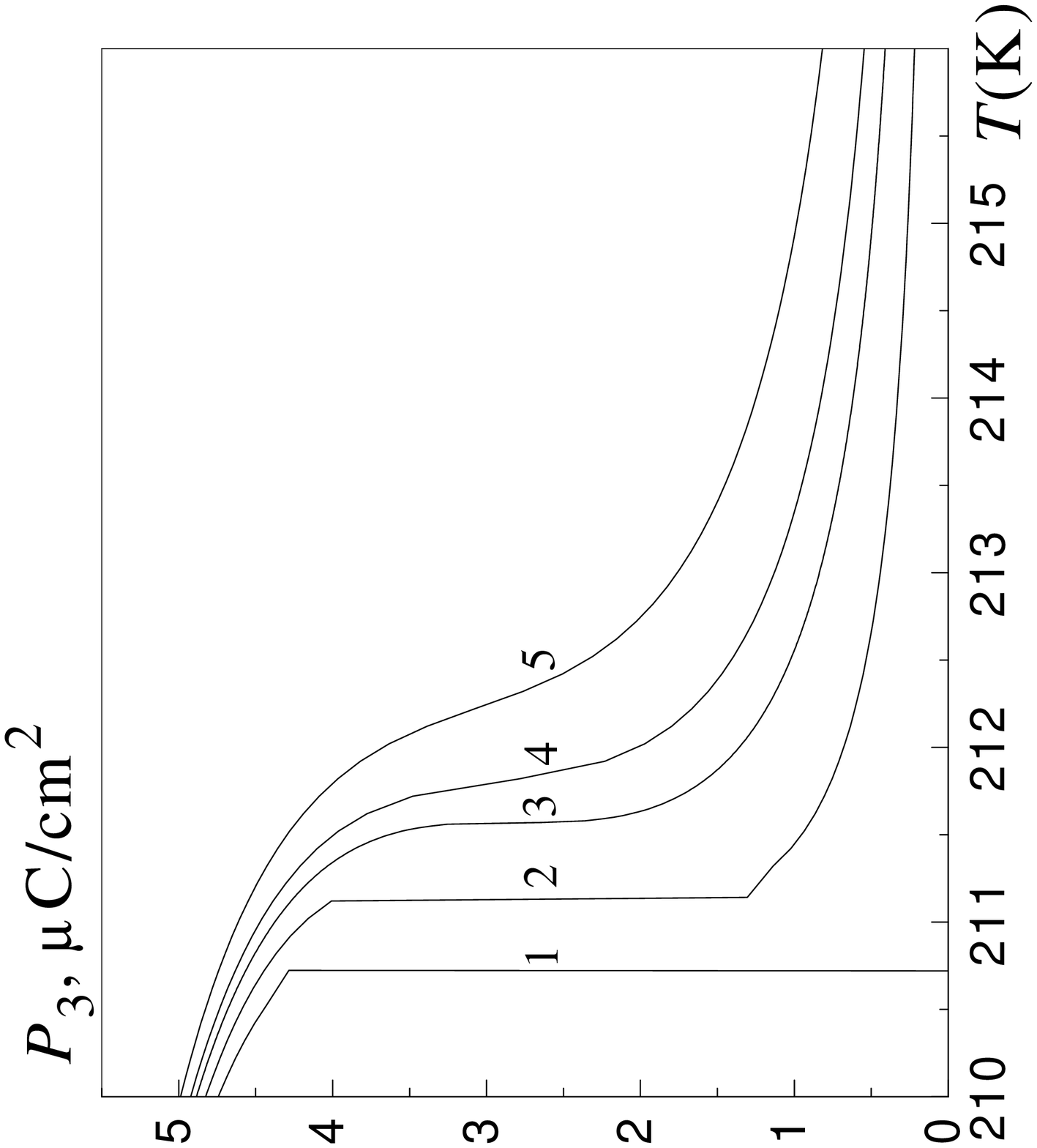}}
\hspace{0.5cm}
\epsfysize=5cm
\rotate[r]{\epsffile{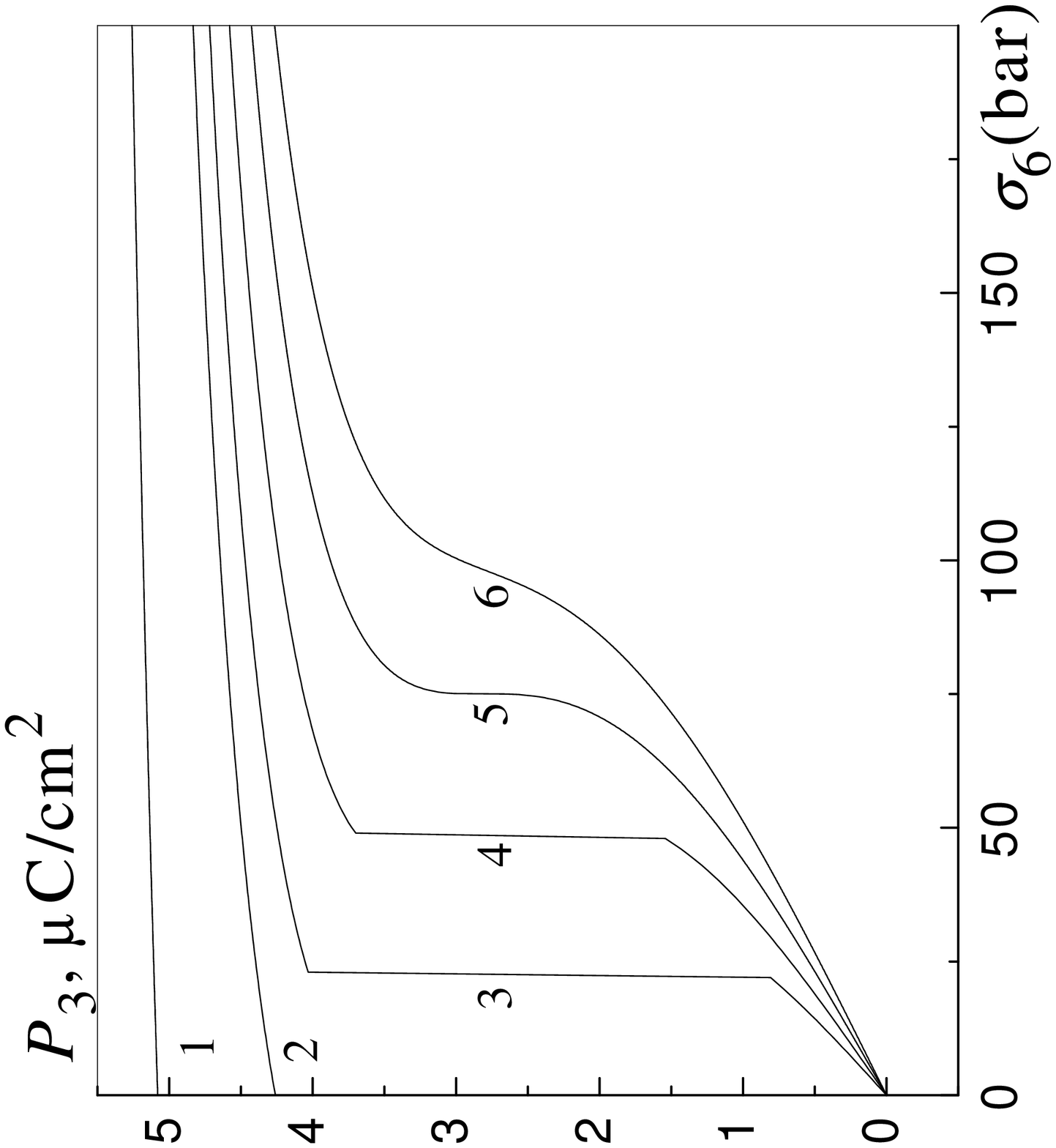}}
\end{center}
\caption{\small
Dependences of polarization $P_3$ on temperature at different
values of stress $\sigma_6$ (bar) (left): 1 -- 0; 2 -- 40; 3 -- $\s^*=75$; 4 --
100;   5 -- 150 and on stress $\sigma_6$ at
different values of temperature $T$(K) (right):  1 -- 209.0; 2 -- $T_{\rm C0}=210.7$; 3 --
211; 4 -- 211.3; 5 -- $T^*=211.57$; 6 -- 211.8.}
 \label{pol} \end{figure}

In fig.~\ref{eps} we present the temperature and stress  dependences of
strain  $\varepsilon_6$ at different values of stress
$\sigma_6$ and different values of temperature. Fig.~\ref{pol} illustrates
the analogous  dependences of polarization $P_3$. As one can
see, these quantities exhibit the identical  variation with temperature or
stress.  The spontaneous strain $\varepsilon_6$ and polarization
$P_3$ at $\sigma_6=0$ in the ferroelectric phase slightly decrease with
temperature and jump to zero at $T=T_{\rm C}$.   When  stress $\sigma_6$ is
applied, the magnitudes of strains $\varepsilon_6$ and polarization $P_3$
increase, whereas the jumps $\Delta \varepsilon_6$ and $\Delta P_3$ decrease;
at $\sigma_6 = \sigma_6^*$ and $T=T_{\rm C}^*$, the jumps vanish.
 At $\sigma_6 \neq 0$
$\varepsilon_ 6$ and $P_3$ differ from zero above the phase transition
and slightly decrease with temperature.  At  $T_{\rm C}<T<T_{\rm C}^*$
values of $\varepsilon_6$ and $P_3$ increase with stress
$\sigma_6$, having upward jumps at  stresses corresponding to the curve of
phase equilibrium with $\Delta
\varepsilon_6$ and $\Delta P_3$ decreasing with stress and vanishing
at $\sigma_6=\sigma_6^*$.

\clearpage

Further increase in
stress slightly raises up the strain
$\varepsilon_6$ and polarization $P_3$. At temperatures $T<T_{\rm C}$ and
$T-T_{\rm C}>3$ K values of $\varepsilon_6$ and $P_3$ exhibit a nearly
linear increase with stress $\sigma_6$.
At changing the sign of the stress  $\sigma_6$, the  absolute values of
the  strain $\varepsilon_6$ and polarization $P_3$ do not change, but
become negative.

Let us discuss the theoretical results for the  stress influence
on the physical characteristics of the  K(H$_{0.11}$D$_{0.89}$)$_2$PO$_4$
crystal. The major stress $\s$ effects on the physical characteristics are
related to the changes induced by this stress in the character of the phase
transition and, therefore, essential only in the vicinity of the transition
point; at $|T-T_{\rm C}|>5$~K, these effects are negligibly small.

\begin{figure}[hbt]
\begin{center}
\leavevmode
\epsfysize=5cm
\rotate[r]{\epsffile{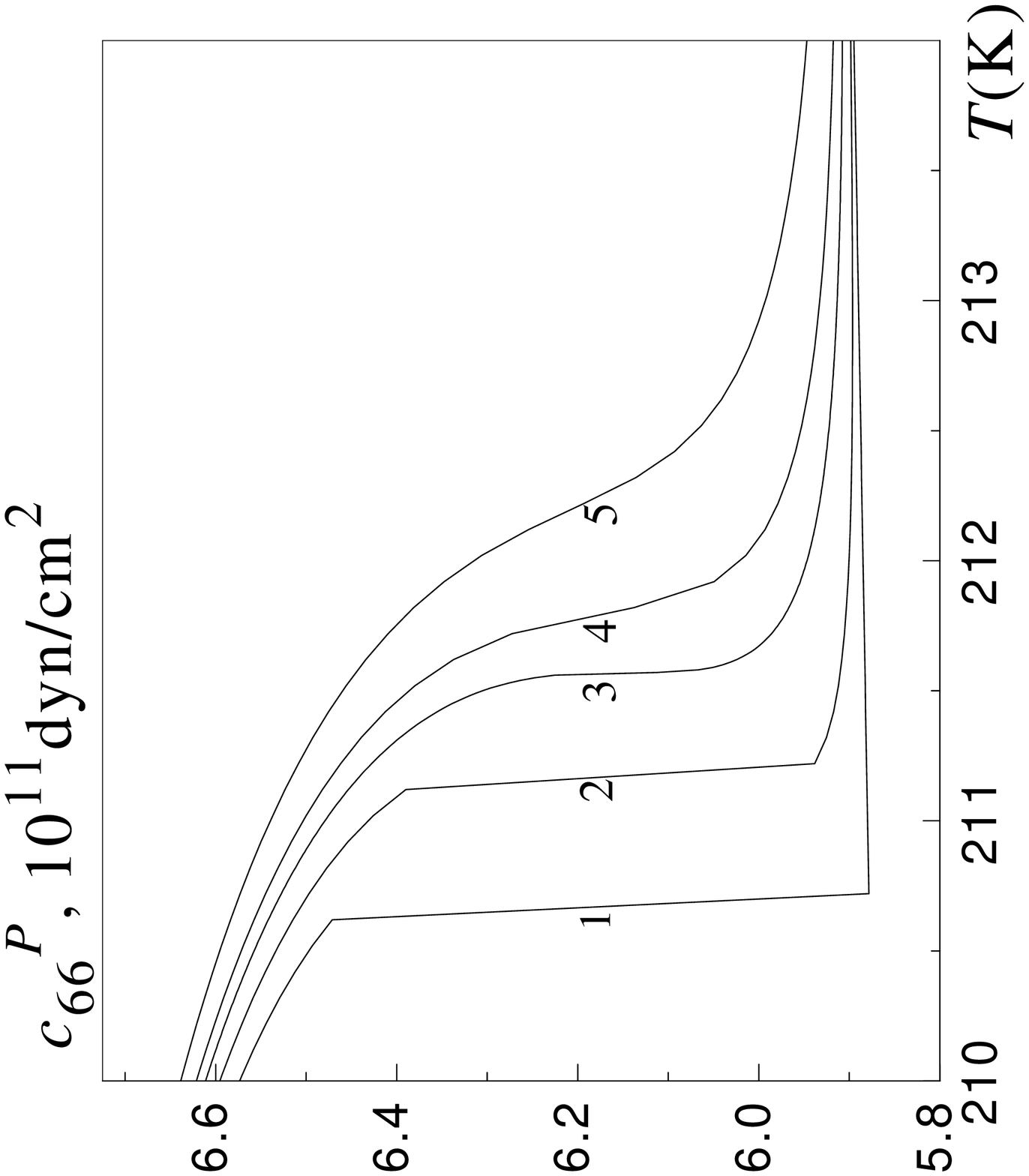}}
\hspace{0.5cm}
\epsfysize=5cm
\rotate[r]{\epsffile{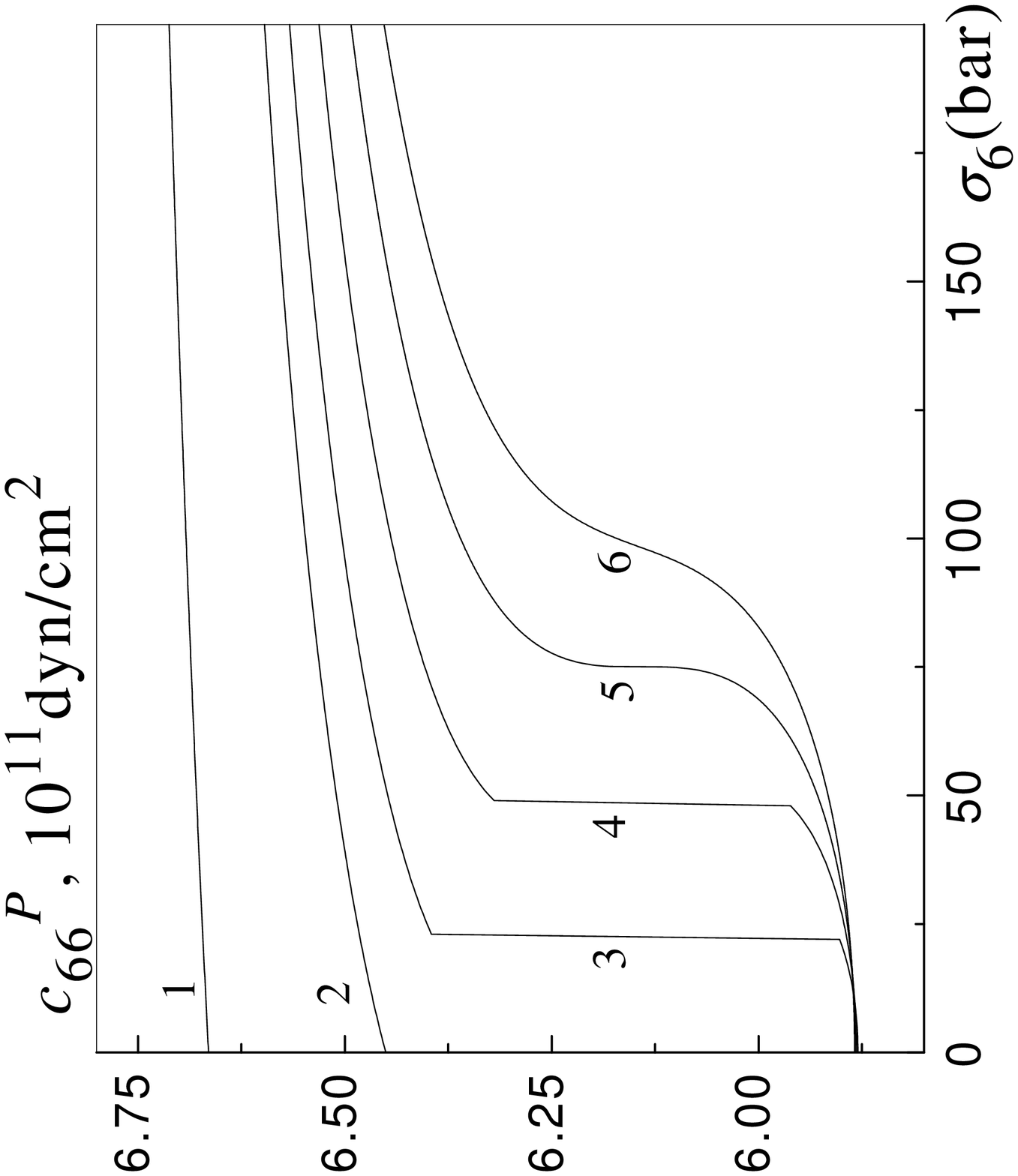}}
\end{center}
\caption{\small
Dependences of elastic constant $c_{66}^P$ on temperature at different
values of stress $\sigma_6$ (bar) (left): 1 -- 0; 2 -- 40; 3 -- $\s^*=75$; 4 --
100;   5 -- 150 and on stress $\sigma_6$ at
 different values of temperature $T$(K) (right):  1 -- 209.0; 2 -- $T_{\rm
 C0}=210.7$; 3 -- 211; 4 -- 211.3; 5 -- $T^*=211.57$; 6 -- 211.8.}
 \label{cp} \end{figure}

In fig.~\ref{cp} we plot the temperature and stress dependences of the
elastic constant $c_{66}^P$ at different values of stress $\sigma_6$ and
temperature. As one can see,  the elastic constant
$c_{66}^P$ change with temperature and stress exactly as the strain
$\varepsilon_6$ and polarization $P_3$ do.


\begin{figure}[hbt]
\begin{center}
\leavevmode
\epsfysize=6.cm
{\epsffile{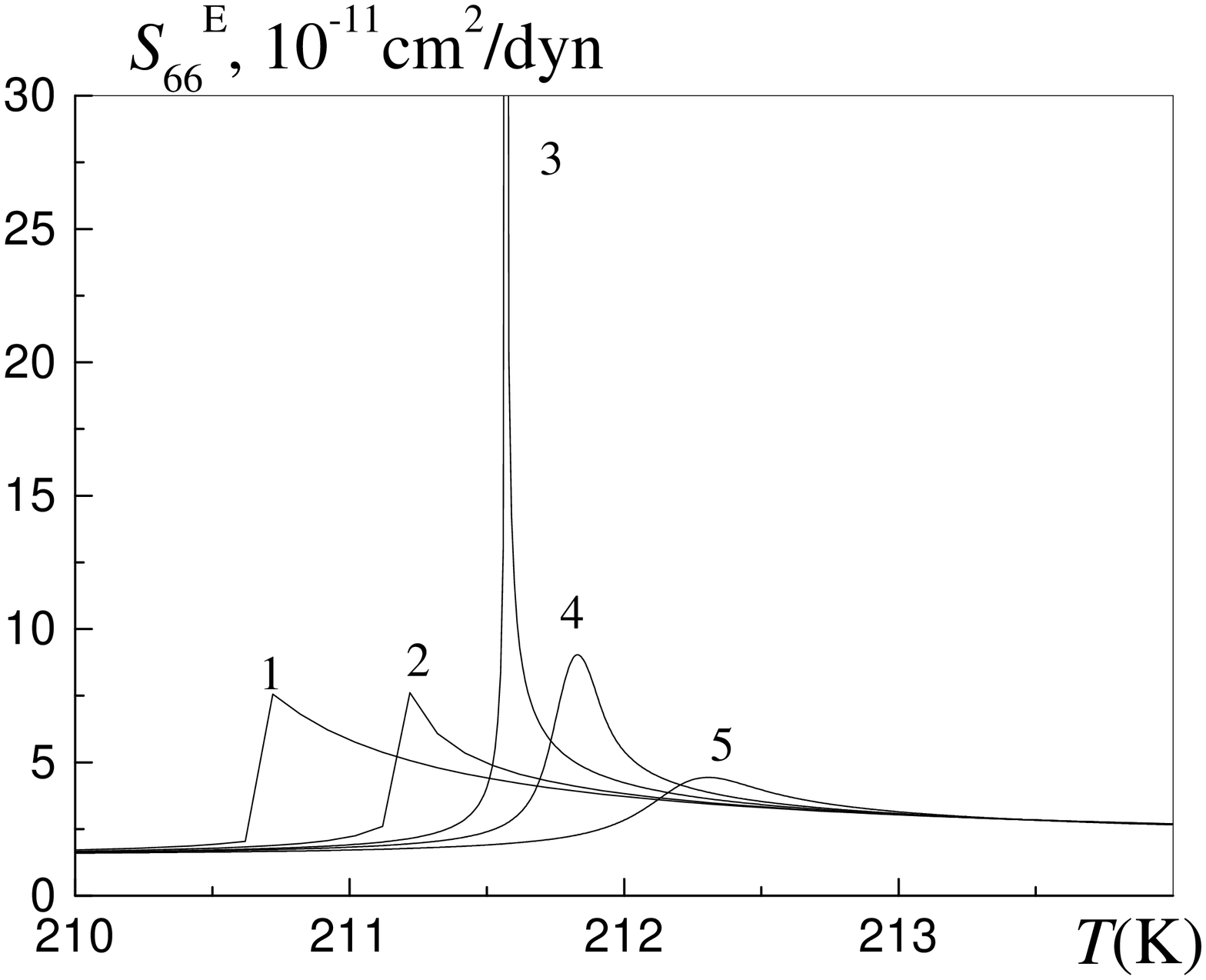}}
\end{center}
\caption{\small
Temperature dependence of compliance $s_{66}^E$ at different values of
$\sigma_6$ (bar):
1 -- 0; 2 -- 40; 3 -- $\s^*=75$; 4 -- 100; 5 -- 150.
} \label{se} \end{figure}

\begin{figure}[hbt]
\begin{center}
\leavevmode
\epsfysize=5cm
\rotate[r]{\epsffile{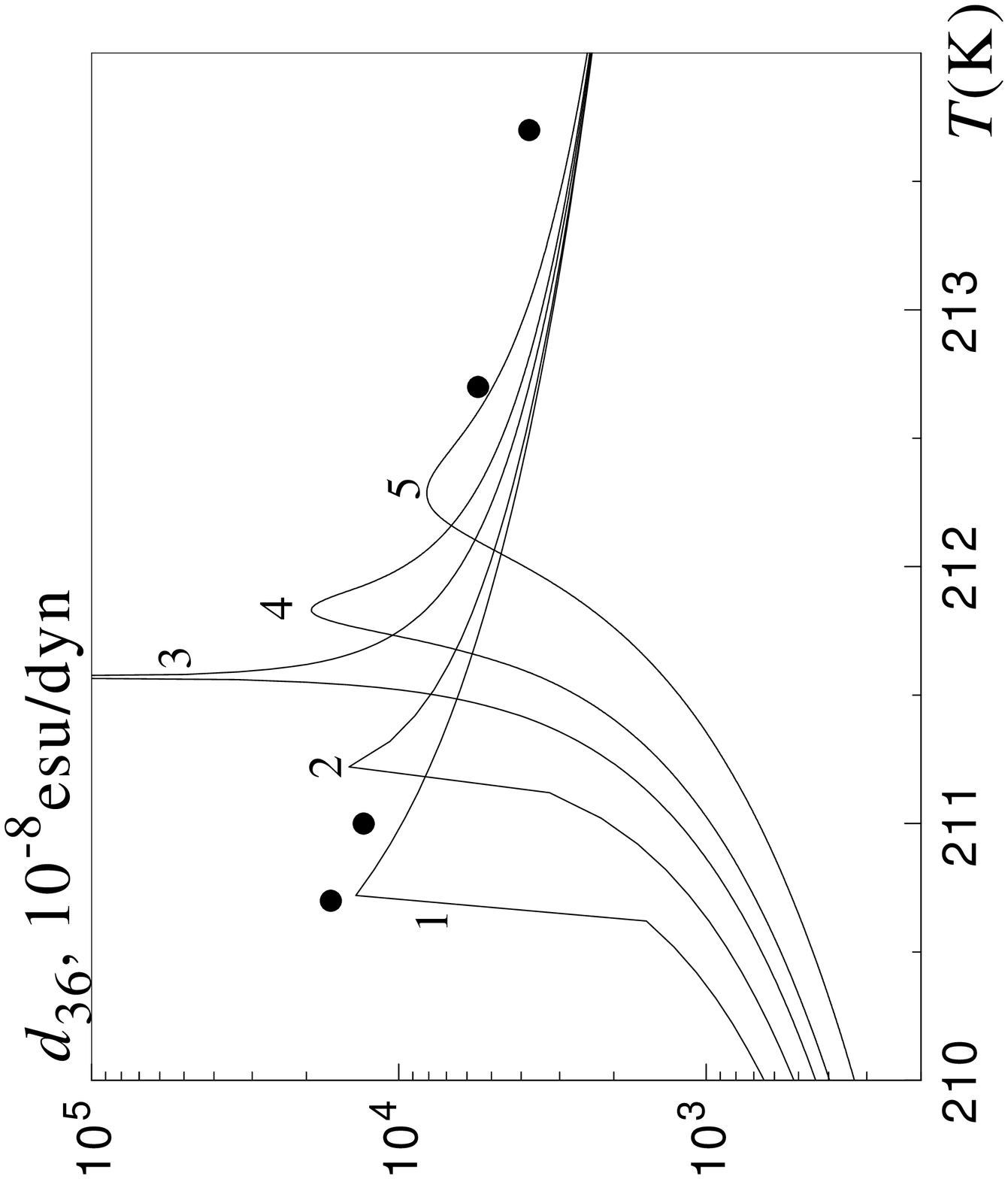}}
\hspace{0.5cm}
\epsfysize=5cm
\rotate[r]{\epsffile{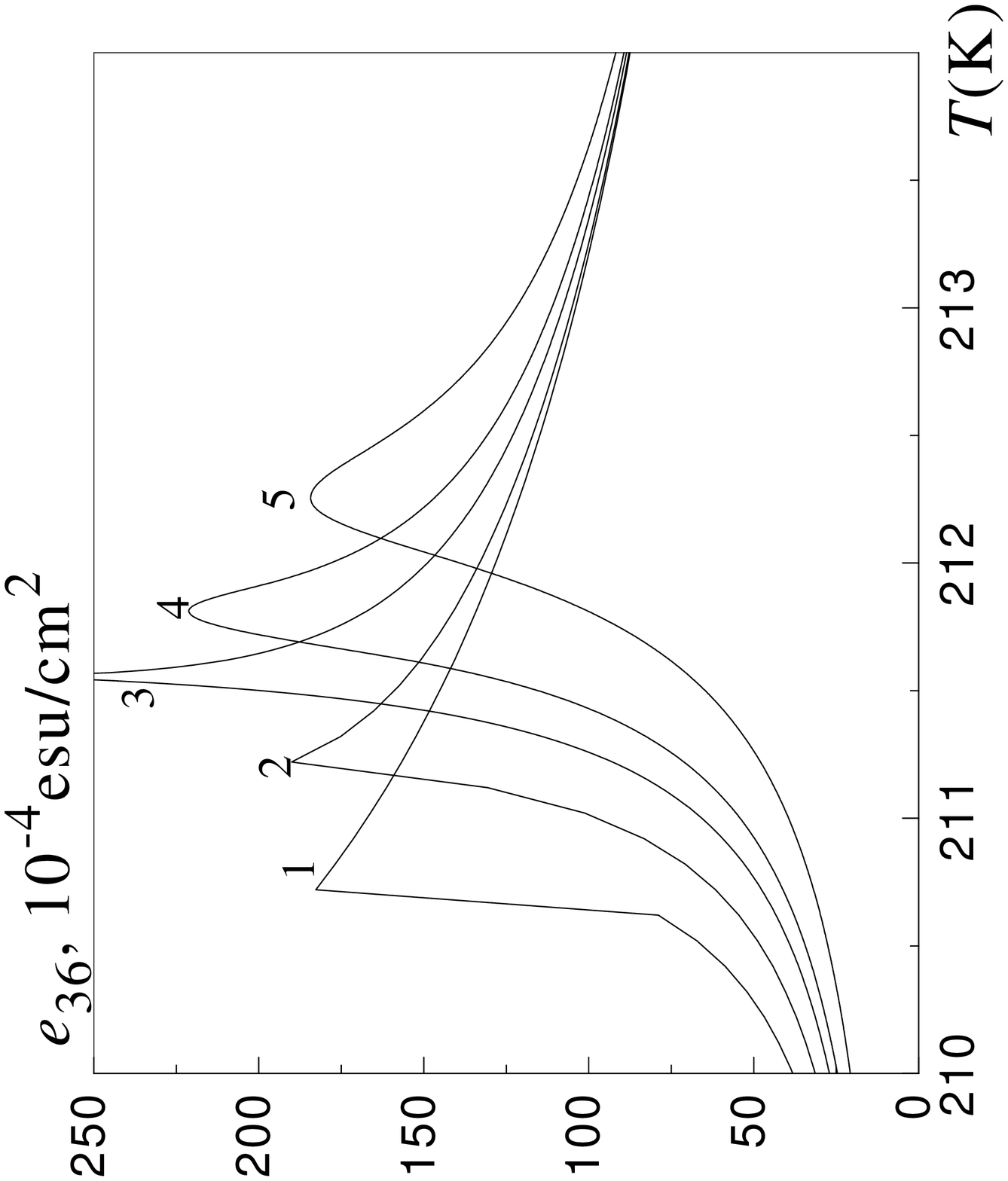}}
\end{center}
\caption{\small
Temperature dependences of coefficient of piezoelectric strain
$d_{36}$  and coefficient of piezoelectric stress $e_{36}$ at different
values of stress  $\sigma_6$ (bar):
1 -- 0; 2 -- 40; 3 -- $\s^*=75$; 4 -- 100; 5 -- 150. }
 \label{d_e} \end{figure}

\begin{figure}[hbt]
\begin{center}
\leavevmode
\epsfysize=6.5cm
\rotate[r]{\epsffile{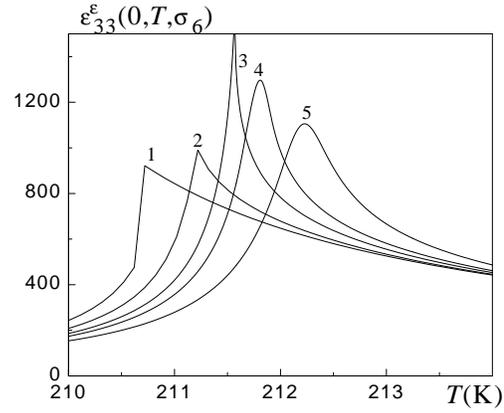}}
\end{center}
\caption{\small
Dependences of dielectric permittivity $\varepsilon_{33}^\varepsilon$
on temperature at different values of stress $\sigma_6$ (bar): 1
-- 0; 2 -- 40; 3 -- $\s^*=75$; 4 -- 100; 5 -- 150. } \label{hi} \end{figure}

Temperature dependences of compliance $s_{66}^E$ (fig.~\ref{se}),
coefficient of the piezoelectric strain $d_{36}$ and
stress $e_{36}$ (fig.~\ref{d_e}), and of the  dielectric
permittivity $\varepsilon_{33}^\varepsilon$ (fig.~\ref{hi}) at different
values of stress $\sigma_6$ are similar.  The maxima of these
characteristics shift to higher temperatures with stress. At
$\sigma_6 < \sigma_6^*$, these characteristics reaches their
maximal values  at the transition temperatures. At $\sigma_6 =
\sigma_6^*$ ÷ $T = T_{\rm C}^*$ the second order phase transition takes
place accompanied by a sharp increase in these quantities.



\begin{figure}[hbt]
\begin{center}
\leavevmode
\epsfysize=5cm
\rotate[r]{\epsffile{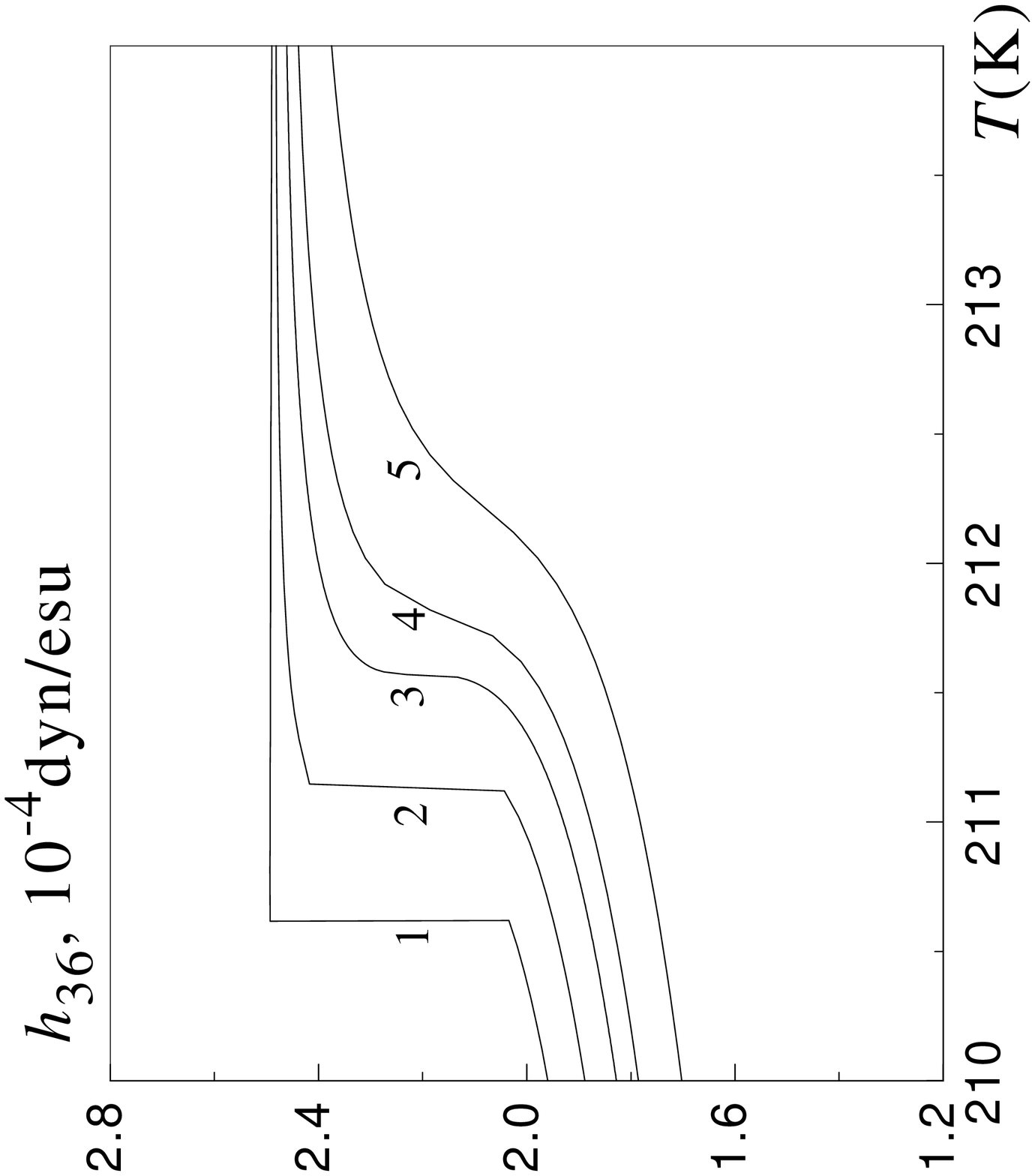}}
\hspace{0.5cm}
\epsfysize=5cm
\rotate[r]{\epsffile{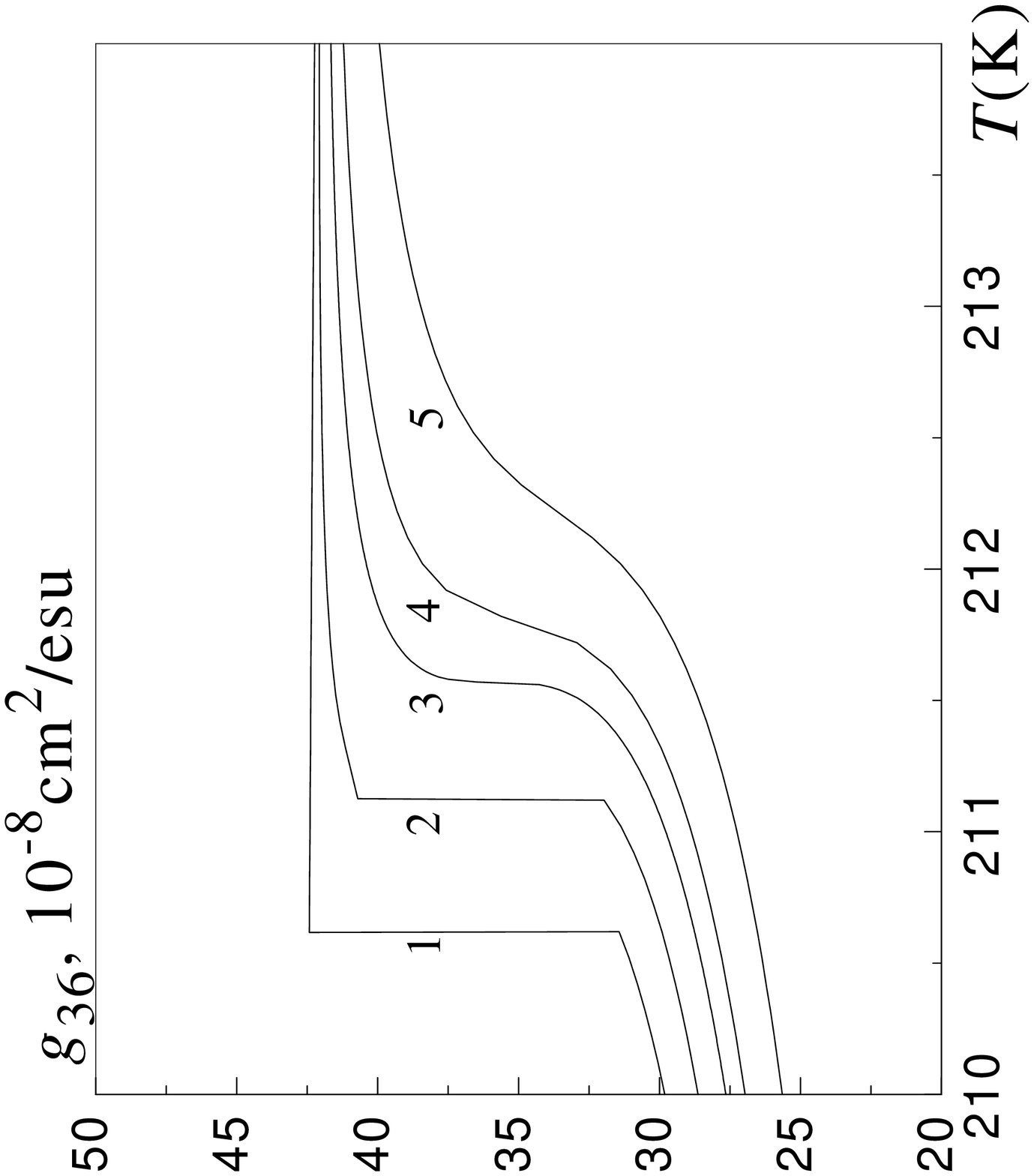}}
\end{center}
\caption{\small
Temperature dependence of constant of piezoelectric
stress $h_{36}$ and constant of piezoelectric strain $g_{36}$  at
different values of stress $\sigma_6$ (bar):  1 -- 0; 2 -- 40; 3 -- $\s^*=75$; 4
-- 100; 5 -- 150. } \label{h_g} \end{figure}

The temperature curves of the constants of  piezoelectric stress
$g_{36}$  and
piezoelectric strain $g_{36}$ (fig.~\ref{h_g}) at different
values of stress $\sigma_6$ are similar to each other. At $\sigma_6 <
\sigma_6^*$, minor jumps in $h_{36}$ and $g_{36}$ are observed, which
vanish at $\sigma_6 = \sigma_6^*$.  At $\sigma_6 > \sigma_6^*$ $h_{36}$ ÷
$g_{36}$ increase with temperature, tending asymptotically to the values
of $h_{36}$, $g_{36}$ at $\sigma_6=0$.


\clearpage

\begin{figure}[hbt]
\begin{center}
\leavevmode
\epsfysize=6cm
\rotate[r]{\epsffile{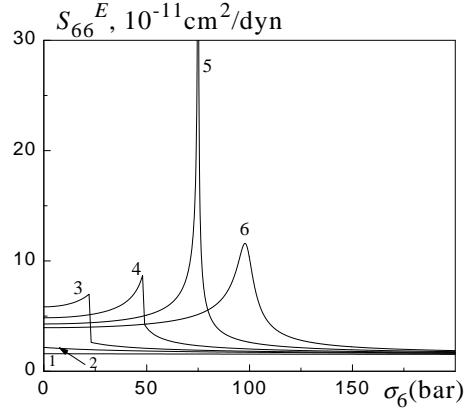}}
\end{center}
\caption{\small
Dependences of compliance $s_{66}^E$ on stress $\sigma_6$ at
different values of temperature $T$~(K): 1 -- 209.0; 2 -- $T_{\rm C0}=210.7$; 3 -- 211; 4
-- 211.3; 5 -- $T^*=211.57$; 6 -- 211.8.
}
 \label{se_p}
\end{figure}

\begin{figure}[hbt]
\begin{center}
\leavevmode
\epsfysize=5cm
\rotate[r]{\epsffile{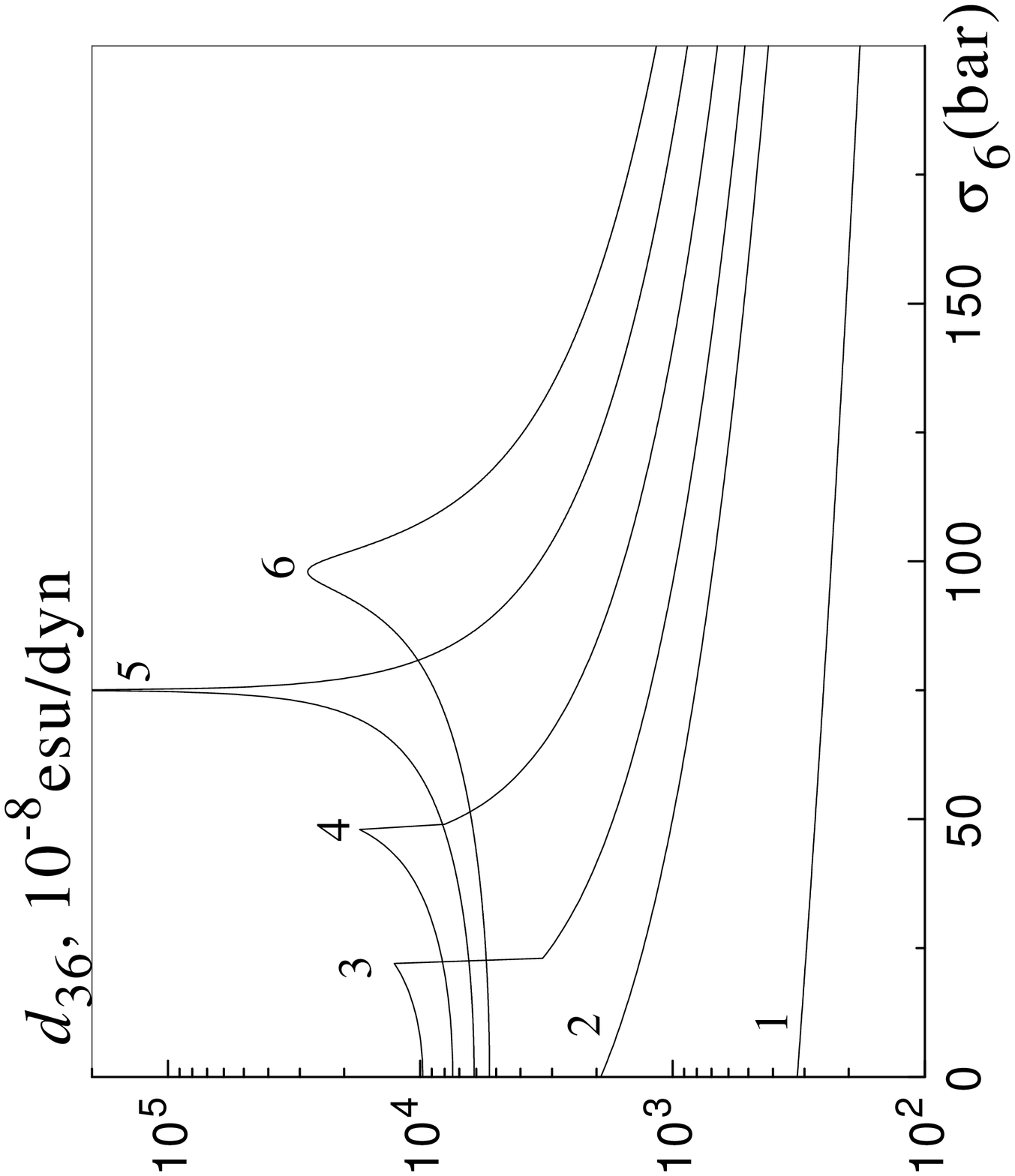}}
\hspace{0.5cm}
\epsfysize=5cm
\rotate[r]{\epsffile{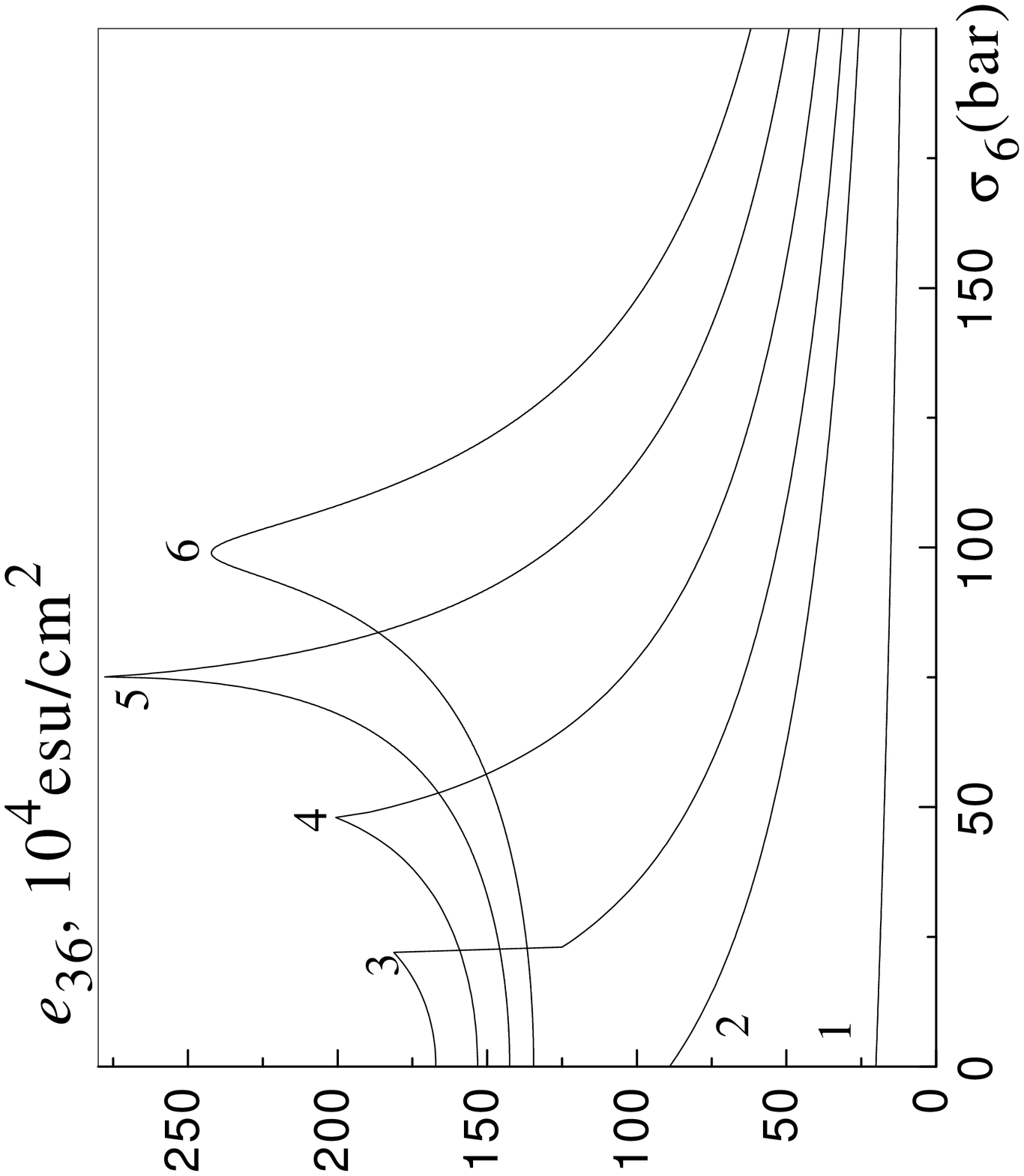}}
\end{center}
\caption{\small
Dependences of coefficient of piezoelectric strain $d_{36}$ and
coefficient of  piezoelectric stress $e_{36}$ on stress $\sigma_6$ at
different
temperatures T~(K):  1 -- 209.0; 2 -- $T_{\rm C0}=210.7$; 3 -- 211; 4
-- 211.3; 5 -- $T^*=211.57$; 6 -- 211.8.
}
 \label{d_e_p}
\end{figure}

\begin{figure}[hbt]
\begin{center}
\leavevmode
\epsfysize=6cm
\rotate[r]{\epsffile{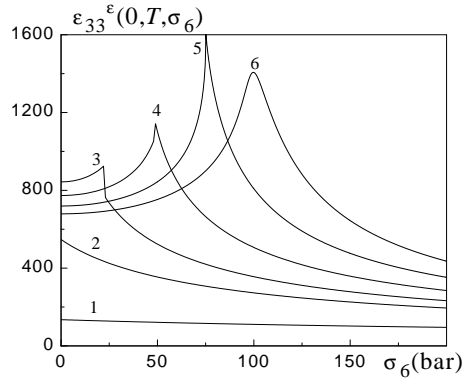}}
\end{center}
\caption{\small
Dependences of dielectric permittivity $\varepsilon_{33}^\varepsilon$
on stress $\sigma_6$ at different values of temperature T~(K):  1 --
210.0; 2 -- $T_{\rm C0}=210.7$; 3 -- 211; 4 -- 211.3; 5 --
$T^*=211.57$;  6 -- 211.8.  } \label{hi_p} \end{figure}

The stress dependences of the compliance $s_{66}^E$, coefficients of
piezoelectric strain $d_{36}$, piezoelectric stress $e_{36}$, dielectric
permittivity $\varepsilon_{33}^\varepsilon$ at different values of
temperature are presented in  figs.~\ref{se_p}, \ref{d_e_p}, \ref{hi_p}.

At temperatures below $T_{\rm C}$, these  characteristics
decrease with stress $\sigma_6$, decreasing being the stronger,
the closer temperature is to $T_{\rm C0}$. At
$T_{C0}< T < T_{\rm C}^*$, the mentioned  characteristics increase with
stress $\sigma_6$,  having downward jumps at  stresses corresponding to
the curve of phase equilibrium, with the jumps vanishing at $T = T_{\rm
C}^*$ and $\sigma_6 = \sigma_6^*$. The further increase in stress lowers
down $s_{66}^E$, $d_{36}$,  $e_{36}$ and $\varepsilon_{33}^\varepsilon$.
At $T > T_{\rm C}^*$, these quantities  exhibit a gradual
increase with stress, reach their maxima, and  decrease on further
increase in stress.
As temperature rises, the maximal  values of the quantities
decrease and shift to higher stresses. At $\Delta T \geq 15$ K, the values
of $s_{66}^E$, $d_{36}$, $e_{36}$, and $\varepsilon_{33}^\varepsilon$
are stress $\sigma_6$ independent.

\clearpage

\begin{figure}[hbt]
\begin{center}
\leavevmode
\epsfysize=5cm
\rotate[r]{\epsffile{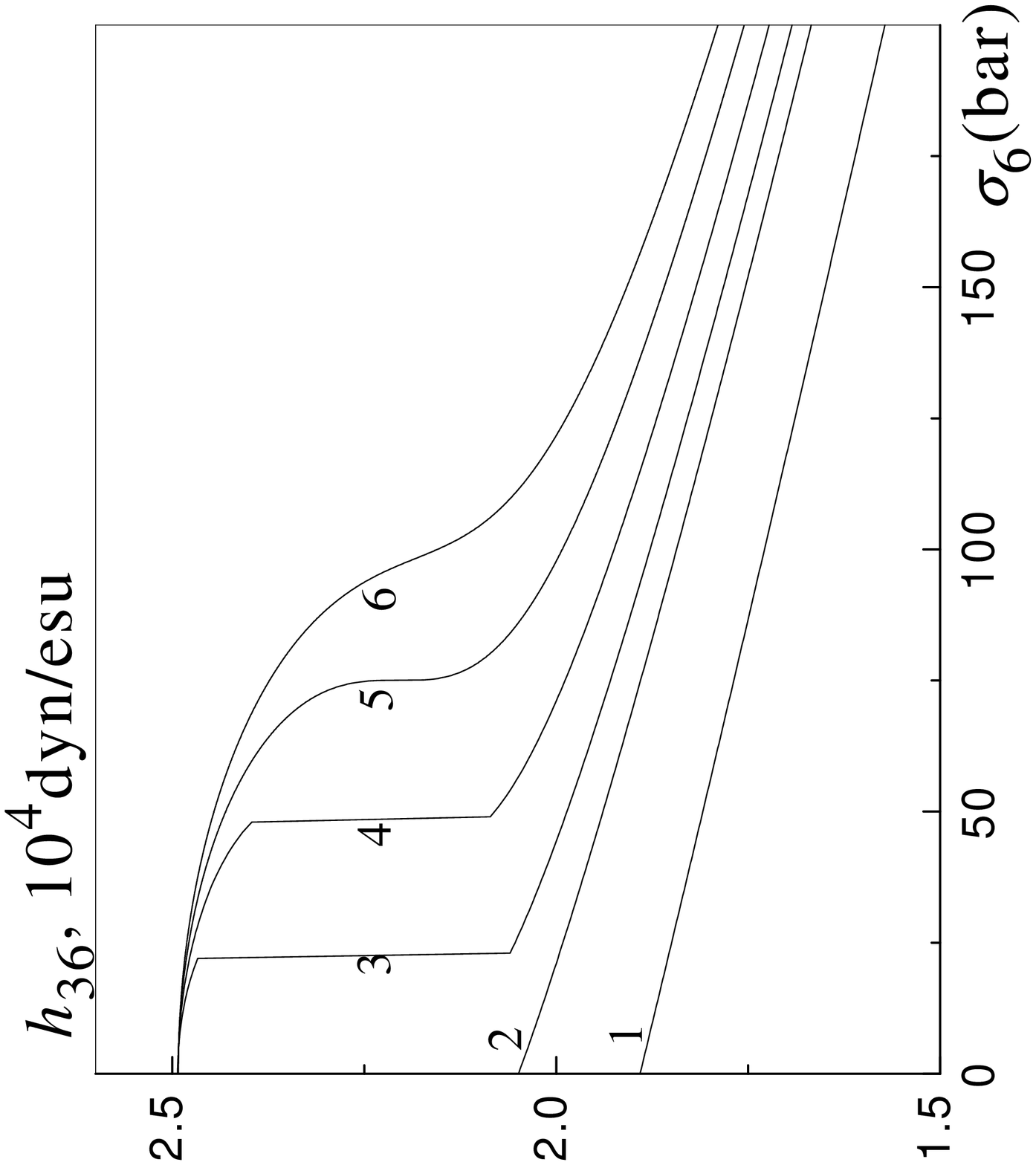}}
\hspace{0.5cm}
\epsfysize=5cm
\rotate[r]{\epsffile{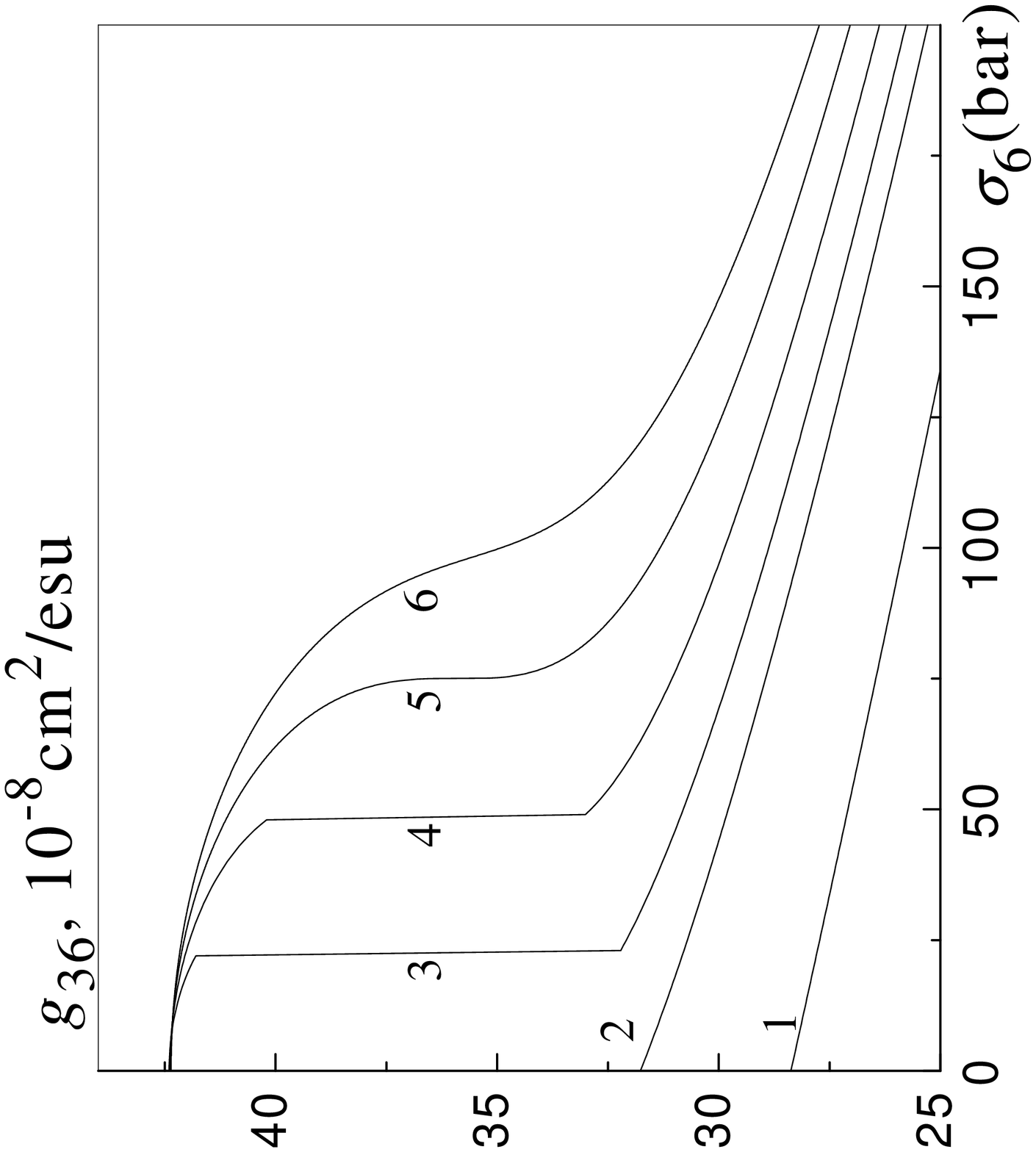}}
\end{center}
\caption{\small
Dependences of constant of piezoelectric stress $h_{36}$  and constant of
piezoelectric strain $g_{36}$  on stress $\sigma_6$ at
different values of temperature T~(K):   1 -- 209.0; 2 -- $T_{\rm
C0}=210.7$; 3 -- 211; 4 -- 211.3; 5 -- $T^*=211.57$; 6 -- 211.8.  }
 \label{h_g_p}
\end{figure}

In fig.\ref{h_g_p} we depict the dependences of constants of
piezoelectric stress $h_{36}$ and piezoelectric strain $g_{36}$ on
stress $\sigma_6$ at different values of temperature. At temperatures $T
\leq T_{\rm C}$ the values of $h_{36}$ and $g_{36}$ decrease with stress
$\sigma_6$.  If $T_{\rm C} < T < T_{\rm C}^*$, the constants $h_{36}$ and
$g_{36}$ are slightly lowered down by stress. At stresses corresponding to
the curve of phase equilibrium they decrease with a jump, and at $\sigma_6
= \sigma_6^*$ the  jump magnitude turns to zero, and the further increase
in stress gives rise to an almost  linear decrease in $h_{36}$ and
$g_{36}$.  At temperatures $T > T_{\rm C}^*$ $h_{36}$ and $g_{36}$ exhibit
a gradual decrease with stress $\sigma_6$

\begin{figure}[hbt]
\begin{center}
\leavevmode
\epsfysize=4cm
{\epsffile{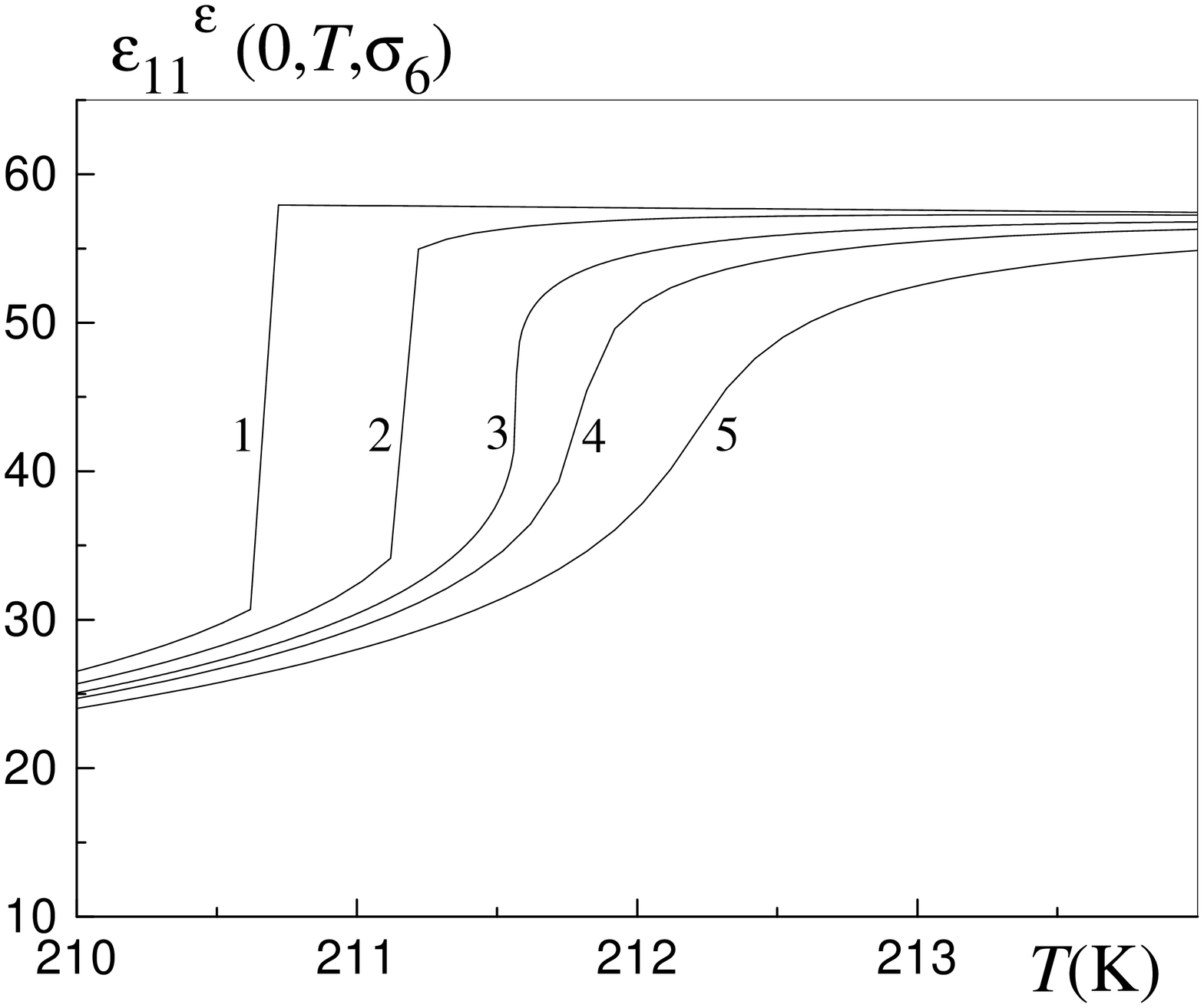}}
\hspace{0.5cm}
\epsfysize=4cm
{\epsffile{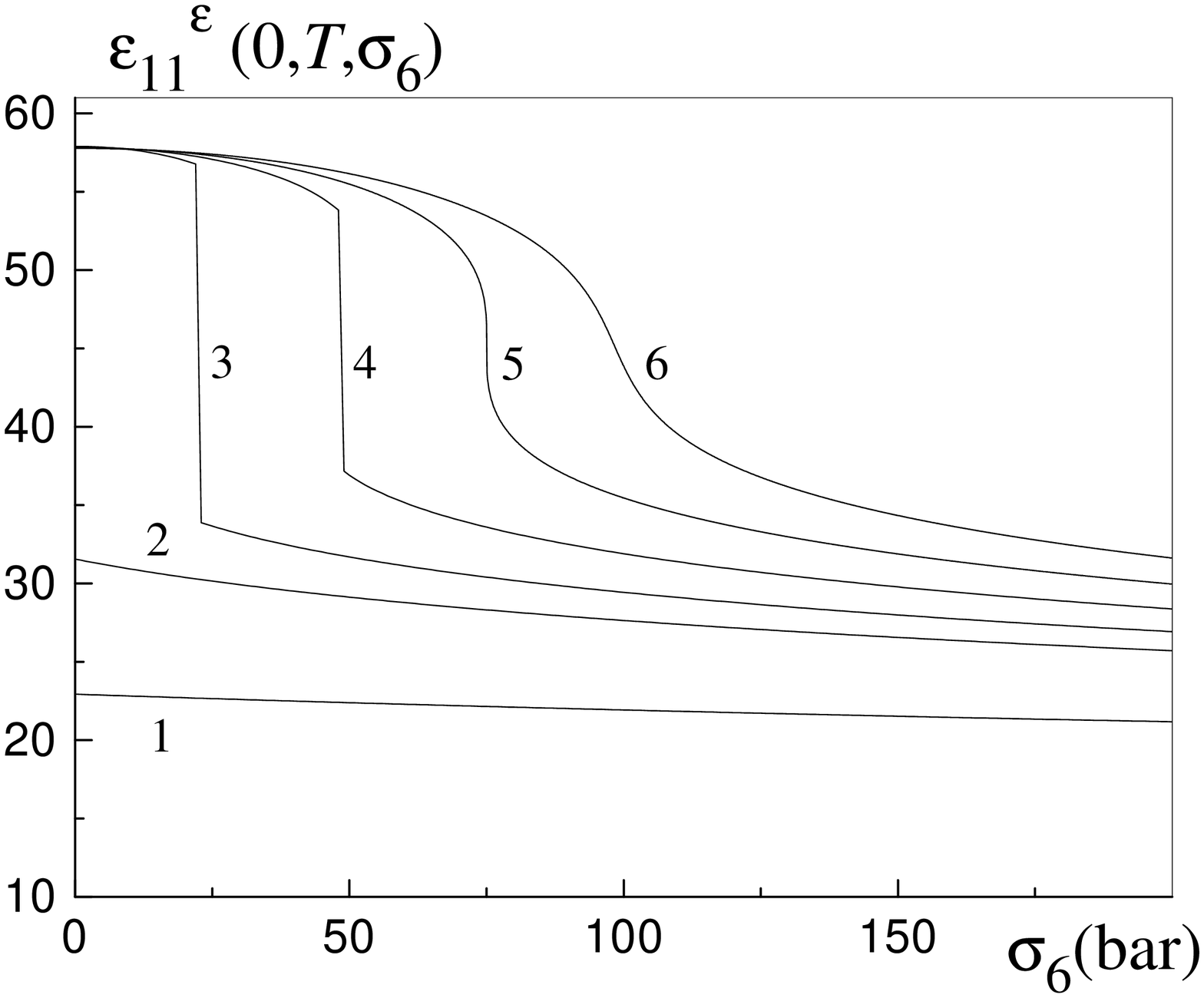}}
\end{center}
\caption{\small
Dependences of dielectric permittivity $\varepsilon_{11}^\varepsilon$ on
temperature at different
values of stress $\sigma_6$ (bar) (left): 1 -- 0; 2 -- 40; 3 -- $\s^*=75$;
4 -- 100;   5 -- 150 and on stress $\sigma_6$ at
different values of temperature $T$(K) (right):  1 -- 209.0; 2 -- $T_{\rm
C0}=210.7$;  3 -- 211; 4 -- 211.3; 5 -- $T^*=211.57$; 6 -- 211.8.  $\bullet$
 -- \protect\cite{a25}. } \label{trans} \end{figure}

Let us consider now the temperature and stress dependence of the
transverse dielectric permittivity  $\varepsilon_{11}$.
Calculations were carried out at $\nu_1 - \nu_3 = 10$K,  $\mu_\perp = 2.79
\cdot 10^{-18}$ esu $\cdot$ cm, $\varepsilon_\infty = 10$. In fig.\ref{trans}
we depict the dependence of $\varepsilon_{11}$ on temperature at different
values of stress $\sigma_6$ and on stress at different values of temperature
along with the experimental points for $\sigma_6=0$.

  In the paraelectric phase, the
experimental data are described by the theory rather well.  As
temperature rises,  $\varepsilon_{11}$ gradually increases in
the ferroelectric phase, has an upward jump at
at the transition  temperature, the jump magnitude being equal to
zero at $\sigma_6 = \sigma_6^*$.
At $\sigma_6 =0$ the values of $\varepsilon_{11}$ slightly
decreases with temperature in the paraelectric phase,
whereas at $\sigma_6 \neq 0$ after the upward jumps it
asymptotically tends to the values of  $\varepsilon_{11}$ at  $\sigma_6
=0$. At $\sigma_6 = \sigma_6^*$ $\varepsilon_{11}$ exhibits only smooth
increase with temperature.

At temperatures $T < T_{\rm C}$, the magnitude of
$\varepsilon_{11}$ is lowered down by stress
$\sigma_6$.  If $T_{\rm C} < T < T_{\rm C}^*$,
$\varepsilon_{11}$ slightly decreases with stress, then
at stresses corresponding to the curve of phase equilibrium it has a
downward jump, the size of which $\Delta
\varepsilon_{11}$ is zero at $\sigma_6 = \sigma_6^*$. At further increase
in stress $\sigma_6$, $\varepsilon_{11}$
decreases linearly. At $T > T_{\rm
C}^*$ the permittivity $\varepsilon_{11}$ exhibits a smooth decrease with
stress $\sigma_6$.

\clearpage

\begin{figure}[hbt]
\begin{center}
\leavevmode
\epsfysize=5cm
\rotate[r]{\epsffile{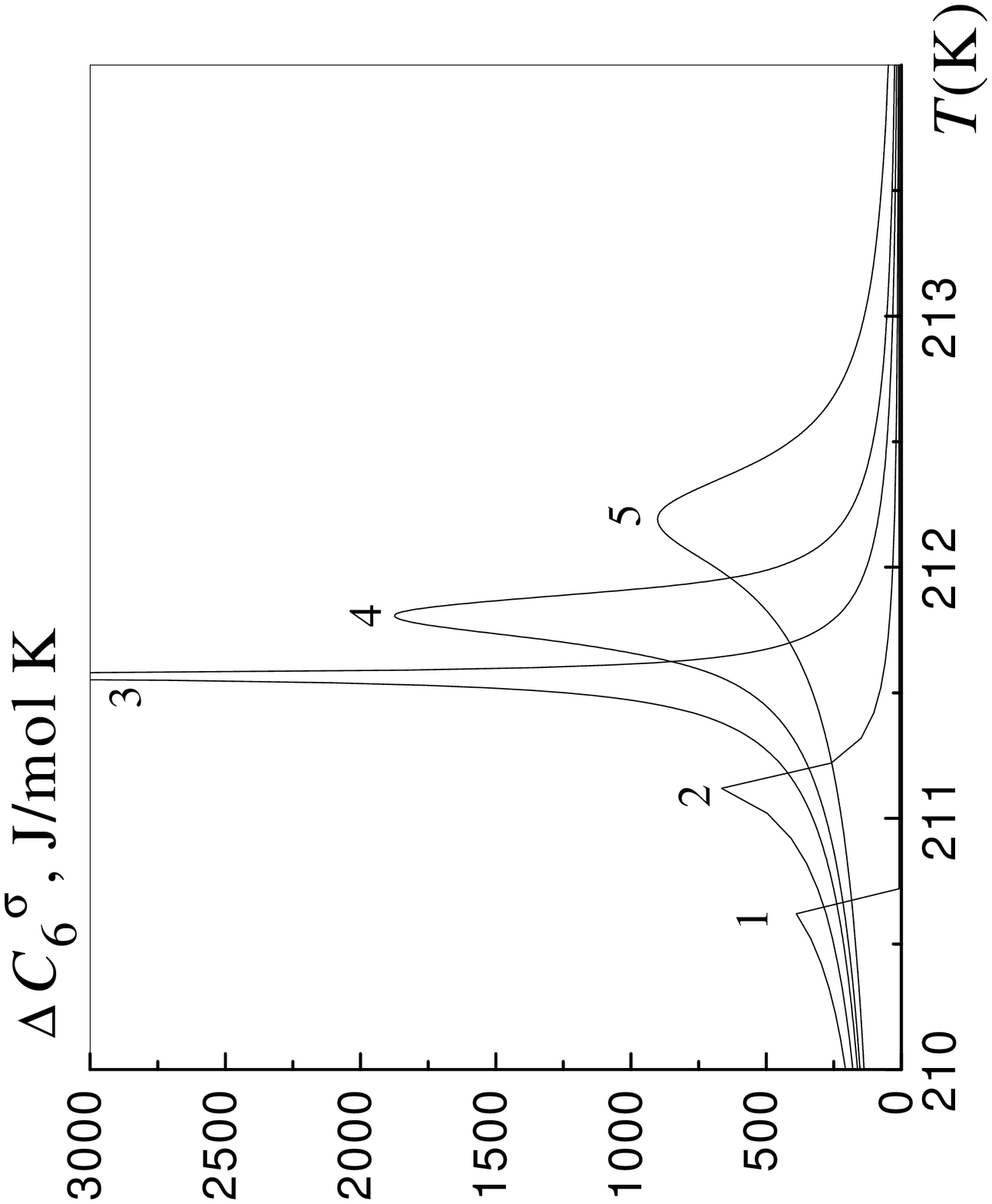}}
\hspace{0.5cm}
\epsfysize=5cm
\rotate[r]{\epsffile{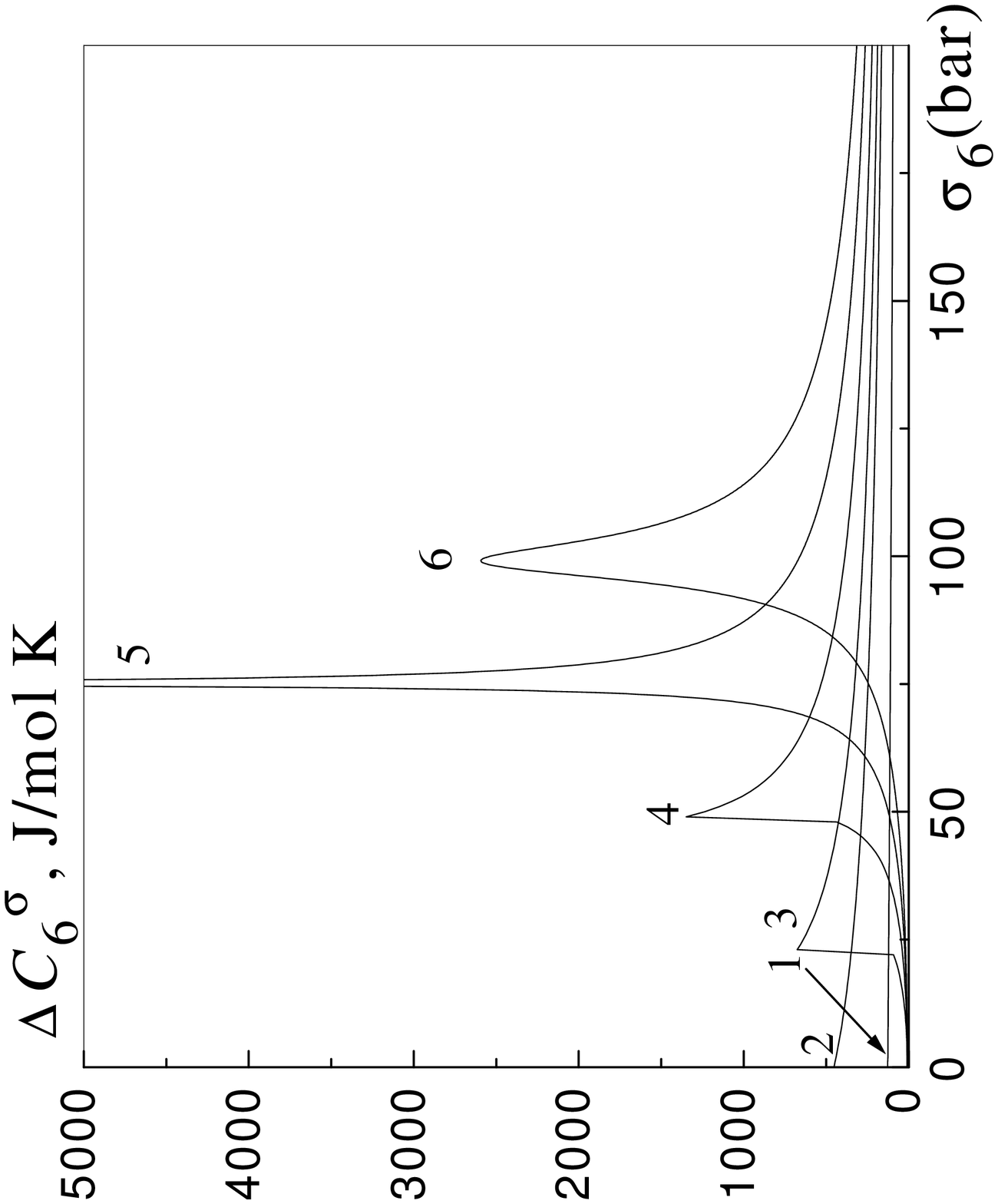}}
\end{center}
\caption{\small
Dependences of specific heat  $\Delta C_6^{\sigma}$ on
temperature at different values of stress $\sigma_6$ (bar) (left): 1 -- 0;
2 -- 40; 3 -- $\s^*=75$; 4 -- 100;    5 -- 150 and on stress $\sigma_6$ at
different values of temperature  $T$(K) (right):  1 -- 209.0; 2 -- $T_{\rm
C0}=210.7$; 3 -- 211; 4 -- 211.3; 5 -- $T^*=211.57$; 6 -- 211.8. }
\label{c} \end{figure}

In fig.\ref{c} we plot the dependence of  specific heat
of a deuteron subsystem of
K(H$_{0.11}$D$_{0.89}$)$_2$PO$_4$ on temperature at different values of
stress $\sigma_6$ and dependence of $\Delta C_6^\sigma$ on stress
$\sigma_6$ at different values of temperature.

In the ferroelectric phase at
$\sigma_6=0$, the magnitude of $\Delta C_6^\sigma$ increases as
temperature approaches $T_{\rm C}$, has a downward jump at the transition
point, and is nearly constant in the paraelectric phase.  On increasing
stress $\sigma_6$ up to $\sigma_6^*$, the maximum of $\Delta
C_6^\sigma$ increases, reaching its maximal value  at $\sigma_6 =
\sigma_6^*$, whereas the further increase in stress
lowers down the maximal value of $\Delta C_6^\sigma$ and shifts it to
higher temperatures.  Under stress
$\sigma_6$ the jump $\Delta C_6^\sigma$ is smeared out.

At temperatures lower than $T_{\rm C}$, the  $\Delta C_6^\sigma$
increases with stress
$\sigma_6$.  At $T < T_{\rm C}^*$  the magnitude of $\Delta C_6^\sigma$
first slightly increases with stress, and then at
stresses corresponding to the curve of phase equilibrium
it exhibits a sharp increase, diverging  at $T = T_{\rm
C}^*$ and $\sigma_6=\sigma_6^*$ $\Delta C_6^\sigma$.  The further increase
in stress gradually decreases $\Delta
C_6^\sigma$. At temperatures $T > T_{\rm C}^*$, the specific
heat $\Delta C_6^\sigma$ smoothly  increases with stress, reaches a
maximum and smoothly decreases.

Temperature and stress dependences of the pyroelectric
coefficient  $p_6^\sigma$ and  of the coefficient of linear
expansion $\alpha_6$ are analogous to those of the specific heat $\Delta
C_6^\sigma$.

\section{Concluding remarks}

In this paper we present the microscopic model for a description of
the stress $\s$ on the phase transition, static dielectric, elastic,
piezoelectric, and thermal properties of deuterated ferroelectrics of the
\dekdp-type. Unlike hydrostatic or uniaxial $p=-\sigma_3$
pressures, the stress $\s$ lowers the symmetry of the
high-temperature phase down to the symmetry of the low-temperature phase.
 As the electric field $E_3$, the stress $\s$ induces in \dekdp-type
 crystals the strain $\e$ and, due to the piezoelectric
 effect, polarization $P_3$.

In our previous papers we showed that an important role in dependences of
the transition temperature and dielectric characteristics of hydrogen
bonded crystals of the \kdp-family on pressures that do not change the
system symmetry is played by the corresponding changes in hydrogen bond
geometry, in particular, in the distance $\delta$ between possible
deuteron sites on a bond.  Most likely, the stress $\s$ does not
perceptibly affects $\delta$.  Instead, it alters the angle
between hydrogen bonds, perpendicular in an unstrained paraelectric
crystal, and distorts the PO$_4$ tetrahedra, splitting thereby energies of 
deuteron configurations.  Another important mechanism of the stress $\s$ 
influence on the phase transition and physical properties of the 
\dekdp-type ferroelectrics is the piezoelectric coupling, giving rise to 
effective fields, action of which, for the symmetry reasons, is equivalent 
to action of an external electric field applied along the ferroelectric 
axis.

As we show, the transition temperature increases with the
stress $\s$, and the order of the phase transition tends to the second
one. In the constructed phase diagram, which has the same topology as the
predicted in the phenomenologic approach $T_{\rm C}-E_3$ diagram, there
are two symmetric critical points where the second order phase
transitions take place, and the curves of phase equilibrium terminate. The
stresses above critical smear off the phase transition and smoothen the
temperature dependences of polarization and strain.  Correspondingly, an
increase in the stress $\s$ raises up the peak values of the physical
characteristics of a crystal having peculiarities in the transition
point (the longitudinal dielectric permittivity, compliance $s_{66}^E$,
piezomodules $d_{36}$ and $e_{36}$, and the specific heat). These
peak values are the highest at critical stresses, the higher stresses
smoothen the temperature dependences of these characteristics and lower
down the peaks.

Taking into account the piezoelectric effect in the developed model allows
one, at the proper choice of the theory parameters, to
quantitatively describe the available experimental data for the
temperature dependences of dielectric, elastic,
piezoelectric, and thermal characteristics of
unstrained \dekdp. Further experimental studies are required to
ascertain the values of the theory parameters and verify its predictions
about the form of the $T_{\rm C}-\s$ phase diagram
 and possible stress $\s$ effects on the physical characteristics of the
 crystals.

\section*{Acknowledgements}
This work was supported by the Foundation for
Fundamental Investigations of the Ukrainian Ministry in Affairs of Science
and Technology, project No 2.04/171.

\end{document}